\documentclass[12pt,letterpaper,epsf,epsfig]{article}

\usepackage{units}
\usepackage{cite}

\usepackage[numbers]{natbib}
\usepackage{graphicx}
\usepackage{amsmath}
\usepackage{amssymb}
\usepackage{afterpage}
\usepackage{graphics}
\usepackage{float}
\usepackage{lineno}
\usepackage{rotating}
\usepackage{xspace}
\usepackage{dcolumn}
\usepackage{bm} 

\usepackage[left=1.0in,top=1.0in,bottom=1.0in,right=1.0in]{geometry}

\title{\textbf{Opportunities for Neutrino Physics at the Spallation Neutron Source: A White Paper}}
\author{
A. Bolozdynya$^{1}$,
F. Cavanna$^{2}$,
Y. Efremenko$^{3,4}$,
G. T. Garvey$^{5}$, 
V. Gudkov$^{6}$, \\
A. Hatzikoutelis$^{3}$,
W. R. Hix$^{4,3}$,
W. C. Louis$^{5}$,  
J. M. Link$^{7}$, 
D. M. Markoff$^{8}$,\\
G. B. Mills$^{5}$,   
K. Patton$^{9}$,
H. Ray$^{10}$,
K. Scholberg$^{11}$,
R. G. Van de Water$^{5}$,\\ 
C. Virtue$^{12}$,
D. H. White$^{5}$,
S. Yen$^{13}$,
J. Yoo$^{14}$\\
\small
\centerline{\it $^{1}$National Research Nuclear University MEPhI, Moscow, 115409, Russia}\\
\small
\centerline{\it $^{2}$Universit\`{a} dell'Aquila and INFN, L'Aquila, Italy}\\
\small
\centerline{\it $^{3}$University of Tennessee, Knoxville, TN 37996, USA}\\ 
\small
\centerline{\it $^{4}$Oak Ridge National Laboratory, Oak Ridge, TN 37831, USA}\\ 
\small
\centerline{\it $^{5}$Los Alamos National Laboratory, Los Alamos, NM 87545, USA}\\ 
\small
\centerline{\it $^{6}$University of South Carolina, Columbia, SC 29208, USA}\\  
\small
\centerline{\it $^{7}$Center for Neutrino Physics, Virginia Tech, Blacksburg, VA 24061, USA }\\ 
\small
\centerline{\it $^{8}$North Carolina Central University, Durham, NC 27707, USA } \\  
\small
\centerline{\it  $^{9}$Physics Department, North Carolina State University, Raleigh, North Carolina 27695, USA} \\  
\small
\centerline{\it $^{10}$University of Florida, Gainesville, FL 32611, USA}\\ 
\small
\centerline{\it $^{11}$Duke University, Durham, NC 27708, USA}\\
\small
\centerline{\it $^{12}$Laurentian University, Sudbury, Ontario, P3E 2C6, Canada}\\
\small
\centerline{\it $^{13}$TRIUMF, Vancouver, British Columbia, Canada}\\
\small
\centerline{\it $^{14}$Fermi National Accelerator Laboratory, Batavia, IL 60510, USA}\\
}

\def\gtwid{\mathrel{\raise.3ex\hbox{$>$\kern-.75em\lower1ex\hbox{$\sim$}}}}
\def\ltwid{\mathrel{\raise.3ex\hbox{$<$\kern-.75em\lower1ex\hbox{$\sim$}}}}

\newcommand{\gsim}{\mathrel{\rlap{\raisebox{.3ex}{$>$}}
    \raisebox{-.6ex}{$\sim$}}}
\newcommand{\lsim}{\mathrel{\rlap{\raisebox{.3ex}{$<$}}
    \raisebox{-.6ex}{$\sim$}}}

\begin{document}

\maketitle

\begin{abstract}
The Spallation Neutron Source (SNS) at Oak Ridge National Laboratory, Tennessee, provides an intense flux of neutrinos in the few tens-of-MeV range, with a sharply-pulsed timing structure that is beneficial for background rejection.
In this document, the product of a workshop at the SNS in May 2012, we describe this free, high-quality stopped-pion neutrino source and outline various physics that could be done using it.  We describe without prioritization some specific experimental configurations that could address these physics topics.
\end{abstract}
\tableofcontents
\section{Executive Summary}

We describe here unique opportunities for physics using the neutrino source
provided by the Spallation Neutron Source (SNS) at Oak Ridge National Laboratory, many of which were discussed at a workshop held at the SNS in May 2012~\cite{snsworkshop}.
Although the SNS is designed as a neutron source,
neutrinos are produced as a free by-product, and this neutrino source is of 
exceptional quality.  The SNS protons on target produce numerous pions, which stop
in the target and decay at rest, yielding monochromatic 30 MeV $\nu_\mu$
from pion decay at rest, followed on a 2.2 $\mu$s timescale
by $\bar{\nu}_\mu$ and $\nu_e$ with a few tens of MeV from $\mu$ decay; there should
be very little contamination from decay-in-flight pions, and hence the neutrino spectral uncertainties are small.  Flavor content uncertainties are also small, as the fraction of neutrinos originating from $\pi^+$ is high.  The expected $\nu$ flux is $\sim$10$^7$ cm$^{-2}$s$^{-1}$ per flavor.
The short-pulse time structure is excellent for neutrino experiments, with 60~Hz of sub-$\mu$s pulses providing a 10$^{-3}$-10$^{-4}$ background rejection factor~\cite{Avignone:2003ep}.

\textbf{A rich program of physics is possible with such a stopped-pion $\nu$ source.}  One possibility is a search for sterile neutrino oscillation, a topic of recent interest which is highly motivated by recent results~\cite{Abazajian:2012ys, oscsns}.
Additional measurements, complementary to the sterile oscillation studies, are also possible (and may potentially share resources)~\cite{Efremenko:2008an}.

\begin{figure}[!htbp]
\centering
\includegraphics[height=2.8in]{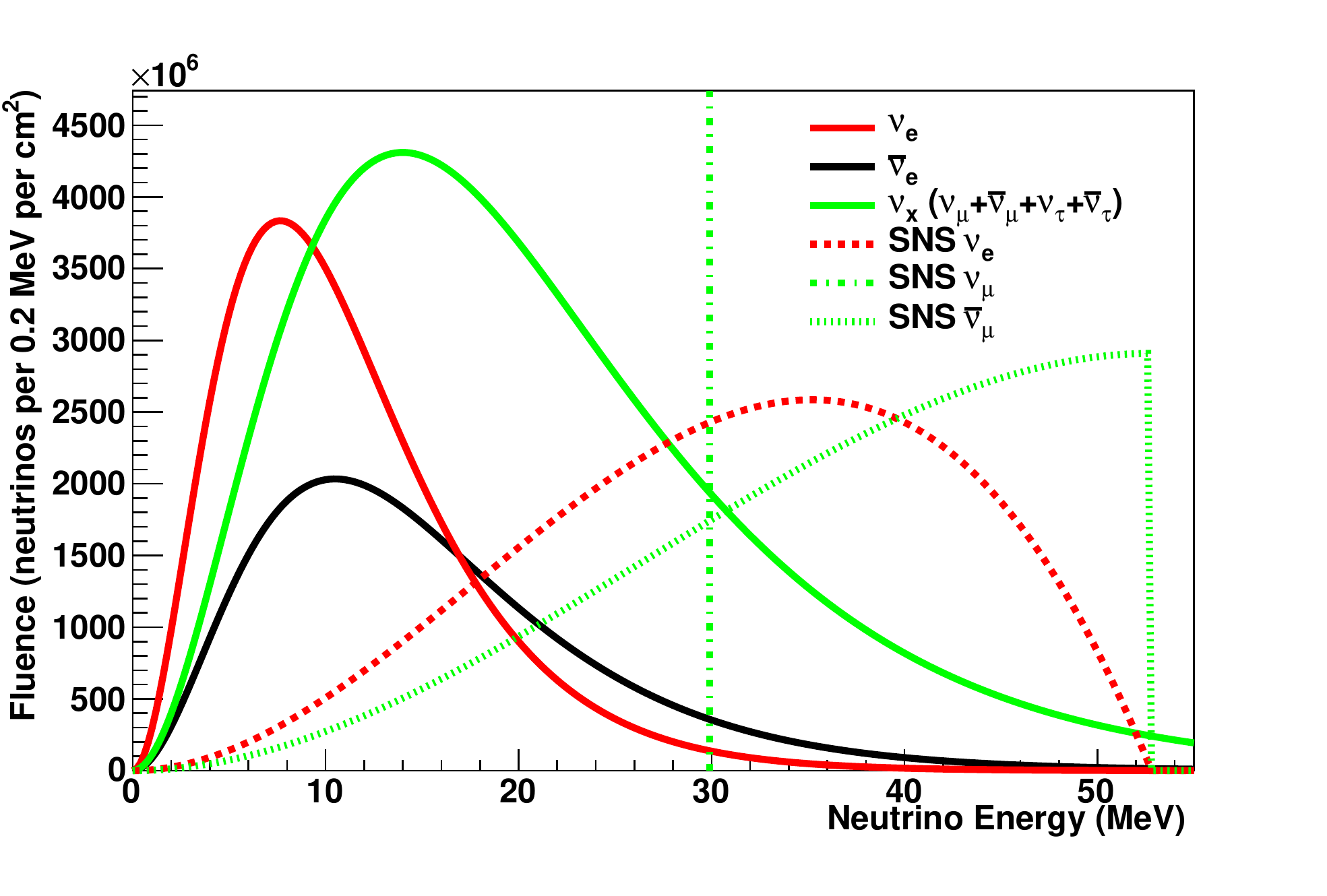}
\caption{Solid lines: typical expected supernova spectrum for different flavors; fluence integrated over the $\sim$15-second burst.  Dashed and dotted lines: SNS spectrum; integrated fluence for one day at 30~m from the SNS target.}
\label{fig:sn_sns}
\end{figure}

In particular, the proposed OscSNS experiment~\cite{oscsns} will directly test 
the Liquid Scintillator Neutrino Detector (LSND) $\bar \nu_\mu \rightarrow \bar \nu_e$ appearance signal and probe oscillation hypotheses involving sterile neutrinos with an 800-tonne scintillator detector. 
The SNS is also ideal for measurements of 
$\nu$-nucleus cross sections in the few tens-of-MeV range in a variety of targets relevant for supernova neutrino physics.  This territory is almost
completely unexplored: so far only $^{12}$C has been measured at the
$\pm 10$\% level~\cite{Auerbach:2001hz,Armbruster:1998gk}. 
The $\nu$ spectrum matches the expected supernova spectrum reasonably well (see Fig.~\ref{fig:sn_sns}); the slightly harder stopped-pion spectrum makes for higher event rates.

Understanding of $\nu$-nucleus interactions in this
regime is vital for understanding of supernovae: core-collapse
dynamics and supernova nucleosynthesis are highly sensitive to
$\nu$ processes.  Neutrino-nucleus cross section measurements will
furthermore enhance our ability to extract information about $\nu$
mixing properties (in particular, mass hierarchy) from the
observation of a Galactic supernova $\nu$ burst, via understanding
of both the supernova itself and of the $\nu$ detection processes.
A number of international neutrino detector collaborations would be ``customers'' for cross section measurements~\cite{Scholberg:2012id}.
The highest-priority targets for which to measure cross sections are 
argon (relevant for current and planned detectors like Icarus, MicroBooNE and the Long-Baseline Neutrino Experiment (LBNE)), lead
(relevant for the new Helium and Lead Observatory (HALO)~\cite{Duba:2008zz}), water (relevant for water Cherenkov detectors) and carbon (relevant for scintillator detectors).  
Such measurements have previously been proposed for the SNS~\cite{nusns}.

Another interesting possibility for a stopped-pion source is the detection of nuclear recoils
from coherent elastic $\nu$-nucleus scattering, which is within the
reach of the current generation of low-threshold detectors~\cite{Scholberg:2005qs}.  This reaction is also important
for supernova processes and detection.  This measurement also has excellent prospects for standard model (SM) tests; even a first-generation experiment has sensitivity beyond the current best limits on non-standard interactions of neutrinos and quarks~\cite{Scholberg:2009ha}.

There have in fact been no neutrino-nucleus interaction measurements in the tens-of-MeV energy regime in the past few years.  However, these measurements are more motivated than ever, given the huge recent progress in core-collapse simulation~\cite{Janka:2012wk}, and new prospects for large underground detectors for supernova neutrinos. 
There has also been as yet no successful detection of coherent neutrino-nucleus scattering, although dark-matter-style detectors, sensitive to low-energy recoils, 
have made enormous technical progress.

The next five years could see the first measurements of charged-current (CC) and neutral-current (NC) neutrino-nucleus interactions in several nuclei, the first detection of coherent elastic neutrino-nucleus scattering,  and new constraints
on (or discovery of) beyond-the-standard-model physics.
There are also interesting prospects for hidden sector experiments.

Beyond the next five years, we could pursue upgrades to the experiments listed above.  For supernova-relevant interactions, the list of potential targets will likely not be exhausted for measurements at better than the 10\% level.\footnote{We note that the occurrence of a nearby core-collapse supernova would increase the urgency for precision measurements of relevance for interpreting the signal.} 
There are furthermore possibilities to test the SM with precision cross section measurements. 
For coherent elastic scattering, second-generation experiments at the $\sim$tonne scale could probe nuclear physics, and potentially measure neutron density distributions.

There is no existing program making use of SNS neutrinos.  Some other similar stopped-pion sources are planned for the future, but none of these has the
combination of high intensity, short-pulse structure, and 
high fraction of decay-at-rest neutrinos. \textbf{The SNS is currently the world's best neutrino source of this nature and will likely remain so for at least a decade.}

The aim of this document is to describe possible physics opportunities at the SNS neutrino source, without prioritization.
In Section~\ref{source} we describe the properties of the SNS neutrino source.  In Section~\ref{physics}, we describe physics motivations and outline general experimental strategies for addressing this physics.  In Section~\ref{experiments}, we describe specific experimental configurations that represent opportunities for making use of the SNS neutrino source.  We note that opportunities are not limited to those described here.

\section{The SNS as a Neutrino Source}\label{source}

The SNS is the world's premier facility for neutron-scattering research, producing pulsed neutron beams with intensities an order of magnitude larger than any currently operating facility. With the full beam power, $10^{14}$ 1-GeV protons bombard the liquid mercury target in 700 ns wide bursts with a frequency of 60 Hz. Neutrons produced in spallation reactions with the mercury thermalize in hydrogenous moderators surrounding the target and are delivered to neutron-scattering instruments in the SNS experiment hall. 

\begin{figure}
\vspace{5mm}
\centering
\includegraphics[width=12cm]{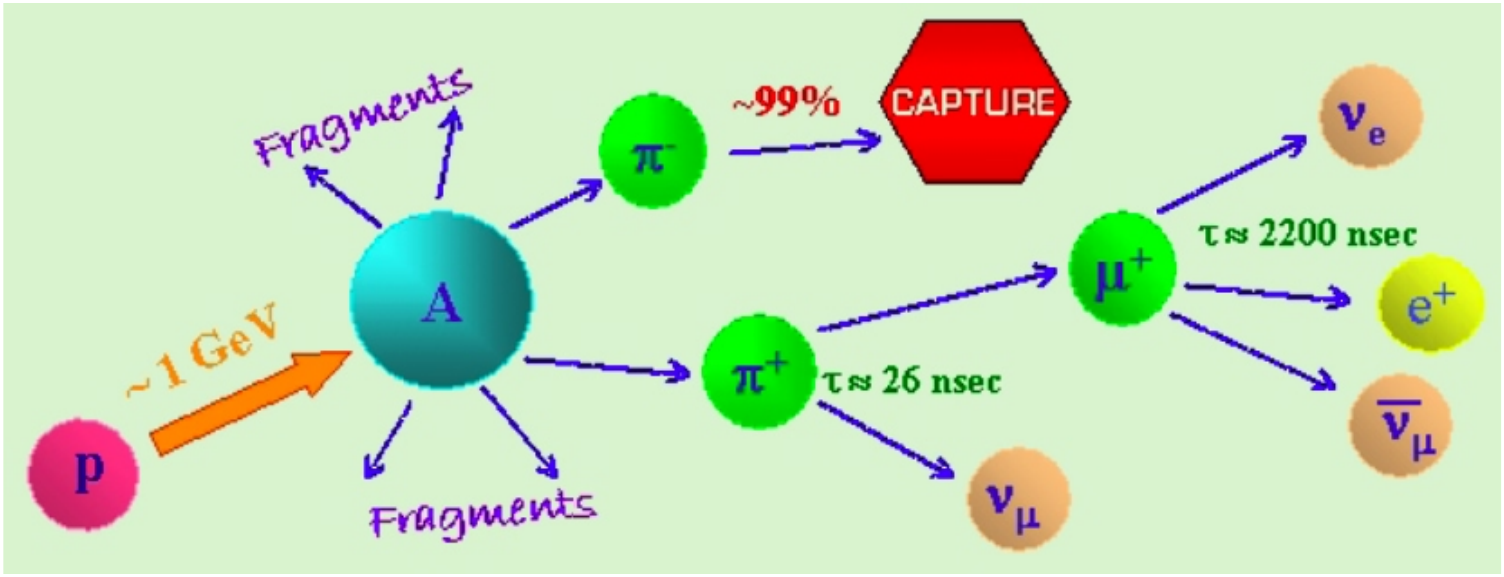}
\vspace{1mm}
\caption{
SNS neutrino production mechanism.
}
\label{fig:snsnu_cartoon}
\end{figure}

As a by-product, the SNS also provides the world's most intense pulsed source of neutrinos in the energy regime of interest for particle and nuclear astrophysics. Interactions of the proton beam in the mercury target produce mesons in addition to neutrons. These stop inside the dense mercury target and their subsequent decay chain, illustrated in Fig.~\ref{fig:snsnu_cartoon}, produces neutrinos with a flux of $\sim 2\times 10^7$ cm$^{-2}$s$^{-1}$ for all flavors at 20 m from the spallation target. This exceeds the neutrino flux at ISIS (where the KARMEN experiment was located) by more than an order of magnitude.

\begin{figure}
\vspace{5mm}
\centering
\includegraphics[width=17cm]{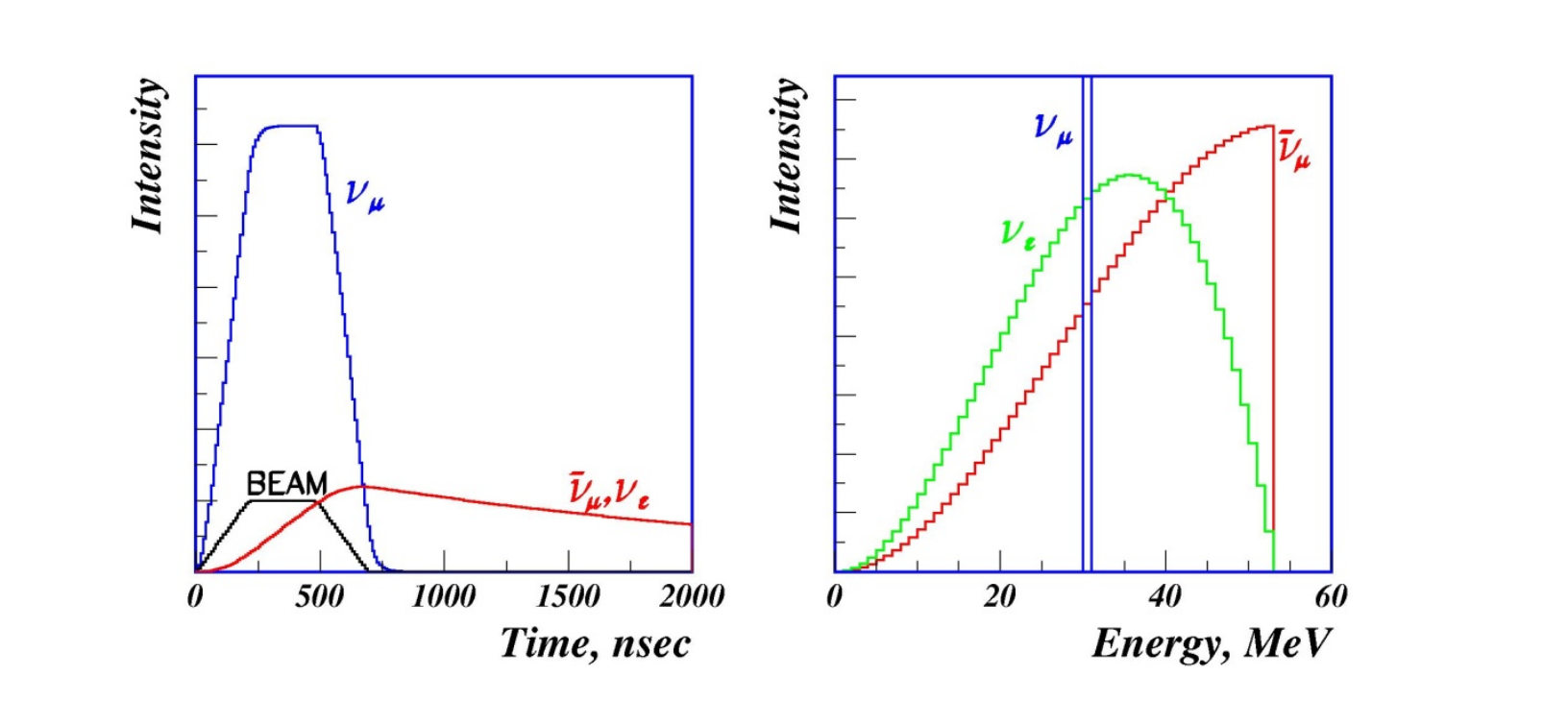}
\vspace{1mm}
\caption{
Time and energy distributions for the different neutrino flavors produced at the SNS.
}
\label{fig:sns_nuplots}
\end{figure}

The energy spectra of SNS neutrinos are shown in the right-hand panel of Fig.~\ref{fig:sns_nuplots}. These spectra are known because almost all neutrinos come from decay-at-rest processes in which the kinematics are well defined. The decay of stopped 
pions produces monoenergetic muon neutrinos at 30~MeV. The subsequent 3-body muon decay produces a spectrum of electron neutrinos and muon antineutrinos with energies up to 52.6~MeV.

\begin{figure}
\vspace{5mm}
\centering
\includegraphics[width=14cm]{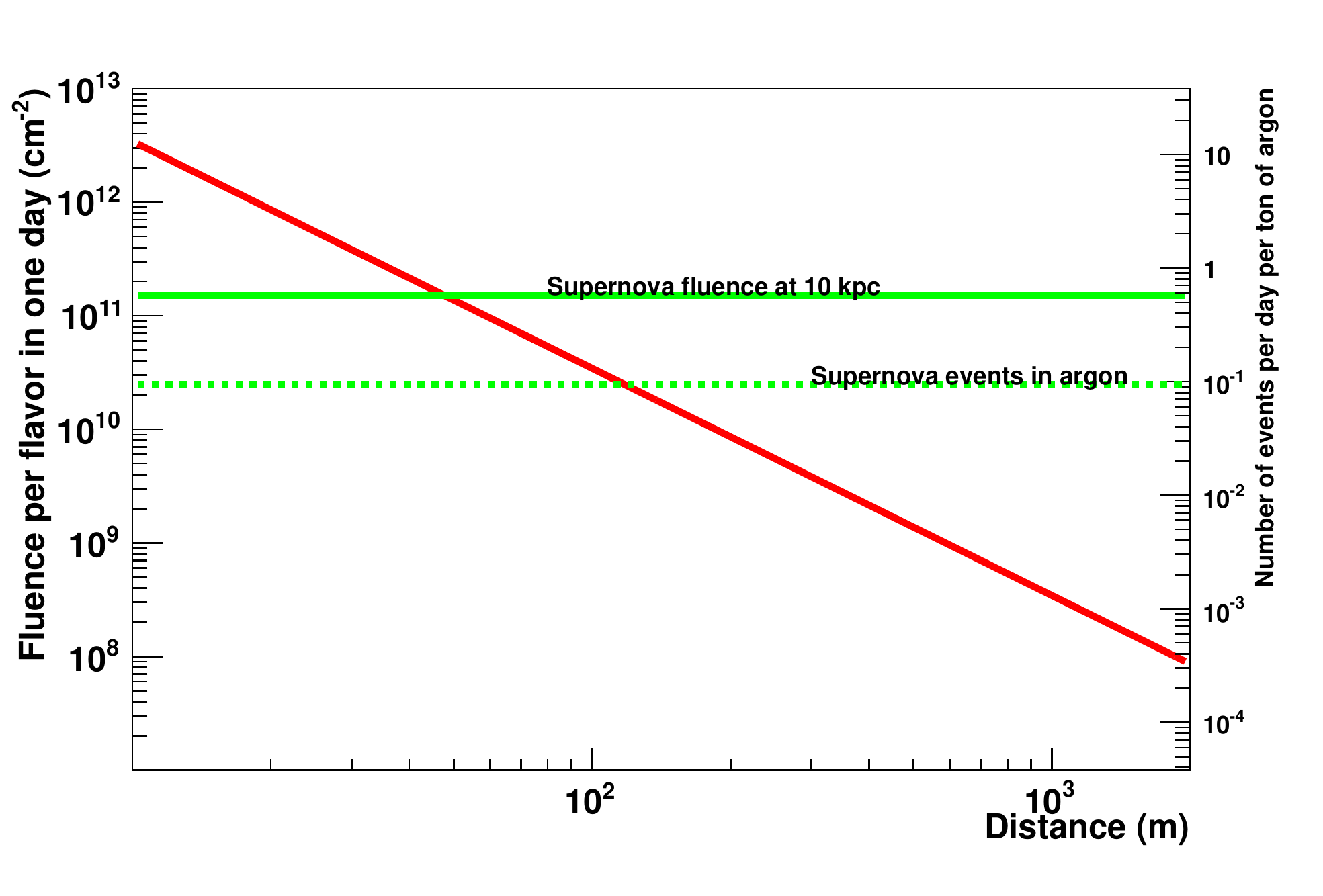}
\vspace{1mm}
\caption{
The red line shows integrated fluence per flavor in one day for the SNS neutrino flux as a function of distance from the source, according to the left axis scale.   
The axis on the right shows events per tonne of argon per day as a function of distance for the red line.  
The green solid line shows approximate  fluence per flavor for a supernova at 10 kpc for the full burst. 
The dashed line shows approximate events per tonne for a supernova at 10 kpc in argon.
}
\label{fig:fluence}
\end{figure}

The time structure of the SNS beam is particularly advantageous for neutrino studies. Time correlations between candidate events and the SNS proton beam pulse will greatly reduce background rates and may provide sensitivity to NC interactions. As shown in the left panel of Fig.~\ref{fig:sns_nuplots}, all neutrinos will arrive within several microseconds of the 60~Hz proton beam pulses. As a result, background events resulting from cosmic rays will be suppressed by a factor of $\sim$2000 by ignoring events which occur too long after a beam pulse. At the beginning of the beam spill the neutrino flux is dominated by muon neutrinos resulting from pion decay, in principle making it possible to isolate pure NC events, since the $\nu_\mu$ in the source have energies below CC threshold. 
Backgrounds from beam-induced 
high-energy neutrons are present, but can be mitigated by
appropriate siting and shielding.
We note that beam-induced neutron backgrounds for CC events
are greatly suppressed for $t\gsim$1~s after the start of the beam spill, while the neutrino production, governed by the muon lifetime ($\tau_\mu \sim$ 2.2 $\mu$s), proceeds for several microseconds. This time structure presents a great advantage over a long-duty-factor facility such as the Los Alamos Neutron Science Center (LANSCE), where the LSND experiment was located.  Figure~\ref{fig:fluence} shows the expected fluence at the SNS, compared to what would be expected for a nearby supernova.

For comparison, Table~\ref{tab:sources} lists characteristics of past, current and planned stopped-pion neutrino sources. In general, one wants high neutrino flux (with flux roughly proportional to proton beam power), sharp pulses to enable rejection of off-beam background, and well-understood neutrino spectra.   Ideally pulses should shorter than than the muon decay lifetime, and separated by at least several $\tau_\mu$.  Proton energies and target configuration resulting in a high fraction of pion decays at rest will lead to a clean decay-at-rest spectrum and well-known flavor composition.

The neutrino flux from the Booster Neutrino Beam (BNB) at Fermilab has a stopped-pion component very far off axis that is potentially usable~\cite{yoo}; however the BNB flux is much smaller than the SNS flux.
The J-PARC Material and Life Science Experimental Facility (MLF)  spallation source could potentially host an experiment~\cite{jparc,Ikeda,nishikawa}, although proton energies are higher, leading to contamination from neutrinos with non-pion-decay-at-rest parents and
therefore a less-well-understood neutrino spectrum. 
The planned European Spallation Neutron Source (ESS), for which there have been some discussions~\cite{3n2mp}, is also a possibility; the planned ESS power is favorably high, but timing is less desirable for neutrino physics. The DAE$\delta$ALUS program now under development also plans cyclotrons dedicated to producing stopped-pion neutrinos for physics including the topics described in this document~\cite{Alonso:2010fs}. However the high duty factors expected for DAE$\delta$ALUS would require additional background mitigation and surface experiments will be challenging.  The future Project X~\cite{Tschirhart:2011yu} program for Fermilab could potentially include stopped-pion neutrino sources, but not within this decade.  Overall, the SNS is the only facility that within the next decade can provide high-intensity, short duty factor, clean decay-at-rest neutrinos.   We note that a second SNS target station may eventually be built; this would provide additional flux, although timing characteristics are as yet unknown.

\begin{table}[!ht]
\begin{centering}
\begin{tabular}{||c|c|c|c|c|c|c||}\hline \hline
 Facility & Location  & Proton & Power & Bunch & Rate & Target\\
 & & Energy & & Structure &  & \\
 & & (GeV) & (MW) &  &  & \\ \hline \hline

LANSCE & USA (LANL) & 0.8  & 0.056  & 600 $\mu$s & 120~Hz  & Various \\ \hline
ISIS & UK (RAL) & 0.8   & 0.16 & 2 $\times$ 200 ns &  50 Hz  & Water-cooled \\  
 &  &   &  & &    & tantalum \\  \hline
BNB & USA (FNAL) & 8  & 0.032  & 1.6 $\mu$s & 5-11 Hz  & Beryllium \\\hline 
SNS & USA (ORNL) & 1.3  & 1  & 700 ns & 60 Hz  & Mercury \\\hline 
MLF & Japan (J-PARC) & 3  & 1  & 2 $\times$ 60-100 ns & 25 Hz  & Mercury \\ \hline
ESS & Sweden (planned) & 1.3  & 5  & 2 ms & 17 Hz  & Mercury \\ \hline
DAE$\delta$ALUS & TBD (planned) & 0.7 & $\sim 7 \times 1$  & 100 ms & 2 Hz  & Mercury \\
\hline \hline 
\end{tabular}
\caption{Characteristics of past, current and planned stopped-pion neutrino sources worldwide.}\label{tab:sources}
\end{centering}
\end{table}

\section{Physics Motivations}\label{physics}

In this section we outline some of the main physics motivations for experiments using the SNS neutrino source: these include measurements of neutrino interactions for astrophysics and SM tests, and also hidden sector experiments.
We describe in this section generically the types of experiments that could address these goals.  Specific experimental designs instantiating these types are described in Section~\ref{experiments}.

\subsection{Light Sterile Neutrinos and Neutrino Oscillations}\label{sterile}

A sterile neutrino is a lepton with no ordinary electroweak interactions except those induced by mixing.  Sterile neutrinos are present in most extensions of the SM and in principle can have any mass.  Very heavy sterile neutrinos are used in the minimal type I seesaw model~\cite{Minkowski:1977sc,Mohapatra:1979ia,Schechter:1980gr} and play a pivotal role in leptogenesis~\cite{Fukugita:1986hr,Davidson:2008bu}.  However, for terrestrial experiments such as those discussed here, we are primarily concerned with relatively light sterile neutrinos ($m_{\nu}<5$~eV) that mix significantly with ordinary (or active) neutrinos.   

The three known neutrinos ($\nu_e$, $\nu_{\mu}$ and $\nu_{\tau}$) have been observed to mix with each other.  This periodic mixing, also referred to as oscillations, is governed by a $3\times3$ unitary matrix which relates the underlying neutrino mass eigenstates to the known flavor eigenstates.  The probability of a neutrino created as one flavor being detected as another flavor is an oscillating function whose amplitude is a product of mixing matrix elements and whose frequency in $L/E$ (distance traveled divided by neutrino energy) space is proportional to the mass squared difference of the mass eigenstates, known as $\Delta m^2$.  In the SM there are known to be only three light active neutrino states.  This constraint comes from LEP's measurement of the invisible $Z^0$ width, which is consistent with exactly three neutrinos (with mass less that $m_Z/2$) that couple to the $Z^0$ boson~\cite{ALEPH:2005ab}.  

\begin{figure}
\vspace{5mm}
\centering
\includegraphics[width=9.5cm,clip=true]{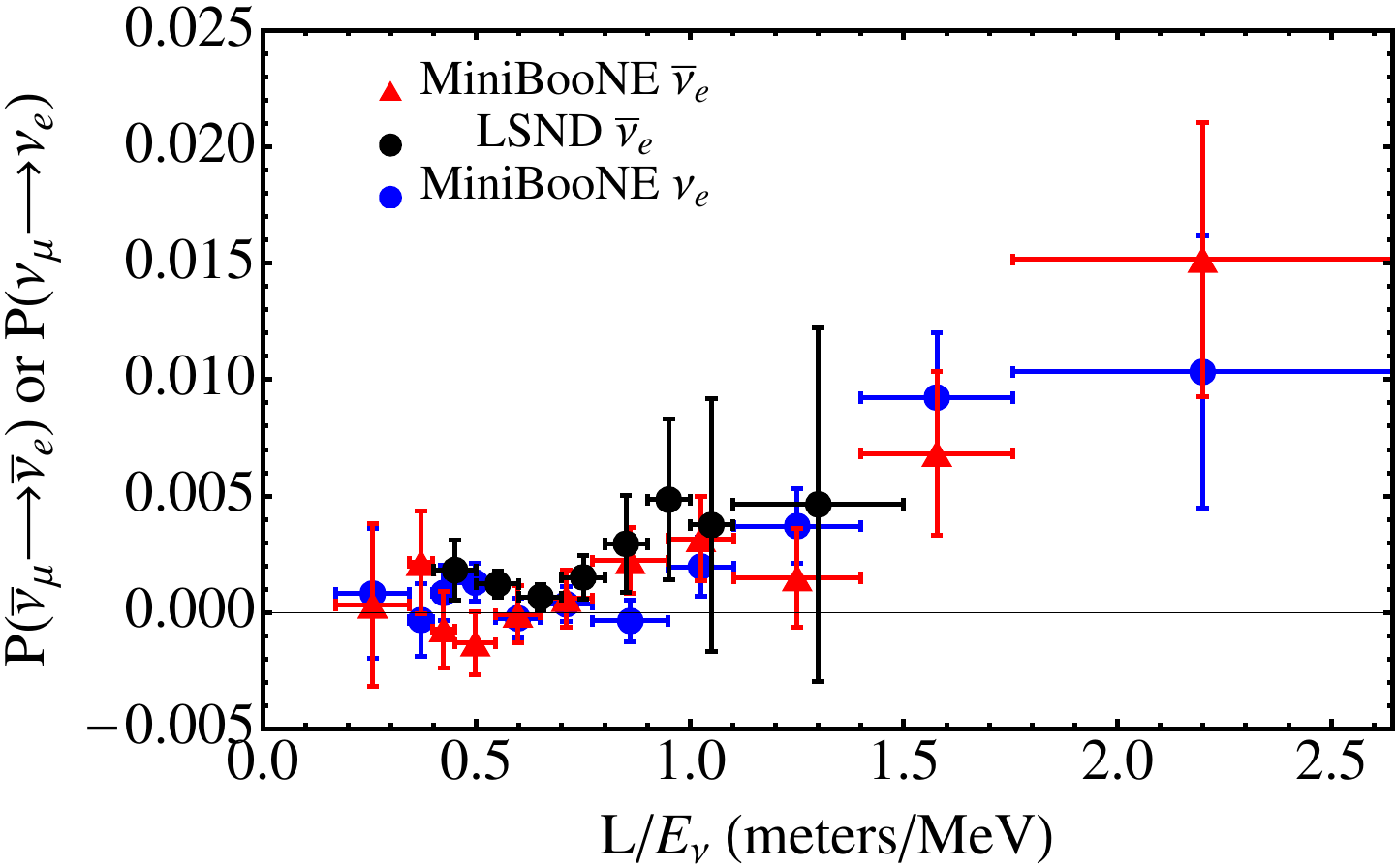}
\vspace{1mm}
\caption{
The probability of $\bar \nu_e$ or $\nu_e$ appearing in a $\bar \nu_\mu$ beam as a function of the 
neutrino proper time, for LSND and MiniBooNE signals.
}
\label{L_E}
\end{figure}

\begin{figure}
\vspace{5mm}
\centering
\includegraphics[width=8cm,angle=90,clip=true]{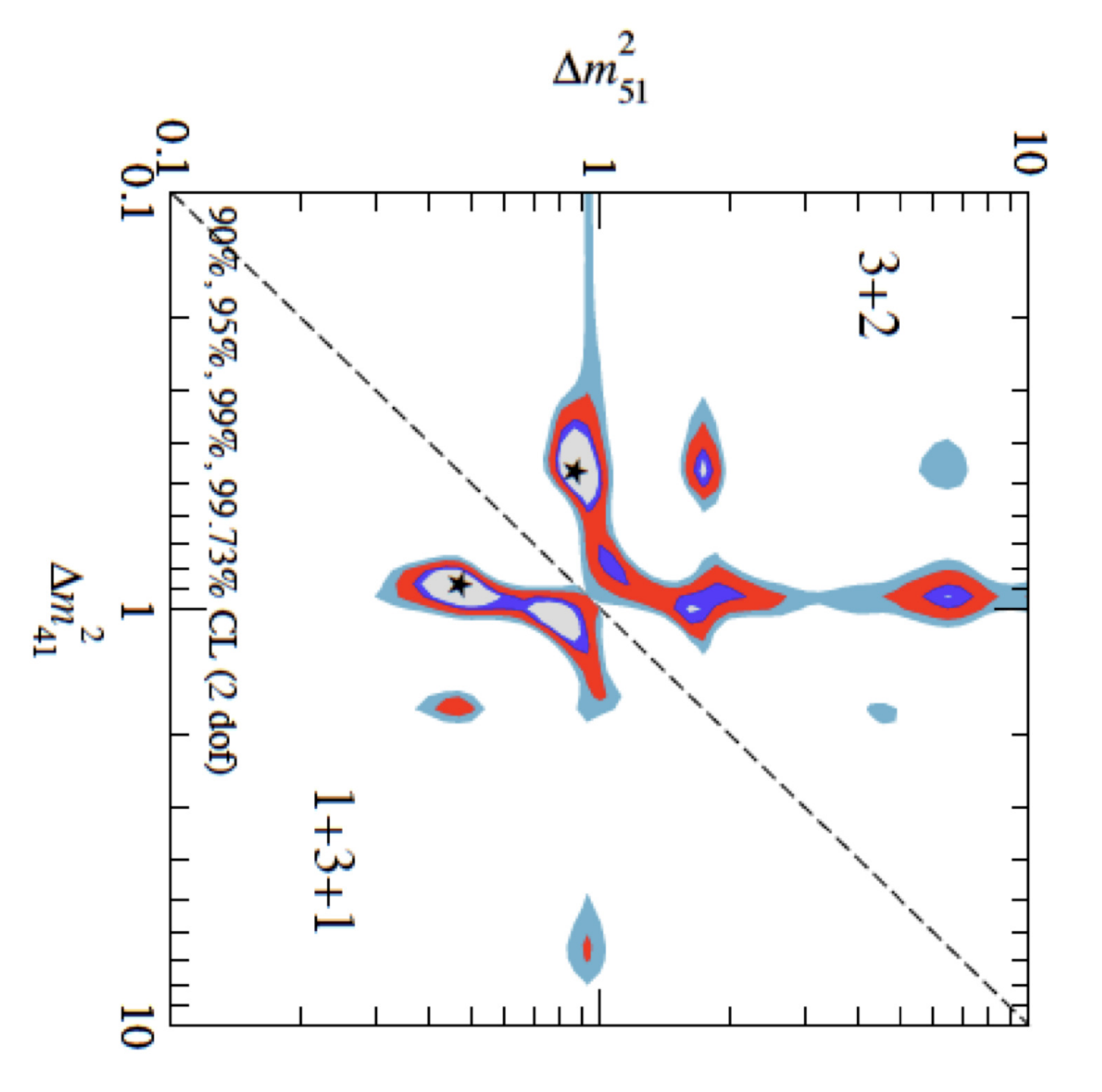}
\vspace{1mm}
\caption{
A global fit to the world
neutrino plus antineutrino data indicates that the world data fit reasonably well to a 3+2 model with
three active neutrinos plus two sterile neutrinos. The global fit is from reference~\cite{Kopp:2011qd}.
}
\label{global}
\end{figure}

Three non-degenerate mass eigenstates allow for two independent $\Delta m^2$ scales, and in fact there are well-established neutrino oscillations at two $\Delta m^2$ scales: the solar scale with $\Delta m^2\simeq8\times10^{-5}\ {\rm eV}^2$ and the atmospheric scale with $\Delta m^2\simeq2.4\times10^{-3}\ {\rm eV}^2$~\cite{Nakamura:2010zzi}.  The question of sterile neutrinos arises because there are persistent hints of a third $\Delta m^2$ scale of order 1~eV$^2$.  A third $\Delta m^2$ scale implies the existence of a fourth neutrino, but because of the LEP determination that only three light neutrinos couple to the $Z^0$ boson, any fourth neutrino must be sterile.  

The original evidence for the third $\Delta m^2$ scale comes from an experiment known as LSND~\cite{Athanassopoulos:1995iw,Aguilar:2001ty}, which observed an excess of $\bar{\nu}_e$ in a $\bar{\nu}_{\mu}$ beam.  LSND used a stopped-$\pi$ beam, in which $\pi^+$ mesons decay at rest producing isotropic fluxes of $\nu_{\mu}$, $\bar{\nu}_{\mu}$ and $\nu_e$, but no $\bar{\nu}_e$.  Then the detection of $\bar{\nu}_e$ in the detector, located 30 meters downstream from the beam stop, is taken as evidence of neutrino oscillations.  The $\bar{\nu}_e$ are detected through the inverse beta decay (IBD) channel, a ``golden mode'' process in which a prompt positron is detected followed by the delayed capture of a neutron on hydrogen.  The detection of this coincident signal unambiguously tags the event as a $\bar{\nu}_e$.  The LSND oscillation signal was strong, but not definitive, at 3.8$\sigma$ significance.    

The MiniBooNE experiment at Fermilab was designed to test the LSND oscillation hypothesis. MiniBooNE used a horn-focused $\pi$ decay-in-flight beam which provided a higher-energy $\nu_{\mu}$ (or $\bar{\nu}_{\mu}$ depending on the direction of the horn-focusing current), to look for $\nu_e$ (or $\bar{\nu}_e$) appearance. The MiniBooNE detector was located about 500~m downstream of 
the pion production target, such that with the higher energy and longer distance the $L/E$ of  MiniBooNE was a match to LSND. At the higher neutrino energy, MiniBooNE lacked a clean golden mode interaction for $\nu_e$ appearance like IBD, but it was able to search for 
$\nu_e$ appearance in both neutrino and antineutrino modes as well as $\nu_{\mu}$ disappearance in both modes. From an analysis of the combined $\nu_e$ and $\bar \nu_e$ appearance data from $6.46 \times 10^{20}$ protons on target in neutrino mode and $11.27 \times 10^{20}$ protons
on target in antineutrino mode, a total excess of $240.3 \pm 34.5 \pm 52.6$ events ($3.8 \sigma$) is observed in the energy range $200<E_\nu^{QE}<1250$~MeV~\cite{AguilarArevalo:2008rc,AguilarArevalo:2010wv,AguilarArevalo:2012va}.  The data are consistent 
with neutrino oscillations in the $0.01 < \Delta m^2 < 1.0$ eV$^2$ range and are also consistent with the evidence for antineutrino oscillations from LSND. In the 
disappearance channels, MiniBooNE saw no evidence of oscillations, but this result can still be consistent with LSND~\cite{AguilarArevalo:2009yj,Mahn:2011ea}.  
Figure~\ref{L_E} 
shows the 
$L/E$ (neutrino proper time) dependence of $\bar \nu_\mu \rightarrow \bar \nu_e$ oscillation probability from LSND and MiniBooNE. 
The correspondence between the two experiments is striking. \footnote{A recent result from ICARUS~\cite{Antonello:2012pq} constrains high $\Delta m^2$ values in a two-flavor fit.}
Fig.~\ref{global} shows fits to the 2011 world 
neutrino plus antineutrino data that indicate that the world data fit reasonably well to a 3+2 model with
three active neutrinos plus two sterile neutrinos.

Other evidence for high $\Delta m^2$ oscillations comes in the form of possible $\nu_e$ and $\bar{\nu}_e$ disappearance.  The first was a observation~\cite{Giunti:2006bj,Giunti:2010zu} that the rate of $\nu_e$ CC interactions on $^{71}$Ga observed during terrestrial source phases of the solar radiochemical experiments GALLEX and SAGE was 15-25\% below the expectation from theory.  In order to ``calibrate" the $\nu_e\,^{71}$Ge cross section, each of these experiments ran with mega-Curie-scale neutrino sources ($^{51}$Cr and $^{37}$Ar),  and the deficit of observed interactions to theory is now known as the ``gallium anomaly".  In a similar vein, new calculations of $\bar{\nu}_e$ fluxes from nuclear reactors~\cite{Mueller:2011nm,Huber:2011wv} seem to indicate that the rate of neutrino interactions observed by all short-baseline reactor experiments is systematically low by about 7\%~\cite{Mention:2011rk}.  This observation is referred to as the ``reactor anomaly".  Combined, the gallium and reactor anomalies provide about 3$\sigma$ evidence for $\nu_e$ disappearance, but both have potentially large theoretical uncertainties which are hard to address.

Outside particle physics, there is evidence for a fourth neutrino state coming from fits to large scale structure, the cosmic microwave background and Big-Bang nucleosynthesis.  These phenomena are sensitive to the energy density of neutrinos in the early universe which is itself proportional to the number of neutrino families.  These fits are sensitive to the number of light degrees of freedom, so the preference for a fourth neutrino type could actually be pointing to a different light state such as an axion.  Also, these fits have large parameter degeneracies which allow parameters to be tweaked to fit three neutrinos.

The SNS as an intense proton beam on a thick target is a spectacular source of stopped $\pi^+$, and therefore is an excellent place to conduct a direct test of the LSND effect.  The LSND signal is the first and most significant evidence for the 1~eV$^2$ $\Delta m^2$ scale and as such calls out for a direct test.  The proposed experiment, known as OscSNS, described in Section~\ref{oscsns}, is perhaps the best way to do a direct test of LSND, where ``by direct test'' means without any $L/E$ scaling which {\it a priori} assumes the simplest neutrino oscillation model involving a fourth, sterile, neutrino flavor.

For a comprehensive review of light sterile neutrinos, see Ref.~\cite{Abazajian:2012ys} and references therein.  

\subsection{Neutrino Interaction Cross Sections}

We organize the discussion of neutrino cross section measurements into two categories:  CC and NC interactions which produce products above about 1~MeV that can be detected using standard neutrino detection techniques, and coherent elastic scattering experiments, which produce very low-energy recoils requiring specialized recoil detection techniques.

\subsubsection{Charged- and Neutral-Current Cross Sections}\label{ccncphys}

A rich program of measurements of neutrino-nucleus interactions in the tens-of-MeV regime is possible with a stopped-pion $\nu$ source, including measurement of 
$\nu$-nucleus cross sections.  Measurements of cross sections and interaction-product
angular and energy distributions are desirable for both CC and NC interactions in a variety of nuclear targets. This territory is almost
completely unexplored: so far only $^{12}$C has been measured at the
10\% level~\cite{Auerbach:2001hz,Armbruster:1998gk}.
Information about these interactions is applicable in a number of contexts.  The following section 
details some of the motivations.  These fall into two main categories: understanding of core-collapse supernovae (process and detection), and SM tests.

\begin{itemize}
\item \textbf{Core-Collapse Supernovae}:
At the end of its life, every massive star (larger than 8-10 solar masses) finds itself with a core composed of iron, nickel and neighboring elements.  In earlier stages in its life,  fusion reactions in the core provide the energy to support the core against gravity; however fusion of iron costs the star energy.  Once the iron core grows too large to be supported by the pressure of degenerate electrons, the core begins to collapse.  The collapse proceeds to supernuclear densities, at which point the core becomes incompressible, rebounds, and launches a shock wave that will ultimately disrupt the star. The core is now a proto-neutron star, radiating away its $10^{46}$~J of binding energy in the form of neutrinos of all flavors at a staggering rate of $10^{57}$ neutrinos per second and $10^{45}$~W.  The effects of this  extremely intense neutrino source are of great importance to the supernova.

Foremost, over the course of several hundred milliseconds, the intense neutrino flux revives the shock wave, that had stalled due to several enervating processes, including nuclear dissociation and neutrino emission.  The revived shock then propagates through the envelope of the star, producing a visual display of $10^{42}$~J which we call a supernova.  These core-collapse supernovae are among the most energetic explosions in our universe, with the supernova shock imparting $10^{44}$~J of kinetic energy as it disrupts nearly the entire massive star and disseminates into the interstellar medium many of the elements in the periodic table heavier than hydrogen and helium. Core-collapse supernovae are therefore a key link in our chain of origins from the Big Bang to the formation of life on Earth and serve as laboratories for physics beyond the SM and for matter at extremes of density, temperature, and neutronization that cannot be reproduced in terrestrial laboratories. 

Beyond re-energizing the supernova explosion, the intense neutrino flux from the proto-neutron star has a variety of important interactions with matter, both in the supernova's ejecta and here on Earth. The effects of these interactions depend on the properties of the neutrinos themselves, their spectra and flavors, and their interaction rates with matter.  Observations of solar and atmospheric neutrinos have shown that neutrinos have tiny rest masses and, therefore, can transform their flavors quantum mechanically.  If such oscillations convert $\mu$ or $\tau$ neutrinos into electron type, this hardens the electron neutrino spectra and increases the efficiency of energy deposition.  Studies of matter-enhanced oscillation by the Mikheyev-Smirnov-Wolftenstein (MSW) effect (see, \textit{e.g.}~\cite{QiFu95b}) show an impact on the terrestrial detection of the supernova signal, but no effect on the mechanism itself.  More recent studies of coherent oscillations (see, \textit{e.g.},~\cite{DuFuQi06,HaRaSi06,Duan:2010bg}) raise the possibility of a strong spectral swap or split that is sensitive to the neutrino mixing angles and to the neutrino mass hierarchy. If this swap or split occurs, well-resolved supernova signals could be used to constrain these properties of the neutrinos.

Neutrino-nucleus cross sections of relevance to supernova astrophysics can be grouped into three categories, those that affect (1) supernova dynamics, (2) supernova nucleosynthesis, and (3) terrestrial supernova neutrino detection, each of which would benefit from experimental study.  In some cases the neutrino capture rate is of direct relevance. In other cases, we use a neutrino capture to study its inverse reaction, electron capture, because time-reversal invariance provides the rate for the inverse from a measurement of the rate for the forward reaction. The SNS produces $10^{15}$ neutrinos per second making it the single most intense source of neutrinos on Earth in this energy range. Most important, the spectrum of supernova neutrinos and the decay-at-rest neutrinos produced at the SNS overlap significantly: see Fig.~\ref{fig:sn_sns}.
The availability of such an intense neutrino source with neutrino energy spectra matching those emanating from distant supernovae seems ``made to order'' for neutrino-nuclear astrophysics research. The combination of such well-matched and intense neutrino fluxes with the spotlight on neutrinos for supernova science makes a compelling case for a neutrino-nuclear astrophysics research program to be developed around the SNS neutrinos.

\noindent \textit{Supernova Dynamics:} 
For more than 20 years, following the realization that the collapsing core was dominated by heavy nuclei that become progressively heavier during the collapse~\cite{LaLaPe78,BeBrAp79}, supernova models used naive electron capture rates based on a simple independent-particle shell model (IPM) for the nuclei in the stellar core~\cite{Brue85}.  In reference~\cite{Full82}, it was realized that electron capture on heavy nuclei would soon be quenched in this picture, as neutron numbers approach 40, filling the neutron $f_{5/2}$ orbital. Calculations using IPM showed that neither thermal excitations nor forbidden transitions substantially alleviate this blocking~\cite{Full82,CoWa84}.  Full shell model diagonalization calculations remain impossible in this regime due to the large number of available levels in the combined $fp + gds$ system~\cite{LaMa00}, but approximate schemes to calculate these rates have been developed~\cite{LaKoDe01}. With one such approach~\cite{LaMaSa03}, electron capture rates calculated for a sample of nuclei with masses from 66 to 112 demonstrated that though the electron capture rates for individual heavy nuclei are smaller than that on protons, they are large enough that capture on the much more abundant heavy nuclei dominates the capture on protons throughout core collapse.  Supernova models using these rates~\cite{HiMeMe03} have demonstrated unequivocally that electron capture on nuclei plays a dominant role in dictating the dynamics of stellar core collapse, which sets the stage for all of supernova dynamics that occurs after stellar core bounce and the formation of the supernova shock wave. The launch radius of the supernova shock wave after stellar core bounce and the stellar core profiles in density, temperature, and composition, are all significantly altered. These differences have ramifications for both supernova dynamics and supernova element synthesis.

\noindent \textit{Supernova Nucleosynthesis:} 
The largest impact of neutrinos on supernova nucleosynthesis occurs deep in the explosion, where the relative neutrino and antineutrino fluxes determine the neutron-richness of the matter.  While this matter is initially dissociated into free nucleons and $\alpha$-particles, as it expands and cools, the nucleons recombine to form iron, nickel and neighboring species.  The specific isotopes of the heavy elements that form are governed by the neutron richness.  Modern simulations point to the neutrinos creating a proton-rich environment \cite{FrHaLi06,PrWoBu05} and \cite{FrMaLi06} (see also \cite{PrHoWo06}) discovered a neutrino-driven flow to proton-rich nuclei above A=64, now called the $\nu$\emph{p}-process.  While the neutrino impact comes largely from captures on free nucleons, the more poorly-known $\nu$ and $\rm e^{\pm}$ interactions with heavy nuclei also contribute \cite{FrMaLi06}.  Still further from the supernova's center, references~\cite{WoHaHo90} and \cite{HeKoHa05} have shown that appreciable amounts of some rarer isotopes can be produced by the combination of neutrino spallation reactions and shock heating.  This includes several isotopes, like $^{19}${F}, $^{138}$La, and $^{180}$Ta, for which this $\nu$-process could be the dominant production mechanism.  

The astrophysical r-process (rapid neutron capture process) is responsible for roughly half of the Solar System's supply of elements heavier than iron, with the remainder originating from the s-process occurring in asymptotic giant branch stars.  While the nuclear conditions necessary to produce the r-process are well established (see, \textit{e.g.}, \cite{KrBiTh93}), the astrophysical site remains uncertain, but most leading candidates occur within a significant neutrino flux. Reference~\cite{QiHaLa97} demonstrated that neutrino-induced reactions can significantly alter the r-process path and hence its abundance yields. In the presence of a strong neutrino flux, $\nu_{\rm e}$ captures on the waiting point nuclei at the magic neutron numbers might compete with $\beta$ decays and speed up passage through these bottlenecks. Neutrinos can also inelastically scatter on r-process nuclei via $\nu_{e}$-induced CC reactions and $\nu$-induced NC reactions, leaving the nuclei in excited states that subsequently decay via the emission of one or more neutrons.  This processing may, for example, alter the shape of the prominent r-process abundance peaks \cite{HaLaQi97}.  

\noindent \textit{Terrestrial Supernova Neutrino Detection:} 
The ability to detect, understand, and ultimately use the detailed neutrino light curve from a future core-collapse supernova in our galaxy is integral to both better understanding supernovae and, in the end, to using precision supernova models together with detailed astronomical observations to constrain fundamental physics that would otherwise be inaccessible in terrestrial experiments. 

The neutrino burst from a core-collapse supernova arriving at Earth will contain neutrinos of all flavors
with energies in the few tens of MeV range.
Because of their weak interactions, the neutrinos are
able to escape on a timescale of a few tens of seconds after core collapse
(the promptness enabling a supernova early warning for astronomers~\cite{Antonioli:2004zb}).  
An initial sharp ``neutronization burst'' of $\nu_e$ 
(representing about 1\% of the total
signal) is expected at the outset, from $p+e^- \rightarrow n + \nu_e$.
Subsequent
neutrino flux comes from NC $\nu \bar{\nu}$ pair production.
Electron neutrinos have the most interactions with the 
proto-neutron star core; $\bar{\nu}_e$ have fewer, because neutrons
dominate in the core; $\nu_\mu$ and 
$\nu_\tau$ have yet fewer, because NC interactions dominate for these. 
The fewer
the interactions, the deeper inside the proto-neutron star the neutrinos
decouple; and the deeper, the hotter.  So one expects generally
a flavor-energy hierarchy, $\langle
E_{\nu_{\mu,\tau}}\rangle > \langle E_{\bar{\nu}_e} \rangle > \langle
E_{\nu_e} \rangle$.   As mentioned previously, MSW oscillation and collective effects may create significant imprints on the signal.
To probe experimentally these effects,  both for core-collapse physics studies and to learn about neutrino properties, measurements giving information about the neutrino spectra for different flavors are essential.
So far the only supernova neutrino observation
is from SN1987A,
and we expect
enormously enhanced information from the next nearby observation.
The detection of supernova neutrinos is reviewed in reference~\cite{Scholberg:2012id}.
Converting an observed neutrino flux into a luminosity requires knowledge of the neutrino-nucleus cross sections for the detector material. For any nearby supernova, astronomical observations will yield distances measured to 10\% or better (see \textit{e.g.}~\cite{ScKiEa94}).  This same level of accuracy should be the goal of any future measurements of neutrino-nucleus cross sections, to prevent the neutrino-nucleus uncertainty dominating the uncertainty in the supernova's luminosity.  Similar constraints apply if we are to make best use of the detection of the diffuse supernova neutrino background, \textit{i.e.} the flux from all supernovae that have occurred in the history of the universe (see, \textit{e.g.}~\cite{Lunardini:2010ab}). From deuterium to lead, a number of nuclei have been proposed (and, in some cases, used) as detector material. In all cases, accurate neutrino-nucleus cross sections are essential.

\begin{figure}[h]
\centering
\includegraphics[width=14cm]{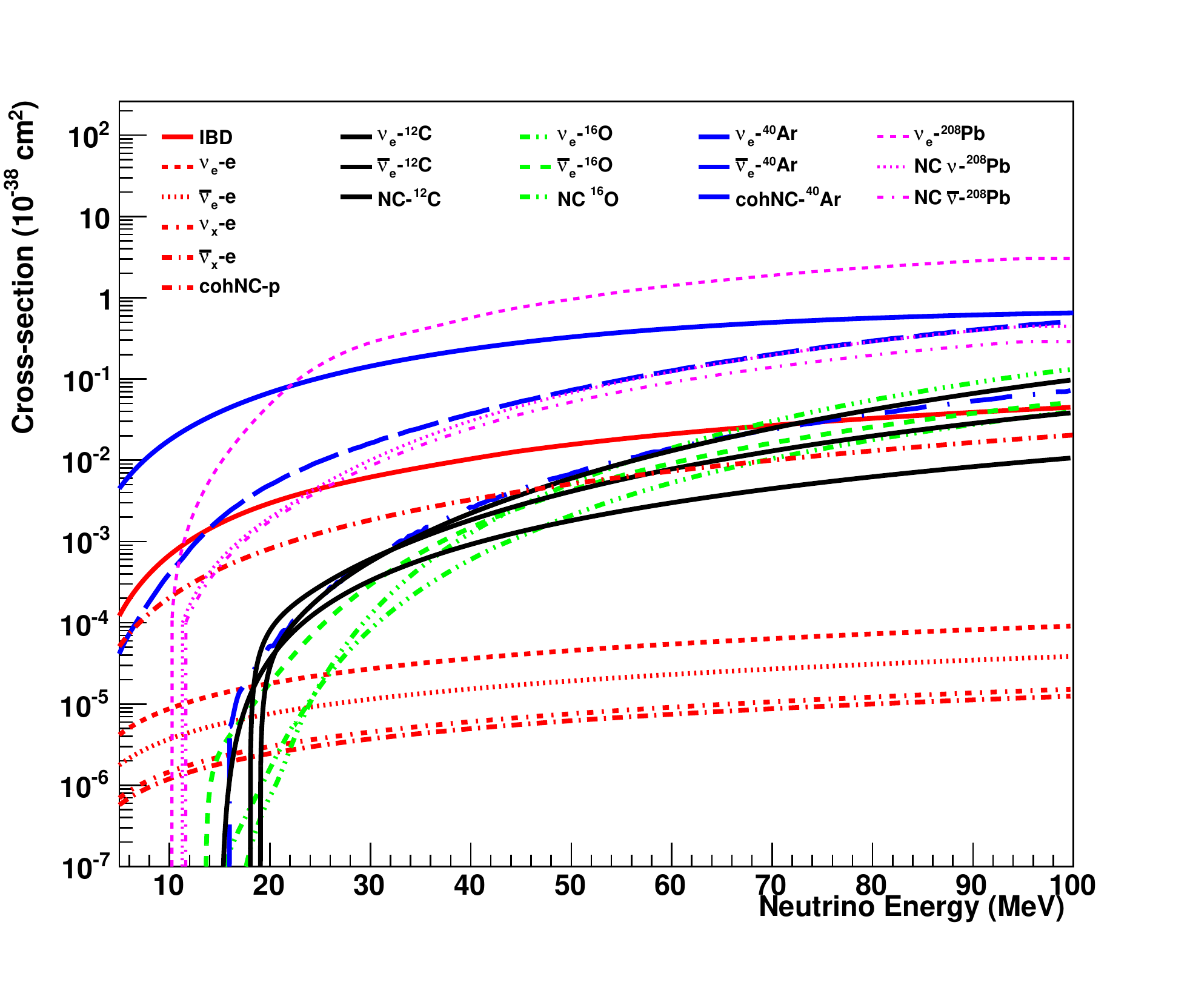}

\caption{Cross sections per target for relevant interactions: see~\cite{snowglobes}  for references for each cross section plotted.}
\label{fig:xscns_all}
\end{figure}

Currently running detectors have sensitivity primarily to the $\bar{\nu}_e$
component of the signal, via IBD,
$\bar{\nu}_e + p \rightarrow n + e^+$, which is the dominant component 
for water Cherenkov and scintillator detectors. The IBD cross section is rather well understood, as is elastic scattering on electrons.  However detectors with broader flavor sensitivity are coming online now and in the near future.
An example is liquid argon (LAr),  proposed for several LAr Time Projection Chamber (LArTPC) detectors such as MicroBooNE and LBNE.  Argon offers unique $\nu_e$ sensitivity and therefore enhanced physics reach~\cite{Akiri:2011dv}.

Calculated cross sections of relevance for supernova neutrino detection are shown in Fig.~\ref{fig:xscns_all}.
We highlight here two possible targets, argon and lead, both of current relevance for supernova neutrino detection (although note that many other targets are of interest as well).  Figure~\ref{fig:eventrates} shows numbers of interactions expected in argon and lead as a function of detector mass and distance.   Section~\ref{ccncmeas} describes some specific possible detector configurations to accomplish these measurements at the SNS.

\begin{figure}[!htbp]
\centering
\includegraphics[height=2.5in]{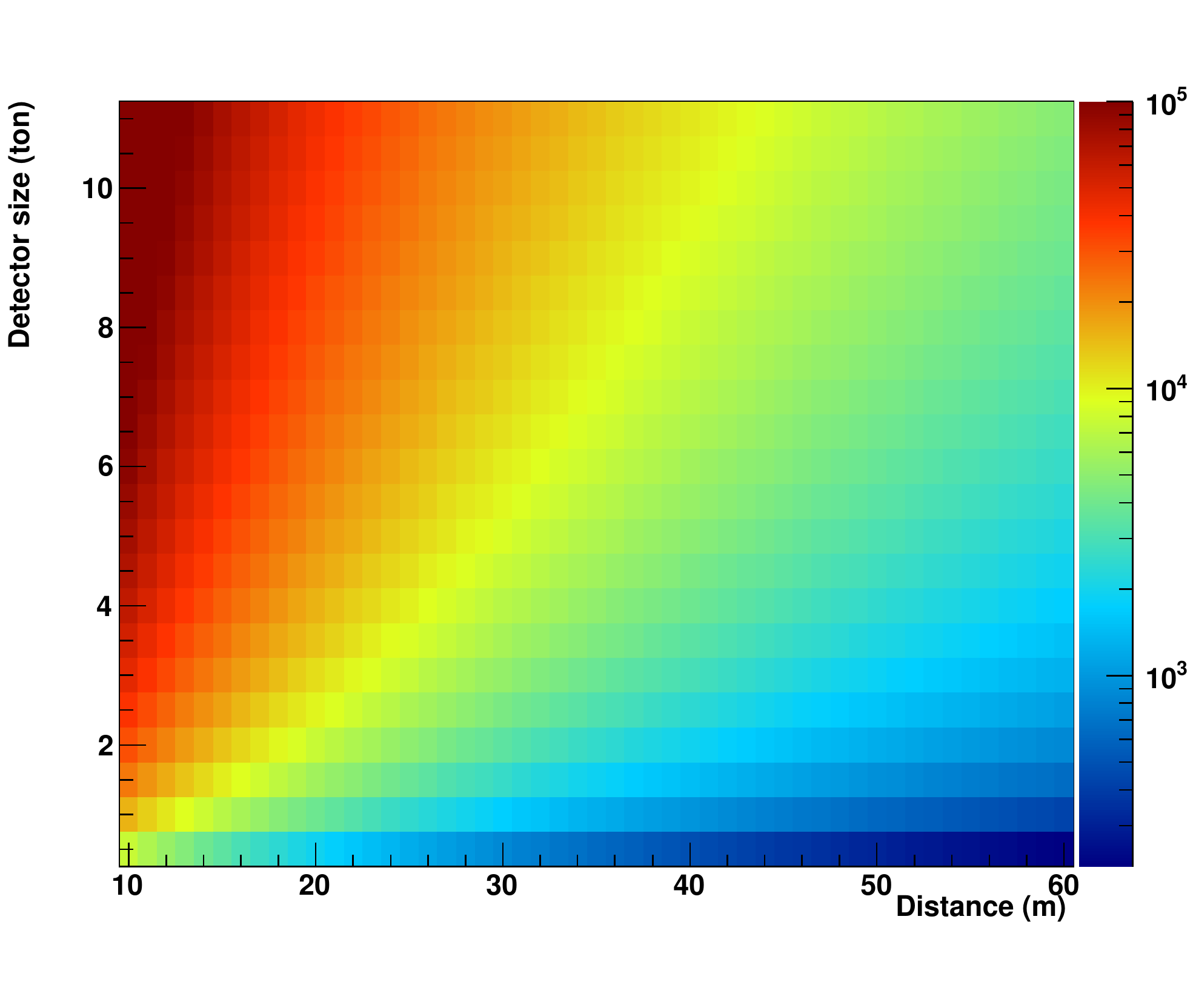}
\includegraphics[height=2.5in]{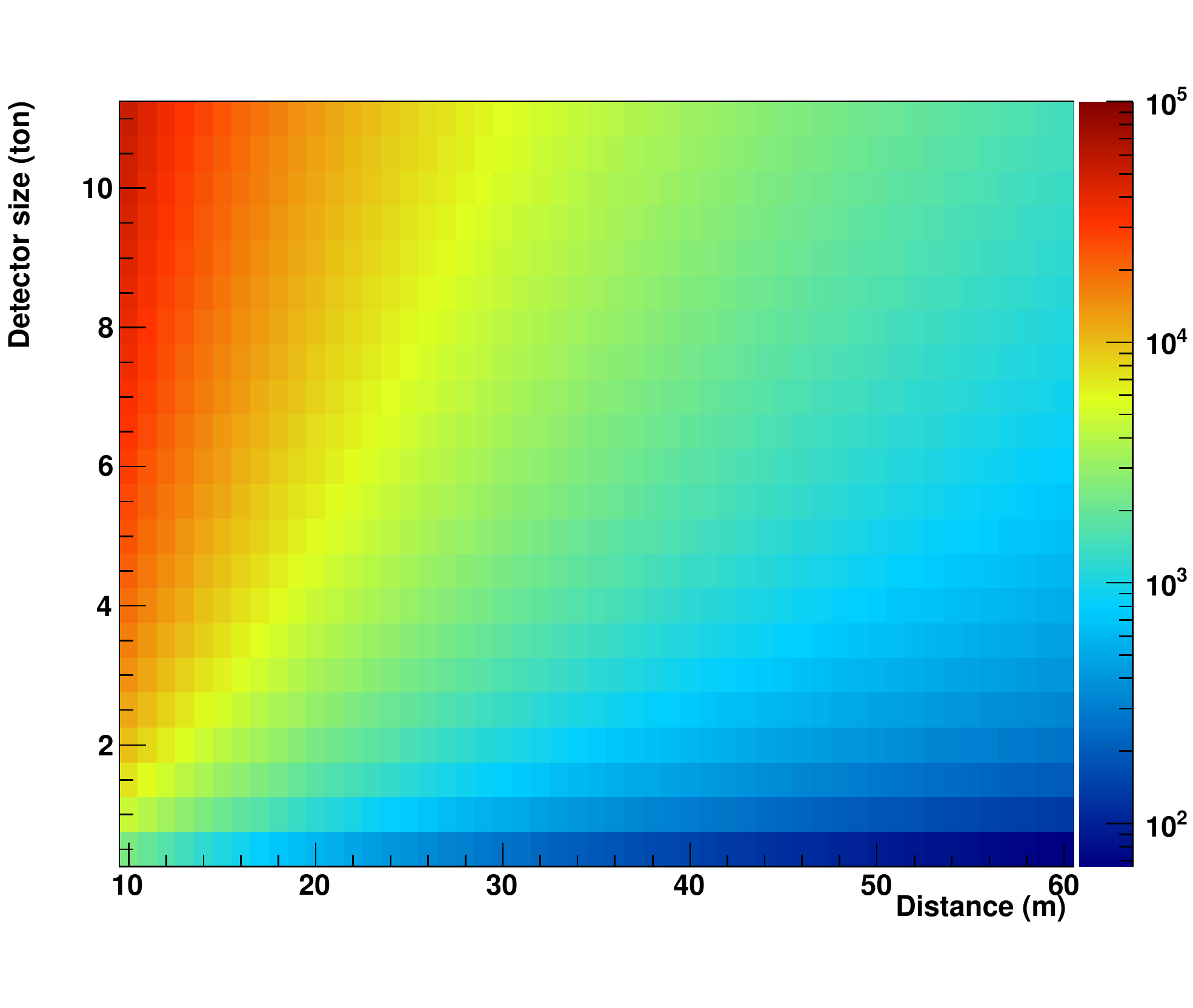}
\hspace{0.8in}
\caption{Expected events per year in argon and lead at the SNS as a function of detector size and distance. The event rate scaling goes simply as the detector mass and as the inverse square of distance from the source.}
\label{fig:eventrates}
\end{figure}

\noindent
\underline{\textbf{Argon}}:  We first consider
the absorption reaction of electron neutrinos on argon nuclei ($^{40}{\rm Ar}$):
\begin{equation}
\nu_e + {}^{40}{\rm Ar} \rightarrow e^- + {}^{40}{\rm K}^*.
\label{eq:absreac}
\end{equation}
This reaction was first proposed in 1986~\cite{Raghavan:1986hv,Bahcall:1986ry} as an alternative signature for real-time observation of low-energy astrophysical neutrino sources (solar  neutrinos and neutrino bursts from gravitational collapse of massive stars).
In these original studies~\cite{Raghavan:1986hv,Bahcall:1986ry}, only the (main) inverse $\beta^-$ decay of $^{40}$Ar ($J^\pi=0^+, T=2$) to the excited isobaric analog state at 4.38 MeV in $^{40}$K ($J^\pi=0^+, T=2$, superallowed Fermi transition) was considered.
The leading-electron detection with LArTPC technology can be accomplished by a distinctive $\gamma$-ray (delayed) coincidence signature from $^{40}$K deexcitation, with powerful background rejection.

Subsequently, shell-model calculations~\cite{Ormand:1994js} (valid for $\nu$-energy up to about 15-20 MeV, \textit{i.e.} for the high-energy component of the solar neutrino spectrum) showed that including several Gamow-Teller (GT) transitions to low-lying $J^\pi=1^+, 
 T=1$ states of $^{40}$K with excitation energies
 between 2.29 to 4.79~MeV leads to a significant enhancement of the overall absorption cross section. The GT matrix elements pertinent to $^{40}$Ar neutrino
absorption were experimentally confirmed~\cite{Trinder:1997xr} from measurement of branching ratios of the $\beta$-decay of $^{40}$Ti to levels in $^{40}$Sc (the mirror nucleus of $^{40}$K).
 In a further work~\cite{Bhattacharya:1998hc}, the
Fermi and GT transition strengths have been measured leading
to excited states up to 6~MeV in $^{40}$K$^*$; the
neutrino absorption cross section in $^{40}$Ar extending into the supernova neutrino range was obtained.
 
The detection and analysis of supernova neutrinos with LAr detectors requires
knowledge of the neutrino-induced cross sections on $^{40}$Ar for neutrinos (and antineutrinos) with
energies up to about 100 MeV, with daughter nucleus excitations
at high energies. An appropriate shell model calculation which describes GT strength
at these energies is currently not available. 
 However, 
 (continuum) random phase approximation (RPA) for
forbidden transitions of the $^{40}{\rm Ar}(\nu_e, e^-)^{40}{\rm K}$
reaction have been performed~\cite{Kolbe:2003ys}, considering allowed and forbidden multipoles up to $J = 6$.   GT transitions dominate the
$(\nu_e, e^-)$ cross section for neutrino energies $E_\nu < 50$ MeV; at higher energies, forbidden (in particular spin-dipole) transitions cannot be neglected.
These are dominated   by the collective response to giant resonance,
so that the RPA model~\cite{Kolbe:2003ys}   is usually considered sufficient to 
describe these non-allowed contributions.
 
The most recent independent  calculation
done for the neutrino absorption cross sections~\cite{SajjadAthar:2004yf} makes use
of the local density approximation (LDA) taking into account
the nuclear medium effects. The Coulomb distortion
of the electron wave function in the field of the final
nucleus is treated with the Fermi function as well
as in the modified effective momentum approximation.

The results from this last model are compared with the
other results available in literature and cited above~\cite{Ormand:1994js,Bhattacharya:1998hc,Kolbe:2003ys} for
the total absorption cross section $\sigma(E_\nu)$ as a function
of neutrino energy E$_\nu$. The respective cross sections are shown in Fig.~\ref{fig:ABS_Xsects}. 
\begin{figure}[ht]
\begin{center}
\includegraphics[width=11cm]{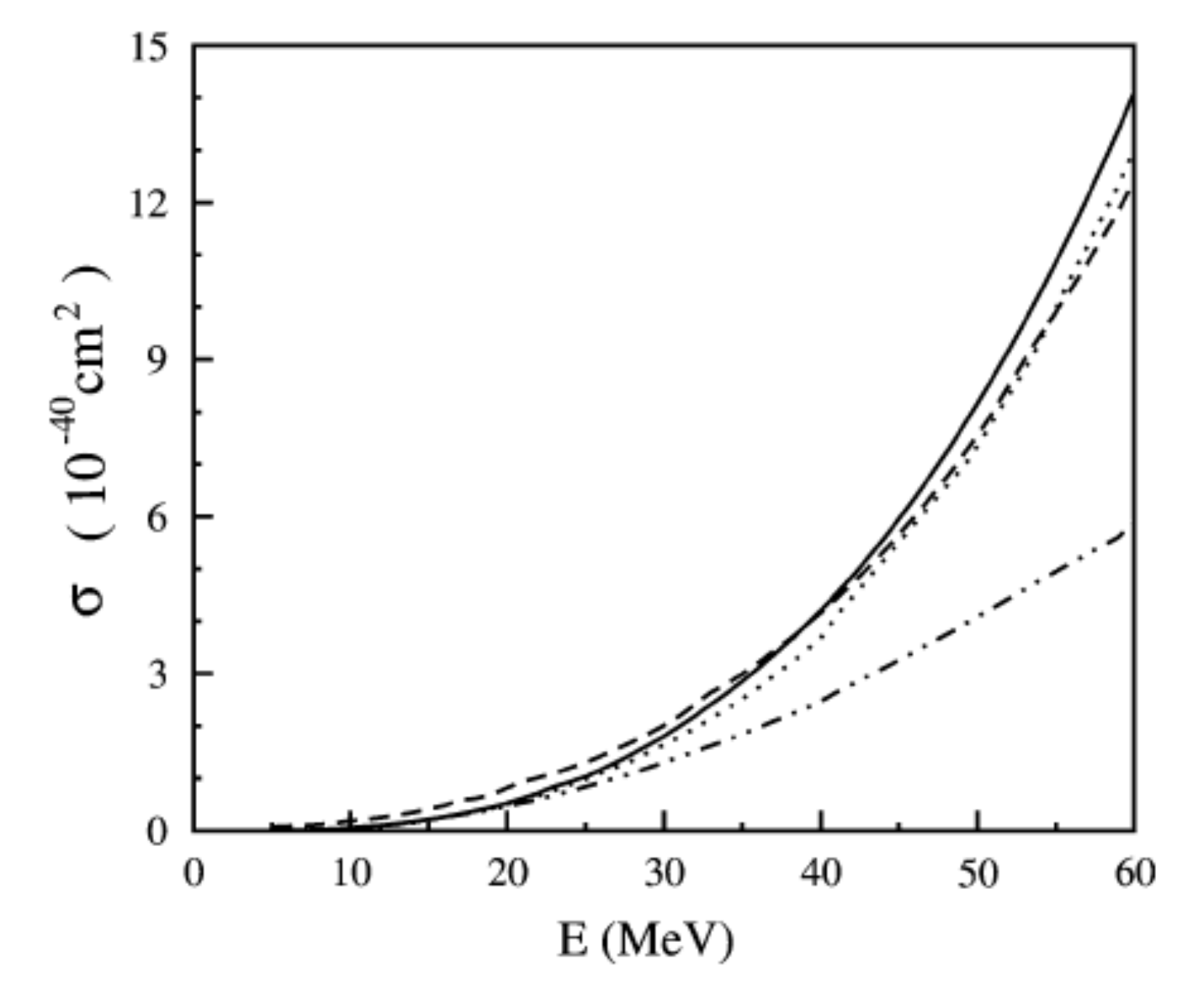}
\caption{{Total cross section $\sigma_{ABS}$ vs. E$_\nu$ for the $\nu_e + {\rm Ar}\rightarrow {\rm K}^{*} + e^-$ reaction for LDA \cite{SajjadAthar:2004yf}
with Fermi function (solid line) and with modified effective momentum
approximation (dashed line), Shell Model \cite{Ormand:1994js} (dashed-double dotted
line) and RPA \cite{Kolbe:2003ys} (dotted line). The plot is taken from \cite{SajjadAthar:2004yf}.}}
\label{fig:ABS_Xsects}
\end{center}
\end{figure}

The results show that the theoretical
uncertainty due to nuclear model dependence for
prediction of total event rates for an argon-based detector
is not too large,
and the absorption reaction of electron neutrinos on argon nuclei from recent theoretical calculations was found to be large (and much larger than previously estimated) in the energy range of interest for supernova neutrinos, with low $Q$-value of few MeV to the lowest-lying levels of the final nuclear state. Note, however, that no direct measurements of $\nu_e(^{40}{\rm Ar},^{40}{\rm K}^*)e^-$ cross section have ever been performed in this energy range.

In the few-hundred-tonne range (\textit{e.g.} for the MicroBooNE Experiment at FNAL with about 100 tonne of LAr), up to some tens of  $\nu_e$ 
events per 100 tonne of (active) mass can be expected in case of a 10 kpc Galactic supernova event. 
This rate largely depends, however, on the choice of the astrophysical parameters -- neutrino temperature, pinching factor, partition, \textit{etc.}--  in the supernova model as well as on the neutrino physical parameters ($\theta_{13}$ - recently determined to be large - and mass hierarchy) and the way neutrino oscillation through the dense stellar medium during the supernova explosion phases is accounted for.
The dependence of the observed  number of events over a reasonable range of input parameters has been evaluated in~\cite{Cavanna:2003fx} and is reported here.
The percentage variation $100\times \delta N/N$
of the number of absorbed $\nu_e$  
under a number of alternative hypotheses is shown in Table~\ref{tab:argon}.

\begin{table}[!ht]
\begin{centering}
\begin{tabular}{ c|c|c|c|c}
${T+\Delta T}$     &  ${T-\Delta T}$ &   ${f\to {1}/{8} }$  &   ${\eta\to 2}$ &  ${P_{ee} \to 0.3}$   \\
\hline
$+51$ \%     & ${- 45}$ \%   &   ${-25}$ \% & ${+15}$ \%  & ${- 16}$ \%  \\
\end{tabular} 
\caption{The first two columns show the effect of 
changing the temperature by $\Delta T=\pm 1.3$ MeV;
the third column describes the effect of non-equipartitioned fluxes;
the fourth column describes the effect of having a pinched (``non-thermal'') spectrum.
The last column assumes $\nu_e\to\nu_2$ due to very small $\theta_{13}$; this hypothesis (and related uncertainty) is no longer valid from the recent indications of the large $\theta_{13}$ value that implies $P_{ee}=0$.}
\label{tab:argon}
\end{centering}
\end{table}
This table shows that the present uncertainty in the 
temperature has a large impact on the expected signal, about $50$\%.
It shows also that a {\em mixture} of 
various phenomena can affect the event rate at the $\sim 20$\% level. 
To separate these effects clearly, it will be important to study several 
properties of the neutrino signal, such as distributions 
in time and energy and to use several reactions with different detectors.

\begin{figure}[ht]
\begin{center}
\includegraphics[width=10cm]{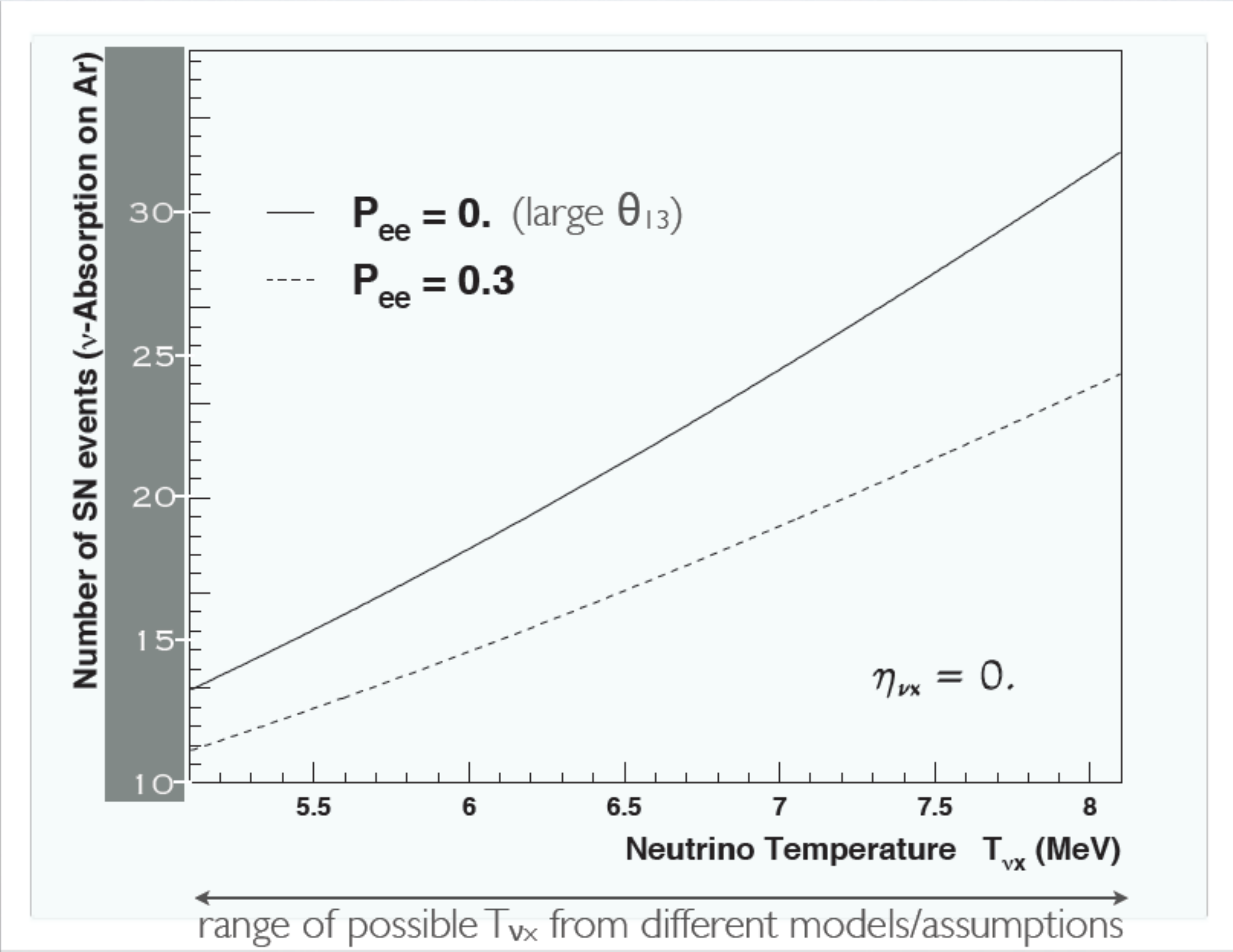}
\caption{Number of expected $\nu_e$ 
events as a function of the effective neutrino temperature.
A reference 100-tonne detector mass (corresponding to $1.5\times 10^{30}$ Ar targets) is considered. 
(For simplicity, an ideal detector without threshold on the final state electron energy
and full detection efficiency is assumed.)
The cross section `hybrid model'  for $\nu_e$-absorption on argon  
is used (shell model \cite{Ormand:1994js} for allowed transitions, 
and RPA \cite{Kolbe:2003ys} for forbidden transitions).
Neutrino fluence is taken with $T$ free to vary and pinching parameter set at $\eta= 0$.
The other supernova parameters are: $D=10$~kpc, $f=1/6$ (strict equipartition), 
and ${\cal E}_B = 3\times 10^{53}$~erg.
Oscillations are separately accounted for: the full line corresponds to the large, now well-established value of 
$\theta_{13}$.}

\label{fig:nevtemp}
\end{center}
\end{figure}
In Fig.~\ref{fig:nevtemp} we show the calculated 
number of expected events for a 100-tonne  LAr mass (on the scale of the MicroBooNE detector) for a wide range of values  of 
the effective neutrino temperature.

The distinctive feature of the argon target relies on the sensitivity to the $\nu_e$ component of the supernova neutrino flux, in contrast to the ``traditional" water Cherenkov or scintillator detector primarily sensitive to the $\bar{\nu}_e$ component. 
The combined information from neutrino and antineutrino detection can provide important additional hints 
about the supernova explosion mechanism, about nucleosynthesis, and about neutrino intrinsic properties.

Figure~\ref{fig:argonspec} shows expected differential event spectra for an argon target at the SNS.  Note that in principle elastic scattering  events may be selected from absorption events using directionality of the signal.

\begin{figure}[!htbp]
\centering
\includegraphics[height=2.5in]{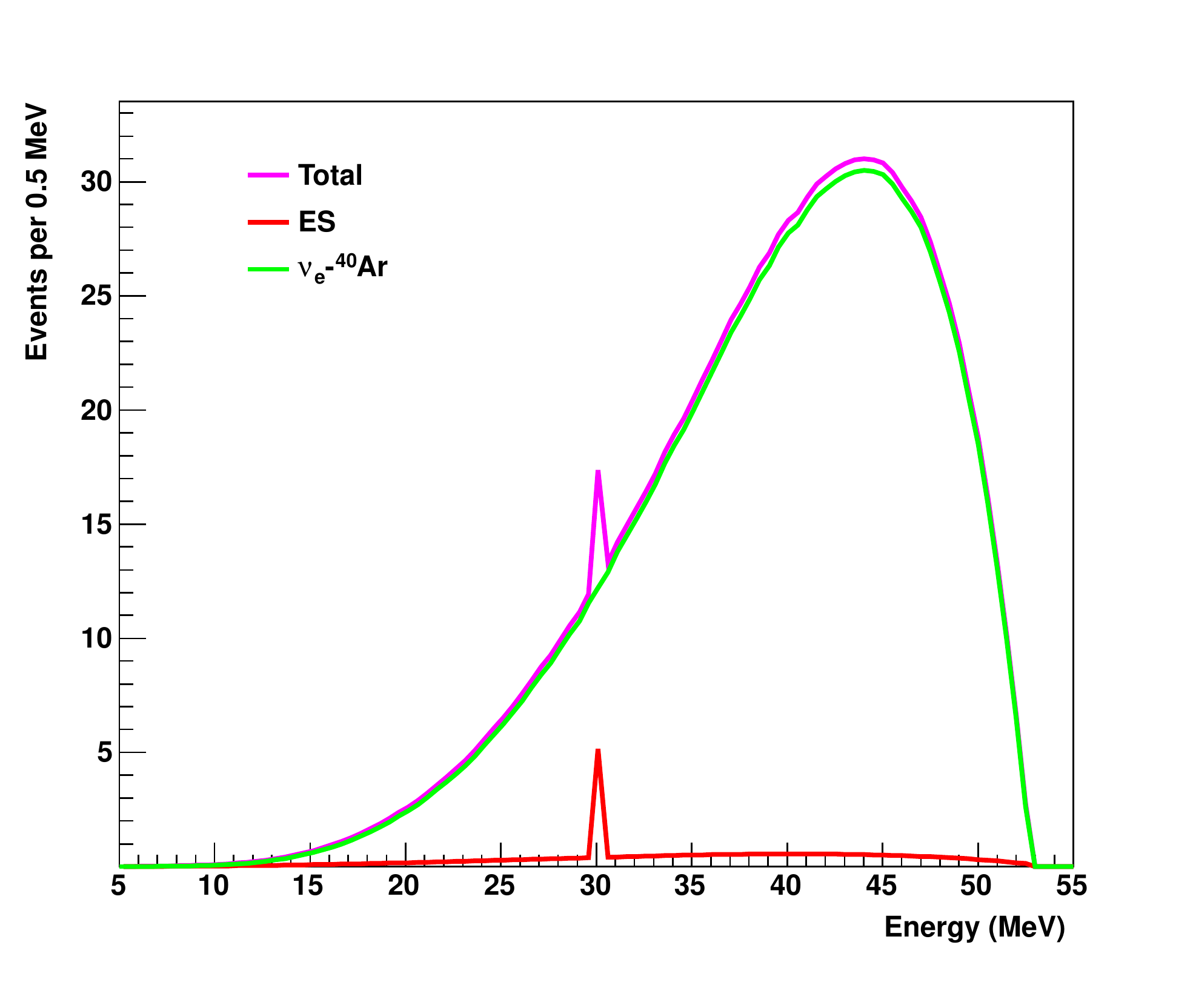}
\includegraphics[height=2.5in]{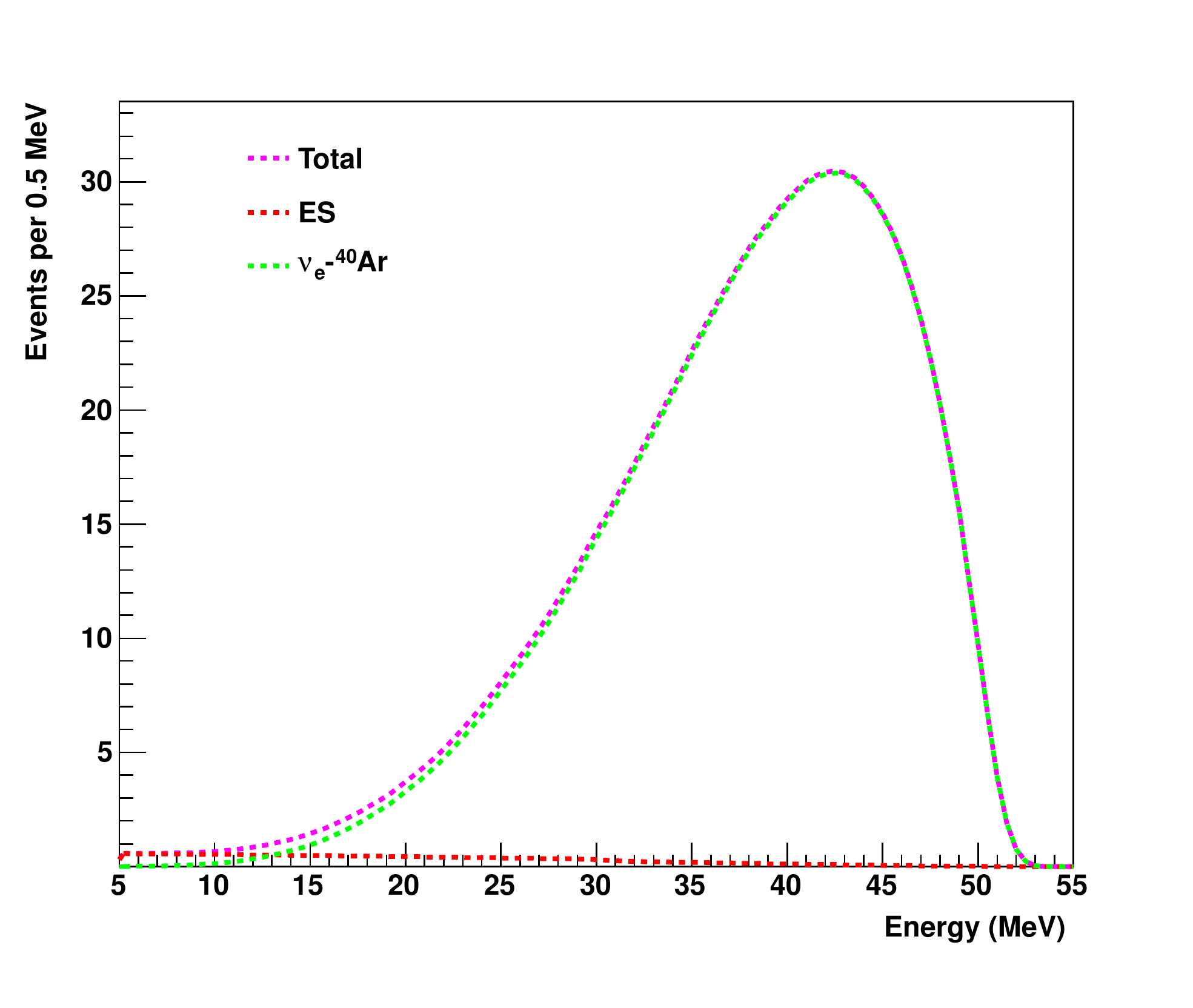}
\hspace{0.8in}
\caption{Expected differential event spectra for different interactions on argon, per tonne per year at 20~m from the source.  ES refers to elastic scattering of neutrinos of all flavors on electrons.  Left: interactions as a function of neutrino energy.  Right: observed energy spectrum smeared with resolution from~\cite{Amoruso:2003sw}.}
\label{fig:argonspec}
\end{figure}

\noindent \underline{\textbf{Lead:}} We now consider lead as another example target.
Heavy nuclei such as iron and lead may yield high interaction rates 
for both CC and NC
interactions of few-tens-of-MeV neutrinos, and furthermore have sensitivity to flavors other than $\bar{\nu}_e$.
Observables include leptons and ejected nucleons.  
Single- and multi-neutron ejections are possible.
The most relevant interactions are
$\nu_e+{}^{A}{\rm Pb}\rightarrow e^{-}+{}^{A}{\rm Bi}^{*}$ 
and 
$\nu_x+{}^{A}{\rm Pb}\rightarrow \nu_x+{}^{A}{\rm Pb}^{*}$; in both cases the
resulting nuclei deexcite via nucleon emission.  Figure~\ref{fig:leadxscns} shows the cross sections for CC and NC single- and double-neutron emission~\cite{Engel:2002hg}, for $^{208}$Pb.  Although natural lead is only about half $^{208}$Pb, other isotopes should have very similar cross sections~\cite{Kolbe:2000np,Engel:2002hg}.  
Because of lead's neutron excess, Pauli blocking strongly suppresses CC $\bar{\nu}_e$ interactions. The 1n- and 2n-emission rates are sharply dependent on neutrino energy.  In particular, 2n-emission rates only turn on for relatively high neutrino energies, resulting in relative 1n- and 2n-emission rates which are very sensitive to the incoming neutrino spectrum.  Secondary emission via $(n,2n), (n,3n),...$ reactions is possible, which will affect observed rates, but it is expected  to be small at the energies involved.

\begin{figure}[htb]
 \centering\includegraphics[width=.65\textwidth]{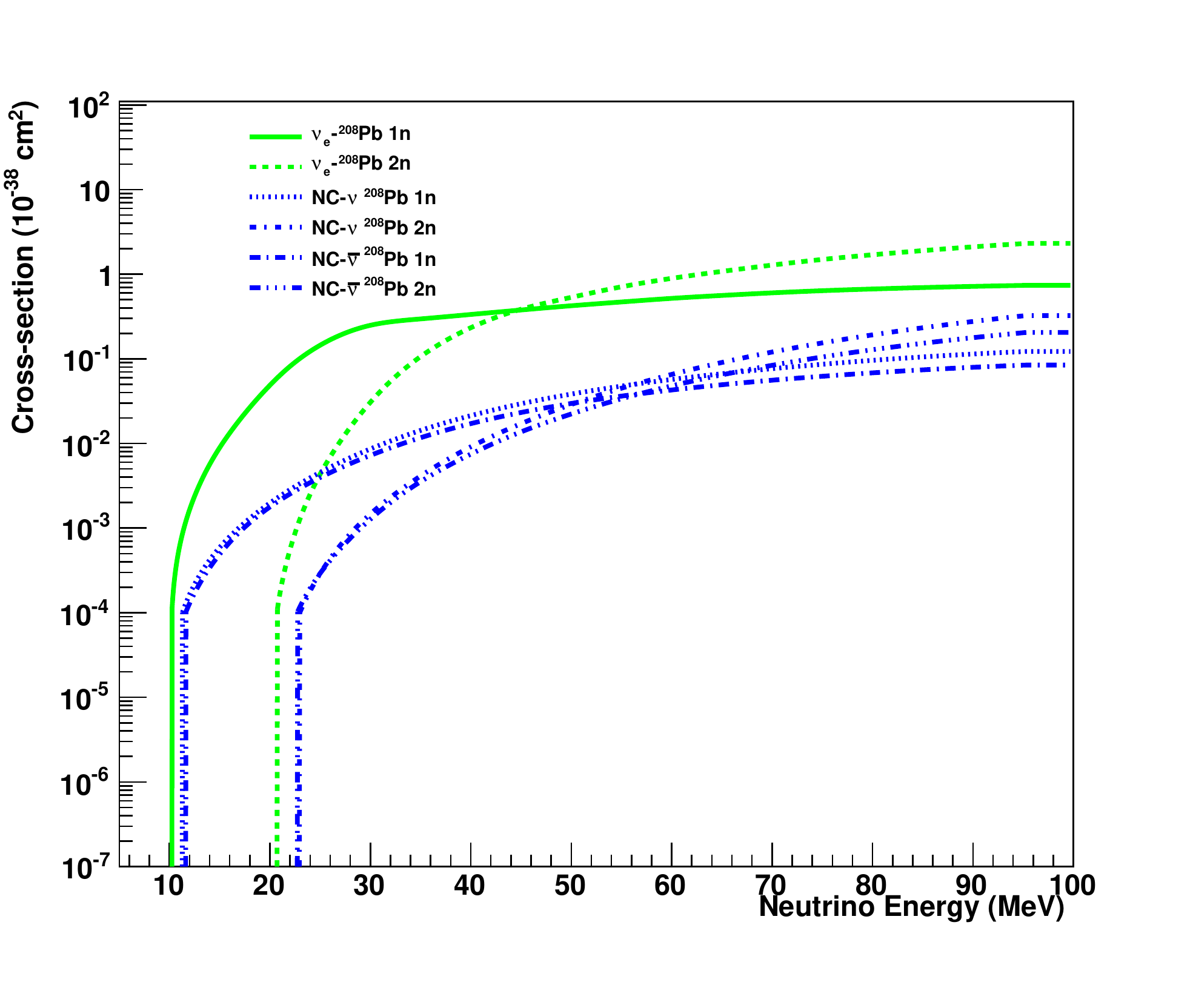}

 \caption{Cross sections for different interactions in lead~\cite{Engel:2002hg}.}

 \label{fig:leadxscns}
\end{figure}

Lead has been proposed by many authors as a supernova detection medium~\cite{Hargrove:1996zv,Engel:2002hg,Fuller:1998kb,Boyd:2002cq,Zach:2002is}.   Physics capabilities are explored in \textit{e.g.} \cite{Vaananen:2011bf}.
Lead has many properties that make it a good choice for a neutrino target.  The element itself is cheap, abundant, and chemically stable, making it straightforward to handle in large quantities.  Further, since its primary isotopes are neutron-rich, their neutron absorption cross sections are orders of magnitude smaller than other materials.
Natural lead has a thermal neutron capture cross section of $\sim$0.15~{barn}, whereas that of $^{56}\mbox{Fe}$ is $\sim$2.5 barn.  This property allows lead to moderate fast neutrons without absorption prior to arrival at a neutron counter.

In a detector like
HALO~\cite{Duba:2008zz,schumaker}, electrons are practically invisible
and only neutrons are observable. 
In practice, single- and double-neutron products from lead can be tagged and reconstructed.  Although
no event-by-event energy information is available, spectral information can be
inferred from the relative numbers of 1n and 2n events.  
Although it would be desirable to detect electrons from CC interactions, as in some proposed detector designs~\cite{Elliott:2000su,Bacrania:2002jq,Zach:2002is,Boyd:2002cq}, instrumentation to enable this adds significant expense.  Nevertheless such capability could be considered for future detectors.
There are theoretical uncertainties in the neutrino-lead cross sections, so it would be highly desirable to measure these cross sections using stopped-pion neutrinos. 
Figure~\ref{fig:leadspec} shows expected rates in lead at the SNS.

\begin{figure}[!htbp]
\centering
\includegraphics[height=2.8in]{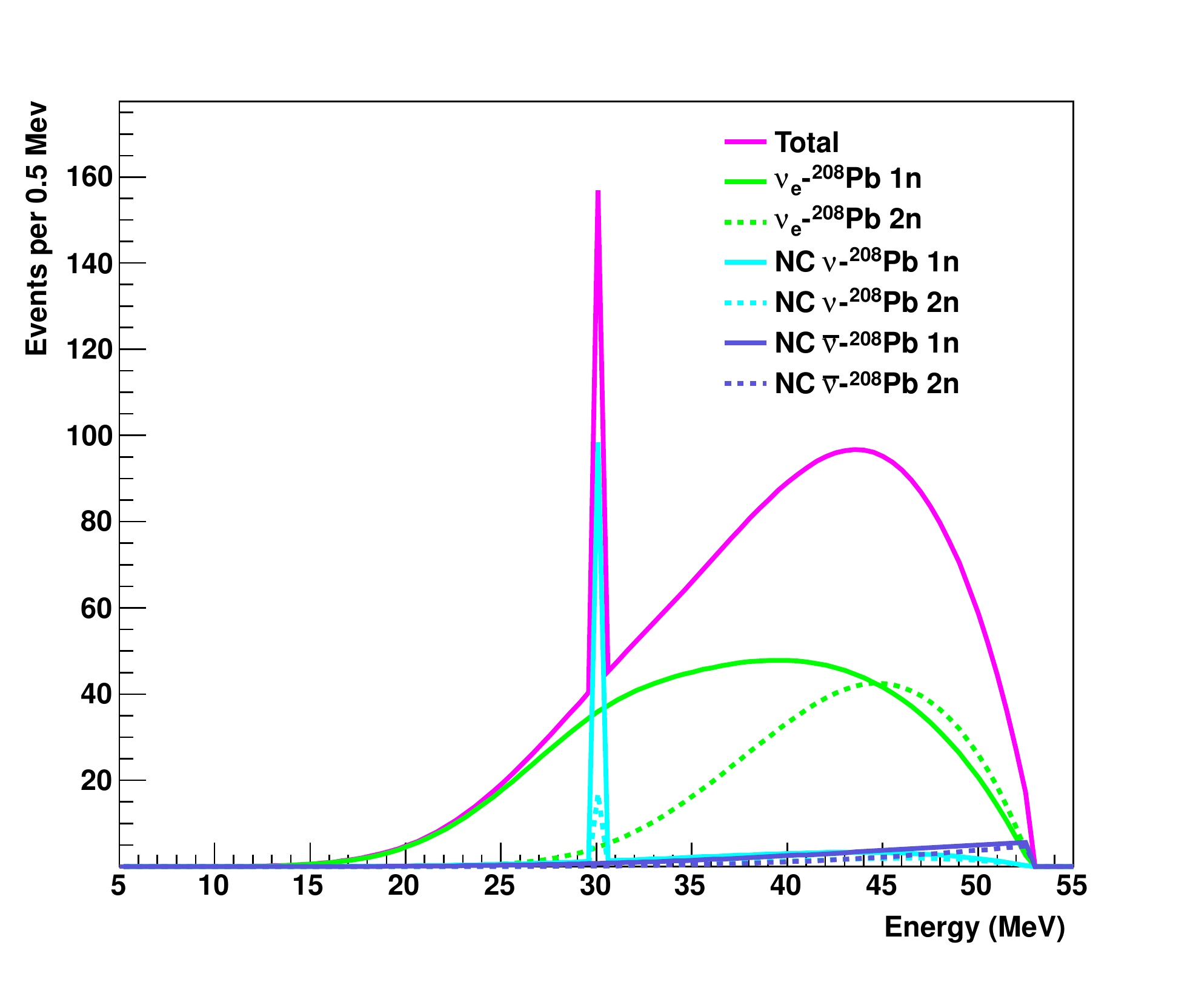}
\hspace{0.8in}
\caption{Interactions in lead as a function of neutrino energy, per tonne per year at 20~m from the source. Note that in a neutron detector like HALO, no event-by-event energy reconstruction is possible.}
\label{fig:leadspec}
\end{figure}

Taken together, the numerous ways in which neutrino properties, especially neutrino-nucleus interactions, affect the physics of core-collapse supernovae strongly supports an experimental program to study their interactions.  In some cases, like calibrating terrestrial detectors, specific measurements of excellent precision are needed to make the most of supernova science.  In other cases, measurements can provide strong constraints on the theory of neutrino-nucleus interactions, theory that can then be applied to calculate the large number of reactions needed in astrophysical circumstances.  In both cases, we are fortunate that stopped-pion facilities provide neutrinos of similar energy to the distant astrophysical sources.

\item 
\textbf{Standard Model Tests from CC and NC Interactions}

Measurements of CC and inelastic NC interactions also offer the possibility of tests of the SM of particle physics.  Below are two examples.

\begin{itemize}
\item 
\textbf{Measurement of the electron spectrum}

The electron spectrum from the reaction $^{12}$C$(\nu_e,e^-)^{12}N_{\rm gs}$ can be used to derive the original electron spectrum from muon decay, taking into account $E = E_e+Q (17.8~{\rm MeV})$, the detector response function, and the $(E_e-Q)^2$ dependence of the differential cross section. A measurement of Michel parameters in muon decay is a direct test of the SM and is a method to search for manifestations of new physics, since these parameters are sensitive to the Lorentz structure of the Hamiltonian of weak interactions. The Michel parameters are related to electron spectrum shape in muon decay. The neutrino spectrum from this decay can be described in terms of similar parameters. The neutrino spectrum can provide complementary information to the set of Michel parameters, increasing the accuracy and reliability in the search for new physics in the lepton sector. 

As pointed out in reference~\cite{Fetscher:1992sk}, the shape of the $\nu_e$ spectrum from $\mu^+$ decay at rest is sensitive to scalar and tensor admixtures to pure $V-A $ interactions due to the parameter $\omega_L$.  This parameter is analogous to the Michel parameter, which determines the shape of the electron spectrum in muon decay.  The $\nu_e$ spectrum can be written as:

\begin{equation}
\frac{dN_{\nu_e}}{dx} = \frac{G_F^2 m_{\mu}^5}{16 \pi^3}Q_L^\nu(G_{0(x)}+G_1(x)+\omega_L G_2(x))
\end{equation}

where $m_\mu$ is the muon mass, $x=2E_\nu/m_\mu$ is the reduced neutrino energy, $Q_L$ is the probability for emission of a left-handed electron neutrino, $G_0$ describes pure $V-A$ interactions, $G_1$ takes into account radiative corrections, and $\omega_LG_2$ includes effects of scalar and tensor components. $G_1$ is very small and can be neglected.

In the SM, $\omega_L$ is exactly zero. The KARMEN experiment~\cite{Armbruster:1998gk} determined an upper limit for $\omega_L$: $\omega_L< 0.113$ (90\% CL). Figure~\ref{fig:electron_spec} shows the calculated electron spectra for the reaction $^{12}{\rm C}(e,e^-)^{12}N_{\rm gs}$ for $\omega_L = 0~(\omega_L = 0.113)$ with a black (blue) line. The largest difference between the two distributions is at the high-energy end of the spectra where the detector resolution is very good and the absolute energy scale can be very accurately calibrated using Michel electrons from stopped cosmic muons. The expected statistical accuracy for a one-year measurement at the SNS with 20~tonne liquid scintillator detector at 20~m is shown by the red points. It should be possible to significantly improve the KARMEN limit on scalar and tensor admixtures to $V-A$ interactions in the lepton sector~\cite{Armbruster:1998gk}.

\begin{figure}
\vspace{5mm}
\centering
\includegraphics[width=9.5cm]{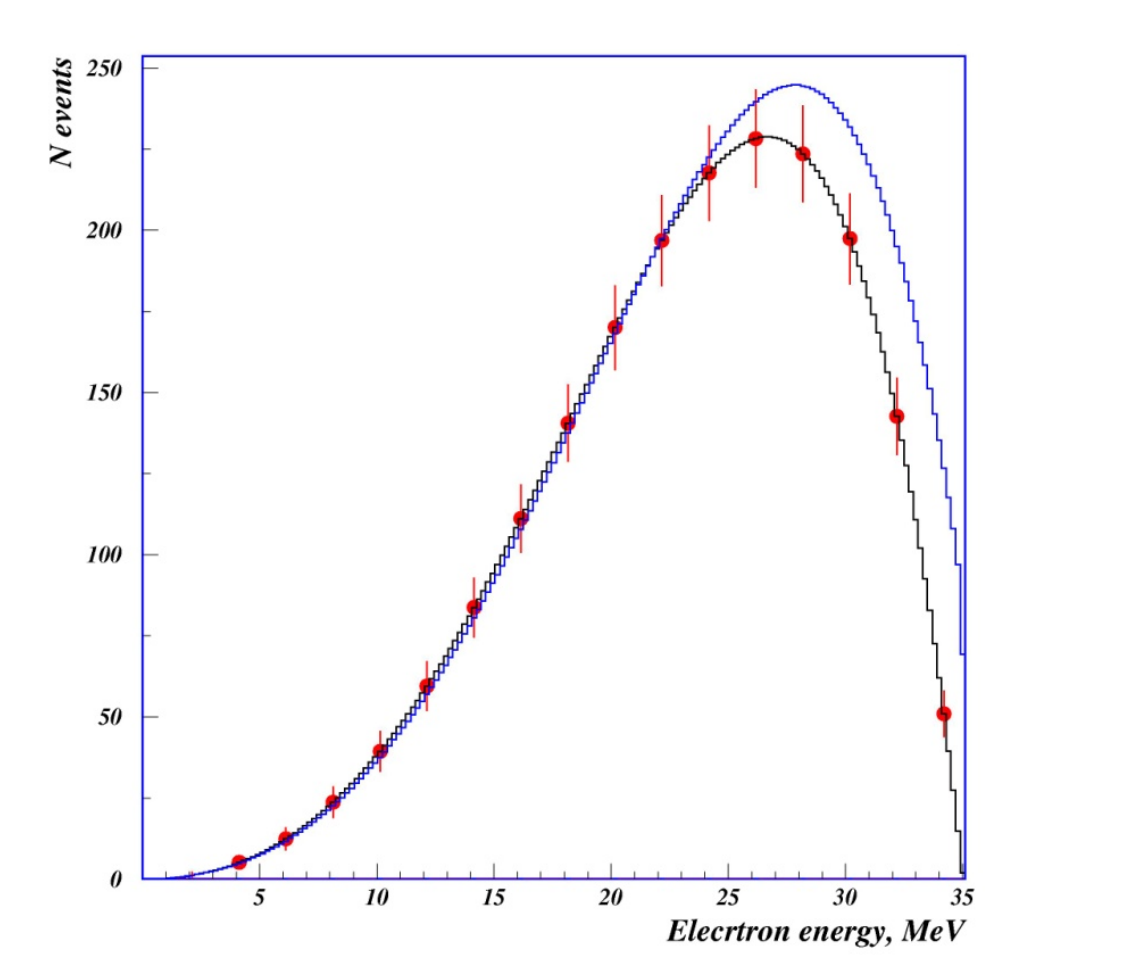}
\vspace{1mm}
\caption{
Electron spectrum from the reaction $^{12}{\rm C}(\nu_e,e^-)^{12}N_{\rm gs}$ caused by neutrinos from muon decay at rest calculated in the framework of the SM is shown with the black line. The blue line shows the upper limit, measured in the KARMEN experiment~\cite{Armbruster:1998gk}, on the distortion of the spectra caused by scalar and tensor admixtures to pure $V-A$ interactions. Red points represent the statistical accuracy of a one-year measurement at the SNS with 20 tonne liquid scintillator detector at 20~m. 
}
\label{fig:electron_spec}
\end{figure}

\item 
\textbf{Test of the SM with neutrino-deuteron cross sections}

The relative simplicity of the decay of the free neutron makes it an
attractive laboratory for the study of possible extensions to the
SM. Measurements of the neutron
lifetime and neutron decay correlations can be used to determine the
weak vector coupling constant, which, in turn, can be combined
with information on strange particle decay  to test such notions as the
universality of the weak interaction or to search for 
nonstandard couplings (see, \textit{e.g.}, \cite{Jackson:1957zz,Holstein:1977qn,deutsch,Abele:2000ar,yeroz,Gardner:2000nk,Herczeg:2001vk,Marciano:2003zr} and references therein). For an unambiguous
search for new physics in neutron decay experiments
and for a precise determination
of fundamental constants, it is necessary
to understand and evaluate all
corrections for neutron decay with higher accuracy than the expected
experimental precision. 
 These include such effects as recoil and
radiative corrections. Recently~\cite{Ando:2004rk} effective field theory (EFT)
has been used to incorporate all SM effects in a consistent
fashion in terms of one parameter (low energy constant, LEC) with an estimated theoretical accuracy on the order of
$10^{-5}$. Because this accuracy is well below that anticipated in the
next generation of neutron decay experiments,
the EFT approach provides an unambiguous model-independent description of neutron decay which gives a unique opportunity  to search for new physics, provided the LEC is known.

However, the only way to obtain the value of EFT low energy constants from experiments is to measure a number of observables in independent processes which are described exactly by the same Lagrangian within a similar energy range.  Thus the low energy neutrino-deuteron ($\nu - d$) reactions are the best candidates to obtain the unknown LEC for neutron radiative corrections, and they can be considered as  ``EFT-complementary'' to the neutron decay process. Indeed, the $\nu  d$ reactions are described by the same Lagrangian and, since the target is two nucleon bound state, the radiative corrections to neutrino-deuteron cross sections can be calculated using EFT. These calculations require significant effort and  have not yet been done.  Therefore, to estimate the required accuracy in the measurement of absolute cross sections for these reactions, one can use existing calculations of radiative corrections given in a conventional approach (see, for example \cite{Towner:1998bh,Kurylov:2002vj,Kurylov:2002vj,Kubota:2004eh,Fukugita:2005hs}). Then, the expected accuracy must be equal to or better than the contribution of strong-interaction model-dependent parts of the corrections to the total cross sections. Using the results of \cite{Kubota:2004eh,Fukugita:2005hs} one can see that these model-dependent parts have a value of about $(2-3)\times 10^{-2}$. 
Therefore, a measurement of neutrino-deuteron absolute cross sections with accuracy of between  $10^{-2}$ and $10^{-3}$ would provide a unique opportunity to test the SM in neutron decay processes in a model-independent way to the level of $10^{-5}$.

\end{itemize}
\end{itemize}

\subsubsection{Coherent Elastic Neutrino-Nucleus Scattering}\label{coherent_phys}

Another interesting possibility for a stopped-pion source is the detection of nuclear recoils
from coherent elastic $\nu$-nucleus scattering, detection of which is within the
reach of the current generation of low-threshold detectors~\cite{Scholberg:2005qs}.  This reaction is also important
for supernova processes and detection.

	Coherent NC neutrino-nucleus scattering has never been observed since its first theoretical prediction in 1974~\cite{Freedman:1977xn}. The condition of coherence requires sufficiently small momentum transfer to the nucleon so that the waves of off-scattered nucleons in the nucleus are all in phase and add up coherently. While interactions of neutrino energy in MeV to GeV-scale have coherent properties, neutrinos with energies less than 50 MeV are most favorable, as they largely fulfill the coherence condition in most target materials with nucleus recoil energy of tens of keV. The elastic NC interaction in particular leaves no observable signature other than low-energy recoils off the nucleus. Technical difficulties involved in the  development of a large-scale, low-energy threshold and low-background detector have hampered the experimental realization of the coherent $\nu$A measurement for more than three decades. However, recent innovations in dark matter detector technology (\textit{e.g.}~\cite{Chepel:2012sj})  have made the unseen coherent $\nu$A reaction testable.  A well-defined neutrino source is the essential component for measurement of the coherent $\nu$A scattering. The energy range of the SNS neutrinos is below 50~MeV, which is the optimal energy to observe pure coherent $\nu$A scattering. 

The cross section for a spin-zero nucleus, neglecting
radiative corrections, is given by~\cite{Horowitz:2003cz},

\begin{equation}
\frac{d\sigma}{dT}(E,T) = \frac{G_{F}^{2}}{2\pi}M \left[ 2 - \frac{2T}{E} +
\left(\frac{T}{E}\right)^{2} - \frac{MT}{E^{2}}\right]
\frac{Q_{W}^{2}}{4}F^{2}(Q^{2})\,.
\label{eq:dsigmadT}
\end{equation}

where $E$ is the incident neutrino energy, $T$ is the nuclear recoil
energy, $M$ is the nuclear mass, $F$ is the ground state elastic form
factor, $Q_w$ is the weak nuclear charge, and $G_F$ is the Fermi
constant.  
The condition for coherence requires that momentum transfer
$Q\lsim \frac{1}{R}$,
where $R$ is the nuclear radius. 
This condition is largely satisfied for neutrino energies up to 
$\sim$50~MeV
for medium $A$ nuclei.
Typical values of the total
coherent elastic cross section are in the range $\sim
10^{-39}$~cm$^2$, which is relatively high compared to other neutrino
interactions in this energy range ($e.g.$ CC
IBD on protons has
a cross section $\sigma_{\bar{\nu}_e p}\sim 10^{-40}$~cm$^2$, and 
elastic
neutrino-electron scattering has a cross section\
$\sigma_{\nu_e e}\sim 10^{-43}$~cm$^2$).

\begin{figure}[!htbp]
\centering
\includegraphics[height=2.5in]{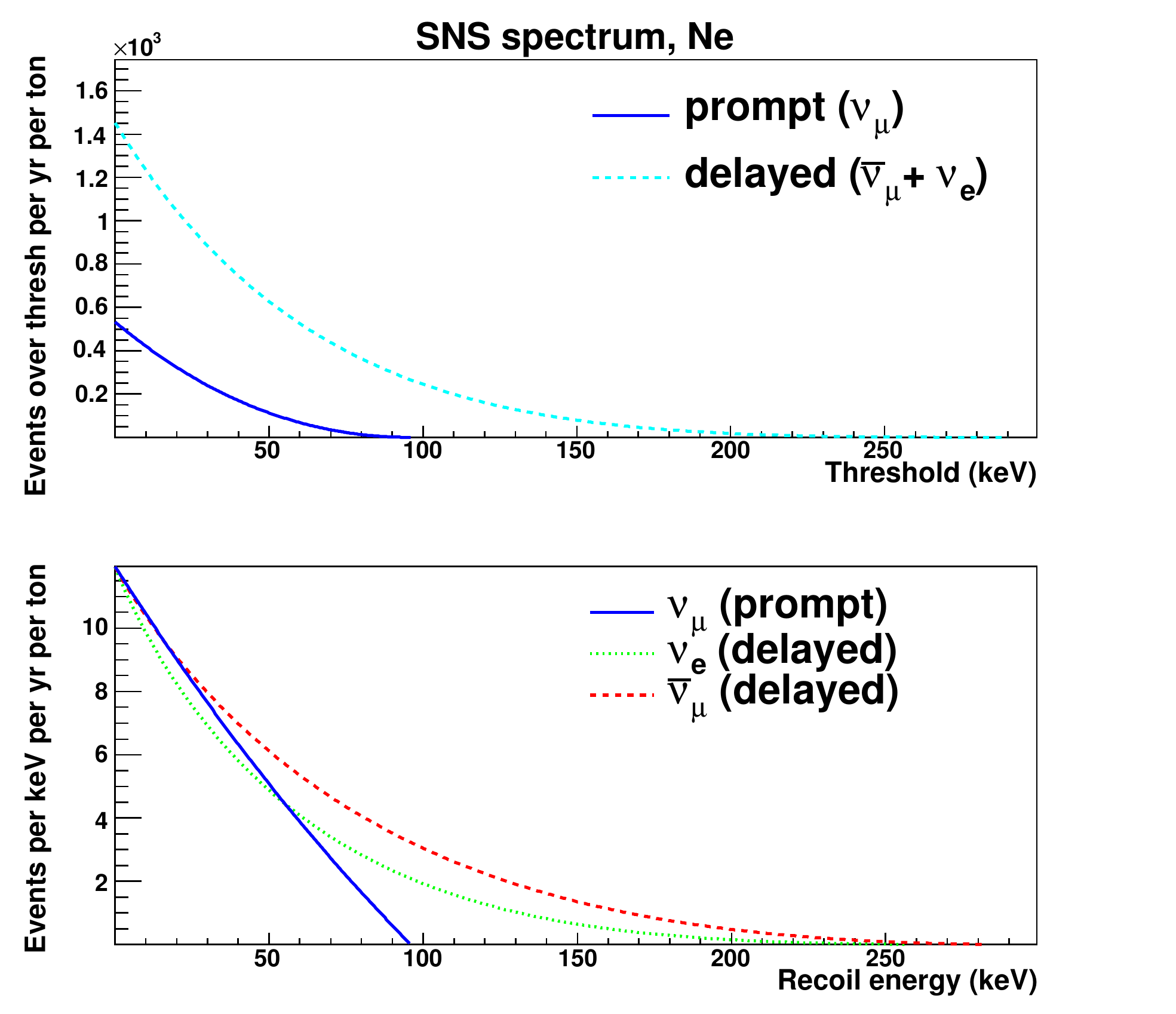}
\includegraphics[height=2.5in]{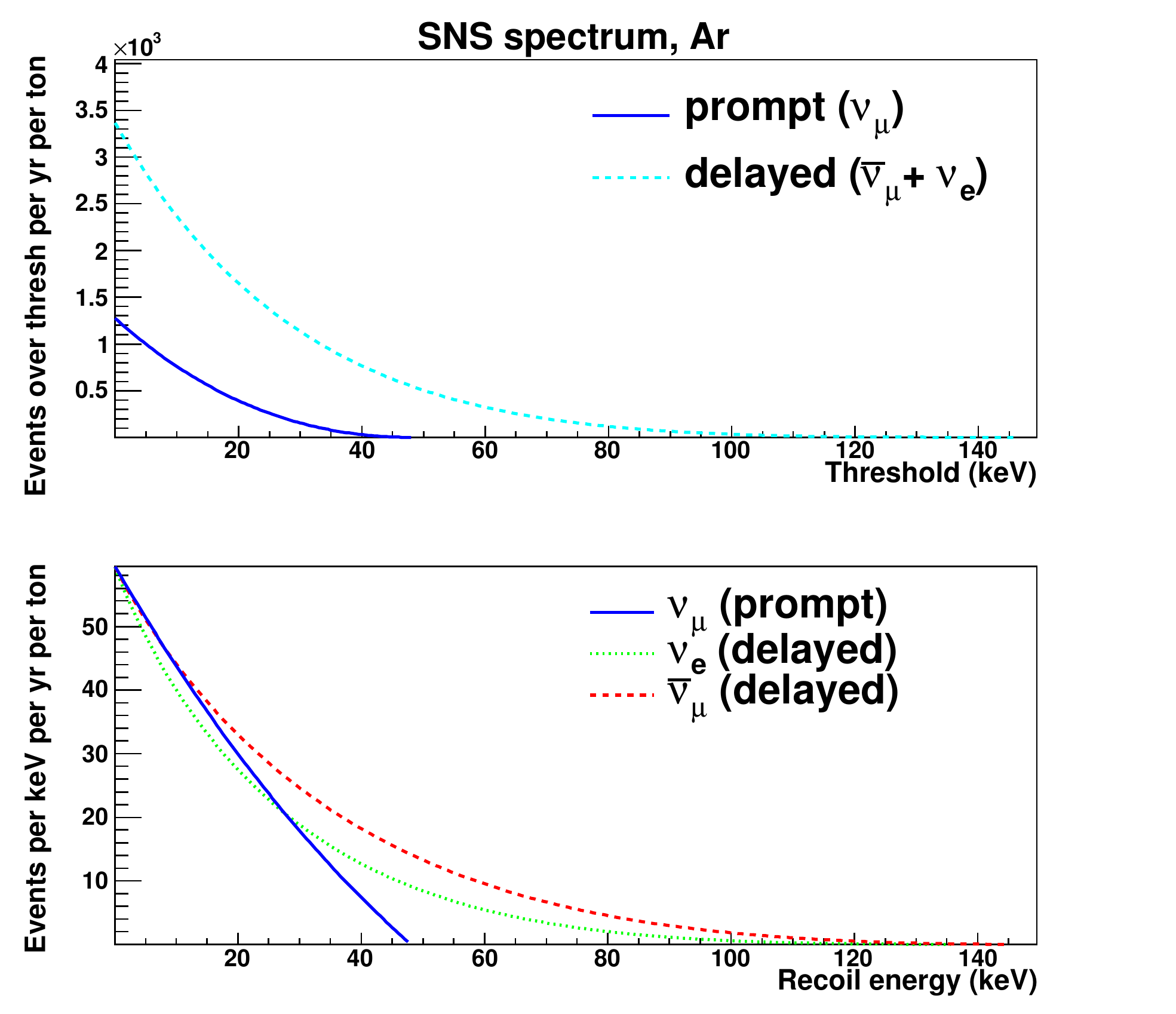}

\caption{Top, left: Number of interactions over recoil energy threshold
in one tonne of $^{20}$Ne for one year of running at the SNS at 46~m from the target
(solid: $\nu_\mu$, dashed: sum of $\nu_e$ and $\bar{\nu}_\mu$),
as a function of recoil energy threshold. 
Bottom, left: Differential yield at the SNS in one tonne of 
$^{20}$Ne
(solid: $\nu_\mu$, dotted: $\nu_e$, dashed: $\bar{\nu}_\mu$) per year 
per keV,
as a function of recoil energy.  Right: same for Ar.}
\label{fig:snsyield1}
\end{figure}

Although ongoing efforts to observe coherent $\nu A$ scattering 
at reactors~\cite{Barbeau:2007qi,Collar:2008zz,Wong:2005vg}
are
promising, a stopped-pion beam has several advantages with
respect to the reactor experiments. Higher recoil energies bring
detection within reach of the current generation of low-threshold
detectors which are scalable to relatively large target
masses. Furthermore, 
the pulsed nature of the source (see Fig.~\ref{fig:sns_nuplots}) allows
both background reduction and precise characterization of the
remaining background by measurement during the beam-off period.
Finally, the different flavor content of the SNS flux means 
that 
physics sensitivity is complementary to that for reactors.
The expected rates for the SNS are quite promising for noble liquids~\cite{Scholberg:2005qs}: see Fig.~\ref{fig:snsyield1}.  

Coherent elastic $\nu$A scattering
reactions are important in stellar core-collapse 
processes~\cite{Freedman:1977xn},
as well as being useful for core-collapse supernova 
neutrino
detection~\cite{Horowitz:2003cz}.  A rate measurement will
have bearing on supernova neutrino physics.
The $\nu$A coherent elastic scattering cross section is
predicted by the SM, and form factor
uncertainties are small~\cite{Horowitz:2003cz}.  Therefore a measured
deviation from prediction could be a signature of new physics.
We also note that successful measurement of coherent neutrino scattering in the energy range of solar and atmospheric neutrinos will be immediately useful for direct dark matter search experiments, for which solar and atmospheric neutrinos will eventually represent a background.

A few of the possible physics measurements are described in more detail below.

\begin{itemize}
\item \textbf{Standard Model Tests With Coherent Scattering}

According to Eq.~\ref{eq:dsigmadT}, the SM predicts a coherent elastic scattering rate proportional to
$Q_w^2$, the weak charge given by $Q_w = N-(1-4\sin^2 \theta_W)Z$,
where $Z$ is the number of protons, $N$ is the number of neutrons, and
$\theta_W$ is the weak mixing angle.  Therefore the weak mixing angle
can be extracted from the measured absolute cross section, at a
typical $Q$ value of 0.04~GeV/$c$.  
A deviation from the SM prediction
could indicate new physics.  If the absolute cross section can be
measured to 10\%, there will be an uncertainty on $\sin^2 \theta_W$ of
$\sim 5 \%$.  One might improve this uncertainty by looking at ratios
of rates in targets with different $N$ and $Z$, to
cancel common flux uncertainties; future use of enriched neon is a possibility.
There are existing precision
measurements from atomic parity
violation~\cite{Bennett:1999pd,Eidelman:2004wy}, SLAC
E158~\cite{Anthony:2005pm} and NuTeV~\cite{Zeller:2001hh}.
However there is no previous
neutrino scattering measurement in this region of $Q$.
This $Q$ value is relatively
close to that of the proposed Qweak parity-violating
electron scattering experiment at JLAB~\cite{vanOers:2007if}.
However coherent elastic neutrino-nucleus scattering tests the SM in a different channel and therefore
is complementary.

In particular, such an experiment  can search for non-standard interactions (NSI)
of neutrinos with nuclei.  Existing and planned precision measurements
of the weak mixing angle at low $Q$ do not constrain new physics which
is specific to neutrino-nucleon interactions.
Reference~\cite{Barranco:2007tz} explores the sensitivity of a 
coherent
$\nu$A scattering experiment on the tonne scale to some specific
physics beyond the standard model, including models with extra 
neutral gauge
bosons, leptoquarks and R-parity breaking interactions.

The signature of NSI is a deviation from the expected cross section.
Reference~\cite{Scholberg:2005qs} explores the sensitivity of an
experiment at the SNS.
As shown in the reference, under reasonable assumptions, if the rate
predicted by the SM is observed, neutrino scattering
limits more stringent than
current ones~\cite{Dorenbosch:1986tb, Davidson:2003ha} 
by about an order of magnitude can be obtained for some parameters.

Searches for NSI are based on precise knowledge of the nuclear form factors, which are known to better than 5\%~\cite{Horowitz:2003cz}, so that a deviation from the SM prediction larger than that would indicate
physics beyond the SM.

\item \textbf{Nuclear Physics from Coherent Scattering}

If we assume that the SM is a good description of nature,  then with sufficient
precision one can measure neutron form factors. 
References~\cite{Amanik:2007ce, Patton:2012jr} explore this possibility, that could
be within reach of a next-generation coherent scattering experiment.
One of the basic properties of a nucleus is its size, or radius, typically defined as
\begin{equation}
\langle R_{n,p}^{2} \rangle^{1/2} = \left( \frac{\int{\rho_{n,p}(r) r^{2} d^{3}r}}{\int{\rho_{n,p}(r) d^{3}r}}\right)^{1/2},
\label{eq:moment}
\end{equation}
where $\rho_{n,p}(r)$ are the neutron and proton density distributions.  Proton distributions in nuclei have been measured in the past to a high degree of precision.  In contrast, neutron distributions are still poorly known.  A measurement of neutron distributions could have an impact on a wide range of fields, from nuclear physics to astrophysics.

Previous measurements of the neutron radius have used hadronic scattering, and result in uncertainties of about $\sim 10\%$ \cite{Horowitz:1999fk}.  A new measurement, being done at Jefferson Laboratory by the PREX experiment, uses parity-violating electron scattering to measure the neutron radius of lead.  The current uncertainty in the neutron radius from this experiment is about $2.5\%$ \cite{Abrahamyan:2012gp}.  An alternate method, first suggested in~\cite{Amanik:2009zz}, is to study the neutron radius through neutrino-nucleus coherent scattering.

Also included in the cross section of equation~\ref{eq:dsigmadT} is the form factor, $F^{2}(Q^{2})$, a function of the momentum transfer $(Q^{2} = 2E^{2}TM/(E^{2} - ET))$.  For a spherical nucleus, the form factor is the Fourier transform of the nuclear densities:
\begin{equation}
F(Q^{2}) = \frac{1}{Q_{W}} \int{ \left( \rho_{n}(r) - (1 -
4\sin^{2}{\theta_{W}}) \rho_{p}(r) \right) \frac{\sin{(Qr)}}{Qr} r^{2} dr }\,,
\label{eq:formFactorIntegral}
\end{equation}
where, as in Eq. \ref{eq:moment}, $\rho_{n,p}(r)$ are the neutron and proton densities.    

The number of scattering events is calculated by folding the cross section with the neutrino spectra:
\begin{equation}
\frac{dN}{dT}(T) = N_{t} C \int_{E_{min}(T)}^{m_{\mu}/2}{ f(E)
\frac{d\sigma}{dT}(E,T) dE }\,, 
\label{eq:dNdT}
\end{equation}
where $N_{t}$ is the number of target nuclei in the detector, $C$ is the flux of
neutrinos of a given flavor arriving at the detector, and $E_{min}(T) =
\frac{1}{2}(T + \sqrt{T^{2} + 2 T M})$ is the minimum energy a neutrino must
have to cause a nuclear recoil at energy $T$.  The neutrino spectrum, $f(E)$, in this case is the Michel spectrum for neutrinos produced from pion decay.

For neutrinos produced by stopped pions, such as at the SNS facility, coherent neutrino-nucleus scattering will produce very low energy nuclear recoils, on the order of $\sim10-100$ keV.  In this case, the form factor can be Taylor-expanded around $Qr\approx0.$  Using the definition of moments of a distribution, we can now write the form factor as 
\begin{eqnarray}
F_{n}(Q^{2}) & \approx & \int{ \rho_{n}(r) \left( 1 - \frac{Q^{2}}{3!} r^{2} + \frac{Q^{4}}{5!}r^{4} -  \frac{Q^{6}}{7!}r^{6}  + \cdots \right) r^{2} dr } \\
 & \approx & N \left( 1 - \frac{Q^{2}}{3!} \langle R^{2}_{n} \rangle +
 \frac{Q^{4}}{5!}\langle R^{4}_{n}\rangle -  \frac{Q^{6}}{7!}\langle R^{6}_{n}\rangle + \cdots \right)\,.
 \label{eq:formFactorExpanded}
\end{eqnarray}

These moments can be calculated theoretically or extracted from an experimental measurement of the scattering curve.  For smaller nuclei such as argon and germanium, it is sufficient to cut the expansion after the fourth moment.  For larger nuclei, such as xenon, the sixth moment is also necessary.  This expansion allows the scattering curve to be parameterized using just two, or three, parameters.  

Detectors of Ar, Ge, and Xe are candidates for the site at the SNS.  Simple Monte Carlo techniques have shown that, using detectors on the tonne-scale, the neutron radius could be measured to a precision of a few percent~\cite{Patton:2012jr}.   In addition, information on higher moments could in principle be obtained.  The factors that influence the size of detector needed include how long data will be collected, how close to the source the detector is placed, and how well the luminosity of the flux is known.  

In Fig.~\ref{fig:kelly1}, a 500~kg detector of $^{40}$Ar was considered.  The detector was assumed to be at a distance of 30~m from the source at the SNS, experiencing a flux of $5.3\times10^{6}$ neutrinos/s/cm$^{2}$ for a full year.  Contours show the 97\%, 91\%, and 40\% confidence levels in the $\langle R_{n}^{2}\rangle^{1/2}$ - $L_{\nu}$ plane, where $L_{\nu}$ is the normalization of the neutrino flux.    With this size of detector, the neutron radius could be measured to $\sim \pm40\%$ at the 91\% confidence.  If the normalization of the neutrino flux were known to $\pm5\%$, a measurement of the neutron radius to $\sim\pm 20\%$ could be achieved.  At this size, $\langle R_{n}^{4}\rangle^{1/4}$ cannot be constrained.   

Figure~\ref{fig:kelly2} shows a result for a 15-tonne detector in the $\langle R_{n}^{2}\rangle^{1/2}$ - $\langle R_{n}^{4}\rangle^{1/4}$ plane. This detector experiences the same neutrino flux as the smaller detector, and also is assumed to run for a full year.  For the larger detector, the neutron radius could be measured to $\sim \pm4 - 6\%$ at 91\% confidence.  With the larger detector, the value of $\langle R_{n}^{4}\rangle^{1/4}$ can also be constrained to $\sim \pm25\%$ at the 91\% confidence level.  

\begin{figure}[htbp]
\centering
\includegraphics[height=2.7in,bb=2 157 647 538 ]{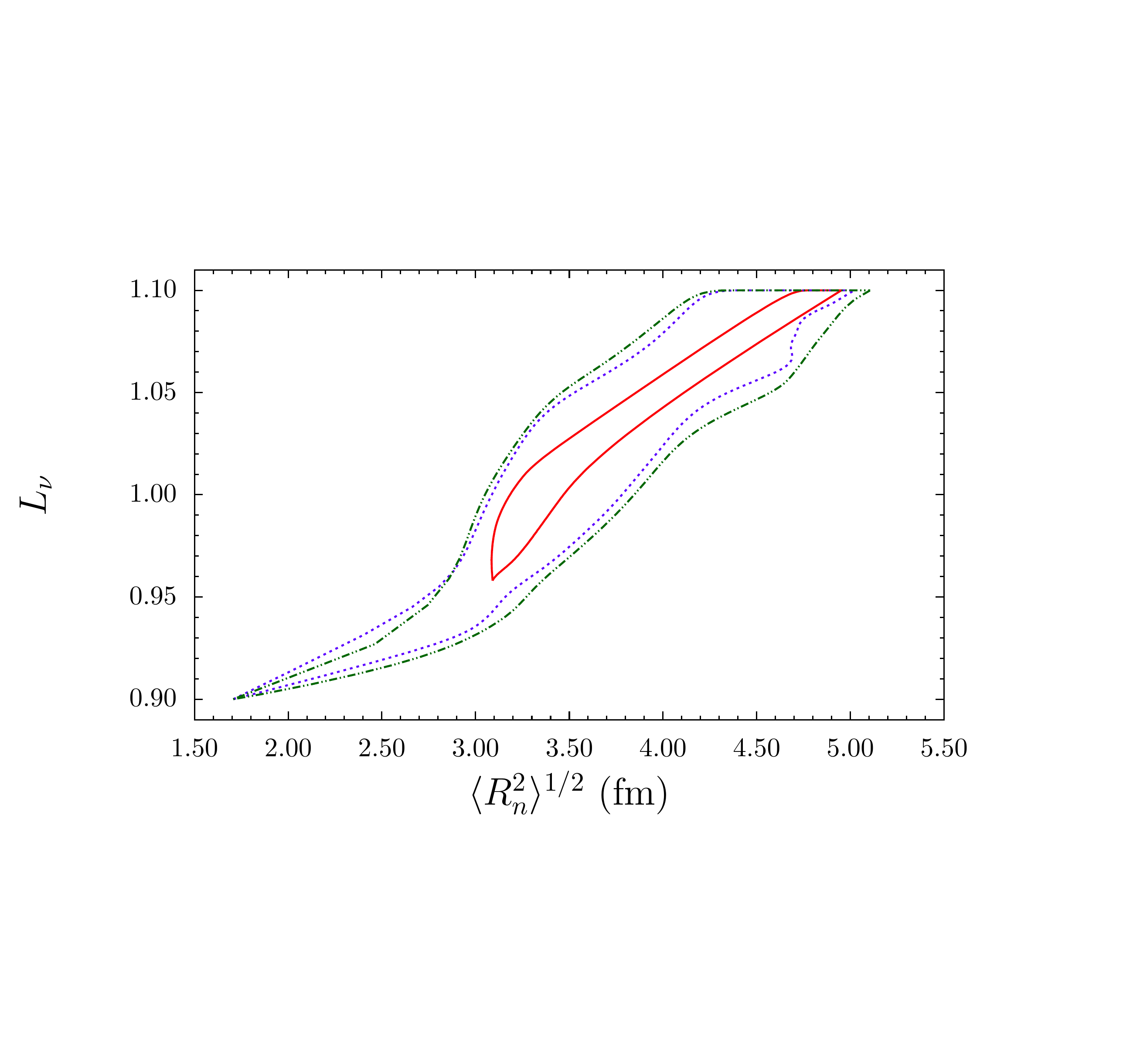}
\caption{Monte Carlo results for a 500 kg $^{40}$Ar detector placed 30 m from the source at the SNS for a full year.  Contour regions show the 40\% (solid red), 91\% (dotted blue), and 97\% (dashed green) confidence regions in the $\langle R_{n}^{2}\rangle^{1/2}$ - $L_{\nu}$ plane, where $L_{\nu}$ is the normalization of the neutrino flux.} \label{fig:kelly1}
\end{figure}

\begin{figure}
\centering
\includegraphics[height=2.7in,bb=2 157 647 538]{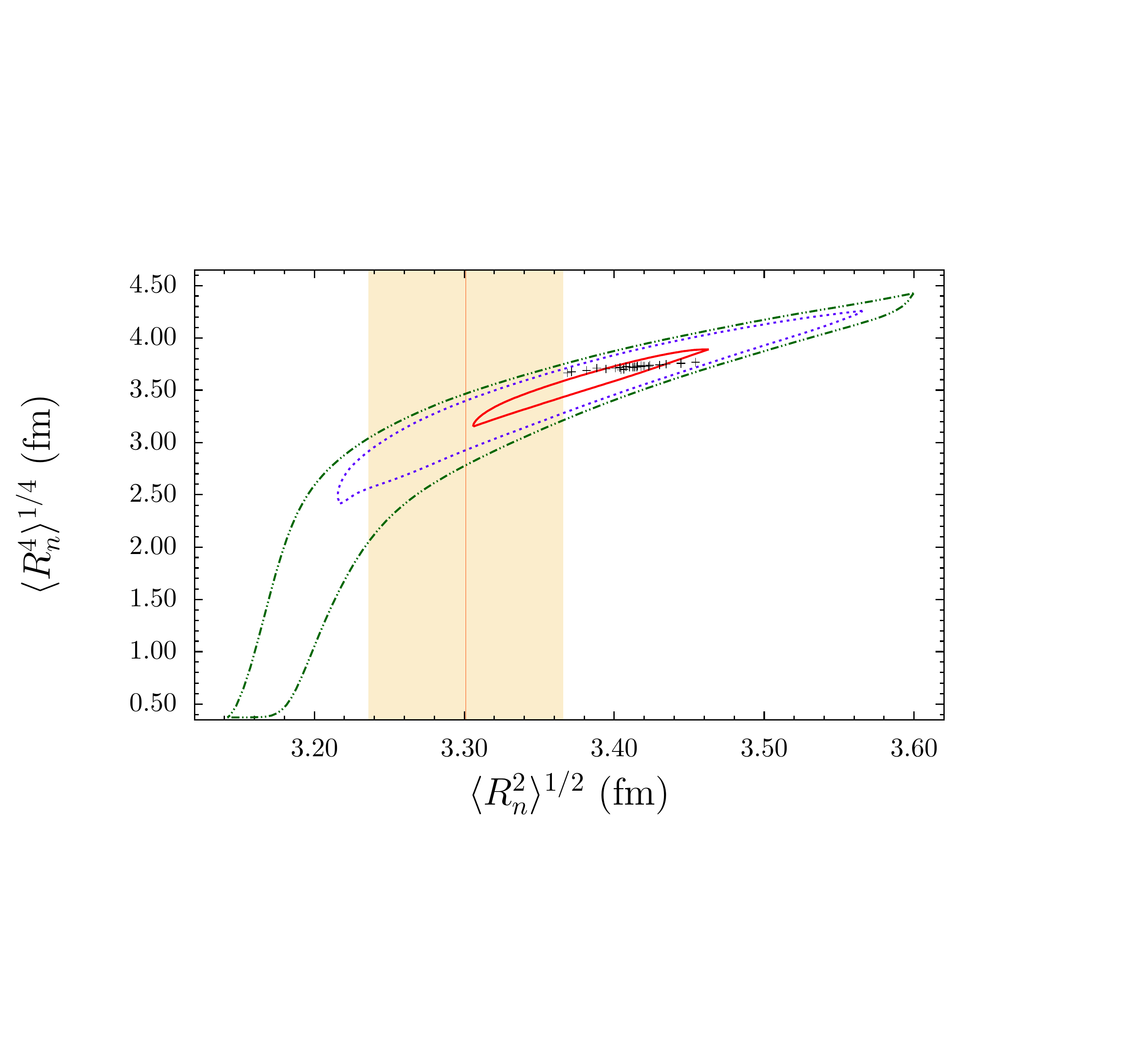}
\caption{Monte Carlo results for a 15 tonne $^{40}$Ar detector placed 30 m from the source at the SNS for a full year.  Contour regions show the 40\% (solid red), 91\% (dotted blue), and 97\% (dashed green) confidence regions in the $\langle R_{n}^{2}\rangle^{1/2}$ - $\langle R_{n}^{4}\rangle^{1/4}$ plane.  Black crosses show theoretical predictions, and the colored band indicates an experimental measurement reported in~\cite{Ozawa:2002av}. }  \label{fig:kelly2}
\end{figure}

\end{itemize}

Some specific possible experiments to measure coherent elastic $\nu$A scattering will be described in Section~\ref{cohmeas}.

\subsection{Hidden Sector Physics}\label{hidden_physics}

Besides the sterile neutrinos that come naturally as an extension of the SMl, there are more exotic light states with  very weak coupling to ordinary matter. Axions and axion-like particles (ALPs),   Hidden Sector or paraphotons (HSPs) that are charge-neutral under  SM forces may not be so under forces
described by ``new physics''.
The community studying the ``Hidden Sector'' (HS) came together during the latest Intensity Frontier workshop~\cite{Hewett:2012ns}.
In the report of the Intensity Frontier workshop, many direct searches for such particles motivated by various models are described. There is a natural division at about 1~MeV of particle mass between optical detection for low-mass cases, and experiments for models predicting particles massive enough to couple to at least an electron-positron pair in collider or fixed target machines.  The SNS potential corresponds to that of fixed-target machines.

Such experiments also have the benefit of rates a million times higher than the corresponding collider experiments~\cite{Hewett:2012ns}. This advantage can compensate for the lower geometric acceptance of a finite detector positioned many meters away from the interaction vertex. 
Prominent examples are neutrino experiments like CHARM, LSND, MiniBooNE, MINOS, MINER$\nu$A, T2K, and those planned for the Project X facility (see~\cite{Bergsma:1985qz,Athanassopoulos:1996ds,AguilarArevalo:2008yp,Ambats:1998aa,Osmanov:2011ig,Abe:2011ks,Tschirhart:2011yu}).
There is international interest in dedicated experiments.

The large numbers of secondary hadrons generated at the target/beam-dump  mostly decay, producing neutrinos flying through the decay volume.  These hadrons may also create a beam of other particles (paraphotons). With new advances in intensity reaching an integrated 10$^{21}$ POT, such a beam has now a chance of creating a substantial exotic  flux within the neutrino beam, which would then generate a measurable signal in a downstream detector. The estimated rate in the detector is proportional to the negative exponential of the ratio of the distance from the target to the decay length of the particle.  Such a high-intensity beam experiment offers an excellent opportunity to search for new, light, weakly-coupled states from the hidden sectors. 
The  potential production channels and the particle's resulting parameters are very strongly model dependent. The models and production channels range from simple elastic scattering of the incoming proton on the nuclei of the target~\cite{deNiverville:2011it},
producing hadron-size HS-particles, to kinetic mixing with photons~\cite{Hewett:2012ns} that produce pions which may decay into ALPs of meson-mass range, and nucleon-nucleon bremsstrahlung~\cite{Giannotti:2005tn}
which may describe the  production of lighter ($>$1~MeV) axions or ALPs in the Sun. Neutrino experiments have the capability of searching a vast parameter space of mass-lifetime (or mass-coupling) by being sensitive to both axions and ALPs and even to light dark matter~\cite{Essig:2010gu}.   

\begin{figure}
\centering
\includegraphics[width = .6\linewidth]{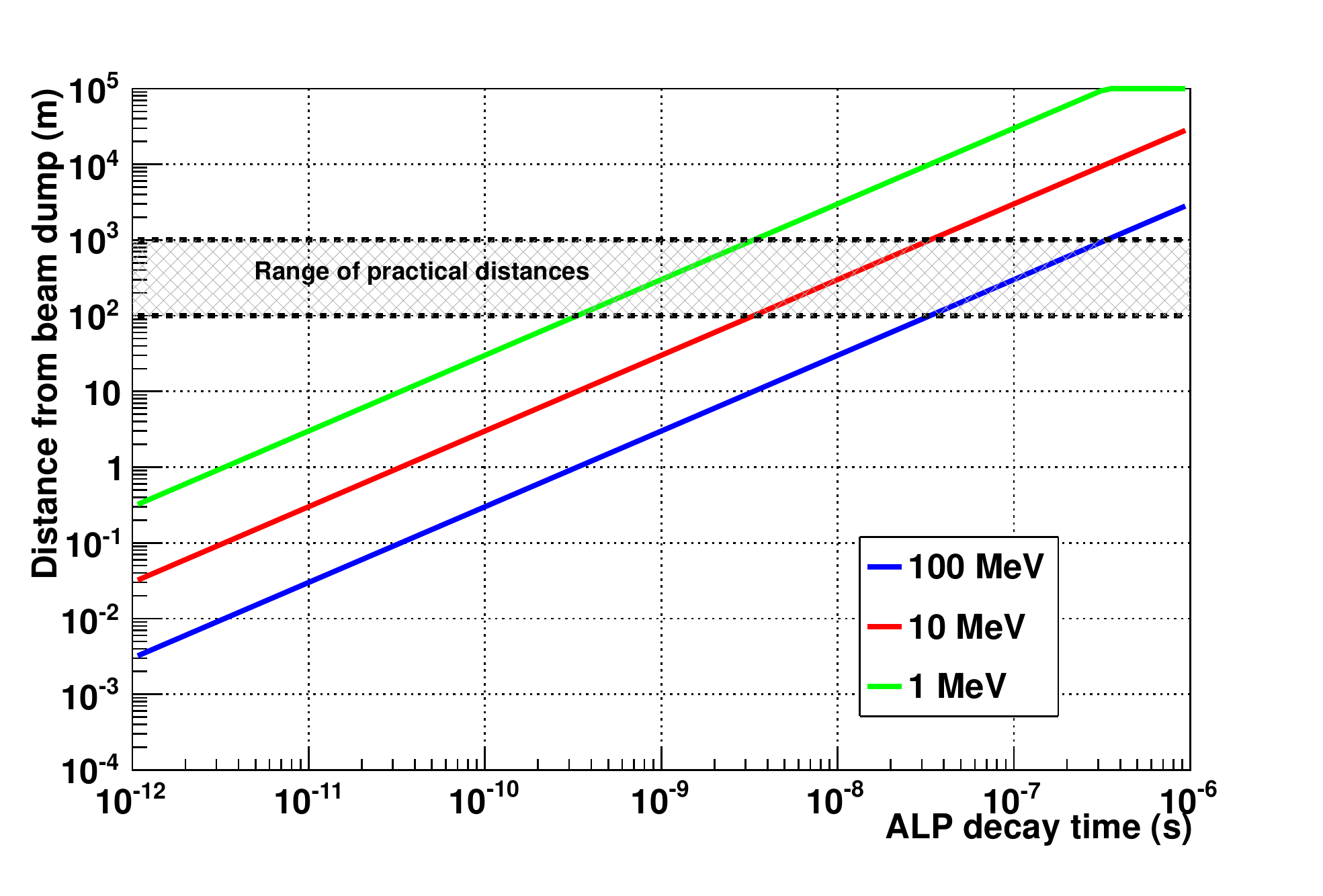}
\caption{Decay path curves for various masses of ALPs from the SNS beam dump, assuming 1~GeV energy.}
\label{fig:hidden1}
\end{figure}

Figure~\ref{fig:hidden1} shows a simple calculation to estimate sensitivity range for detection of a produced ALP, assuming that the ALP takes most of the $E=1$~GeV energy of the incoming proton. 
We plot the distance of decay as a function of particle rest lifetime $\tau$, $d=c\gamma \tau$, where $\gamma=E/m$, for three assumed particle masses.
The eventual sensitivity range will depend also on the physical parameters of the experiment (not the particle); for example it will depend on the background  along the decay path.
The minimum distance to the beam dump at which a detector can be placed is beyond where the secondary particles ($\sim$GeV pions and muons) have mostly decayed; this is within few hundred meters.   The flux and the geometrical acceptance of a detection system, decreases with distance.   With a maximum practical distance at 1~km we have bracketed the values of mass and lifetime of the ALP that the experiment is sensitive to.  Although the axion has been excluded in the middle range of 10~MeV  for lifetimes to which this experiment is sensitive~\cite{Kim:1986ax,deBoer:2005kf},
ALPs that do not follow the coupling constant-mass relation may be allowed and some combinations could be detectable if produced. Furthermore, there are no exclusions in the literature for ALPs as heavy as 1~GeV. 
Such a proposed experimental strategy aims to have a modest-size and -cost detector (few meters in cross section) that would cover multi-ranges of ALP mass. 
As mentioned, neutrino experiment near detector designs are well suited for such searches. The T2K experiment has a near detector at 280~m from the target and NO$\nu$A has one at 900~m.   An example detector setup for the SNS is described in Section~\ref{hidden_experiments}.

\section{Experimental Opportunities}\label{experiments}

In this section we describe some specific proposals for experiments that could address the physics goals described in the previous section.  Some of these have been proposed in the past and have relatively well-fleshed-out designs; others are at a more conceptual stage.  We organize these into four main categories, although there is some overlap between physics motivations (some experiments could be used for both sterile searches and cross section measurements, for example), as well as complementarity.  The categories are: sterile neutrino oscillations, CC and NC cross section measurements, coherent elastic $\nu$A scattering, and hidden sector searches.

\subsection{Sterile Neutrino Oscillation Searches}\label{sterileexp}

Sterile neutrino oscillation searches can be done using flavor-blind NC interactions, either using inelastic nuclear excitations or coherent elastic scattering (\textit{e.g.}~\cite{Anderson:2012pn}).  The SNS muon flavor flux is particularly suitable for NC disappearance experiments, as CC interactions are kinematically inaccessible to 30-MeV $\nu_\mu$. 
Direct tests of the LSND appearance oscillation are also possible.

\subsubsection{OscSNS}\label{oscsns}

The motivation for sterile neutrino oscillations is described in Section~\ref{sterile}.
The OscSNS experiment~\cite{oscsns} can carry out a unique and decisive test of the 
LSND appearance $\bar \nu_\mu \rightarrow \bar \nu_e$ signal that has not been executed to date. The 
existence and properties of light sterile neutrinos are central to understanding the creation of the heaviest 
elements and verifying our understanding of the nuclear reactions involved in nuclear reactors. An 800-tonne
Scintillator-oil OscSNS detector with $\sim 3000$ 8-inch phototubes, based on the LSND and MiniBooNE detectors,
can be built for $\sim \$20$M (or $< \$20$M 
if the MiniBooNE oil and phototubes are reused).

In addition to the general advantages offered by the SNS for neutrino oscillation physics (known neutrino spectra, well-understood neutrino cross sections for reactions on protons and electrons, low duty cycle for cosmic ray 
background rejection, low beam-induced neutrino background, and a very high isotropic neutrino flux), there are some specific advantages for an LSND oscillation test.
Note that 
$\pi^-$ and $\mu^-$ mostly capture in the Hg target before they have a chance to decay, so that hardly any neutrinos 
are produced from either $\pi^- \rightarrow \mu^- \bar \nu_\mu$ or $\mu^- \rightarrow e^- \bar \nu_e \nu_\mu$
decay. The rapid capture of the negatively-charged meson in the Hg environment is an advantage over the LSND 
experiment, where the production target was more open with greater possibility of $\pi^-$ decay in flight and 
the resulting $\mu^-$ decaying to $e^- \bar \nu_e \nu_\mu$.

\begin{figure}
\vspace{5mm}
\centering
\includegraphics[width=8.5cm,clip=true]{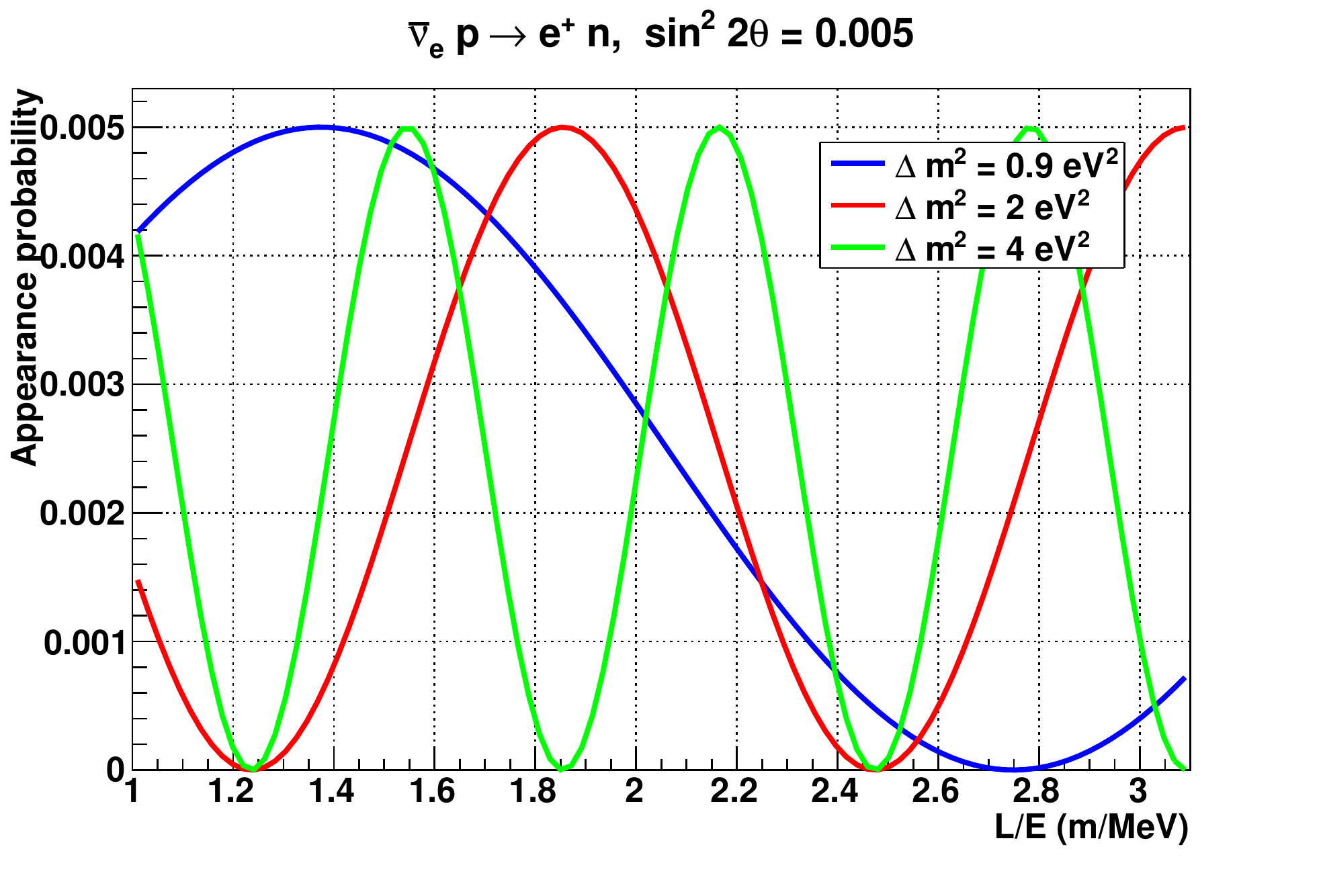}
\vspace{1mm}
\caption{
Probability of $\bar \nu_e$ appearance oscillations as a function for $L/E$, for a range relevant for a 10-m-long OscSNS detector volume and for values of
$\Delta m^2$ around 1 eV$^2$.
}
\label{nuebar_osc}
\end{figure}

\begin{figure}
\vspace{5mm}
\centering
\includegraphics[width=8.5cm,clip=true]{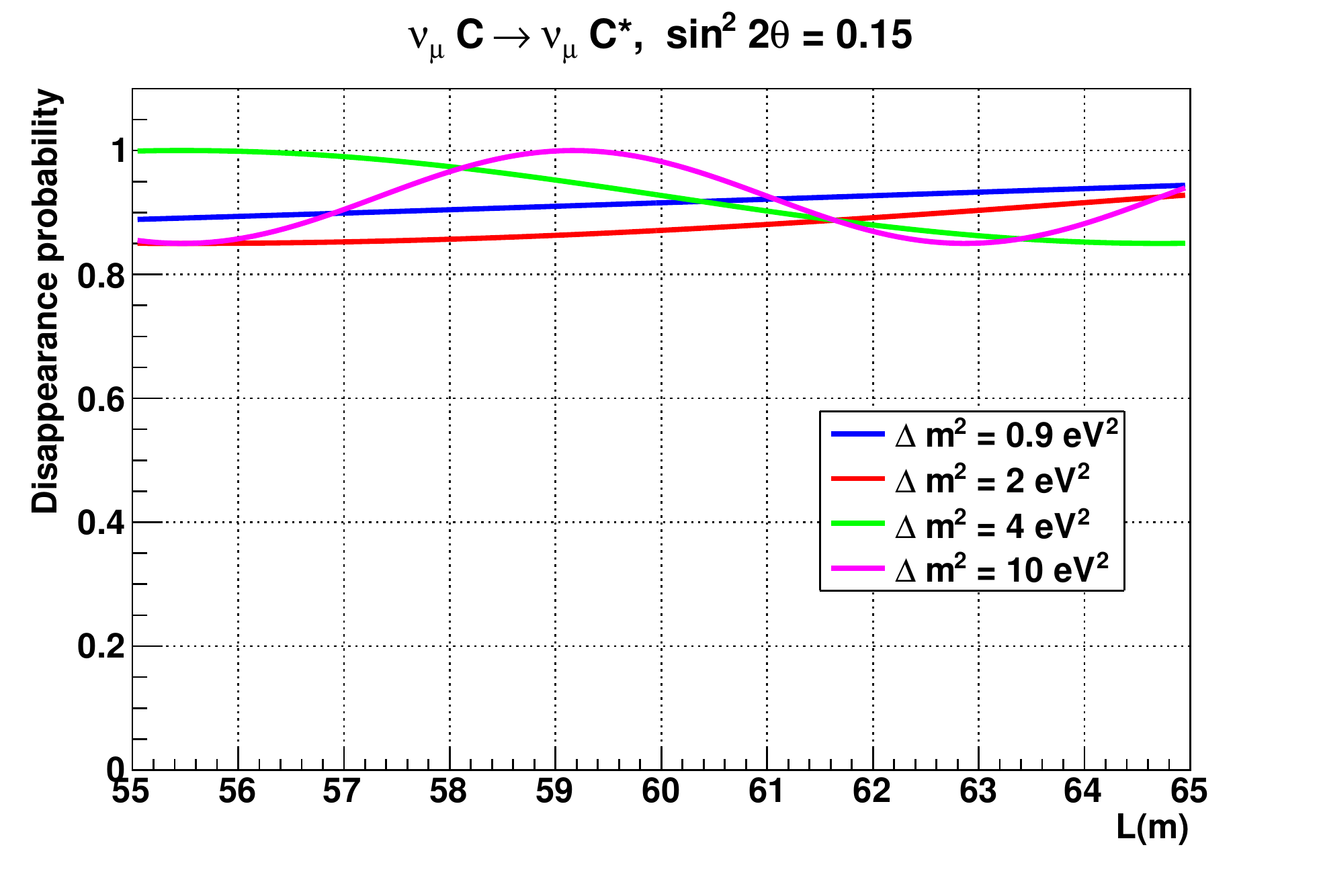}
\vspace{1mm}
\caption{
Probability of monoenergetic $\nu_\mu$ disappearance oscillations as a function of $L$ in a 10-m-long OscSNS detector volume for values of
$\Delta m^2$ around 1~eV$^2$.
}
\label{numu_osc}
\end{figure}

\begin{figure}
\vspace{5mm}
\centering
\includegraphics[width=8.5cm,clip=true]{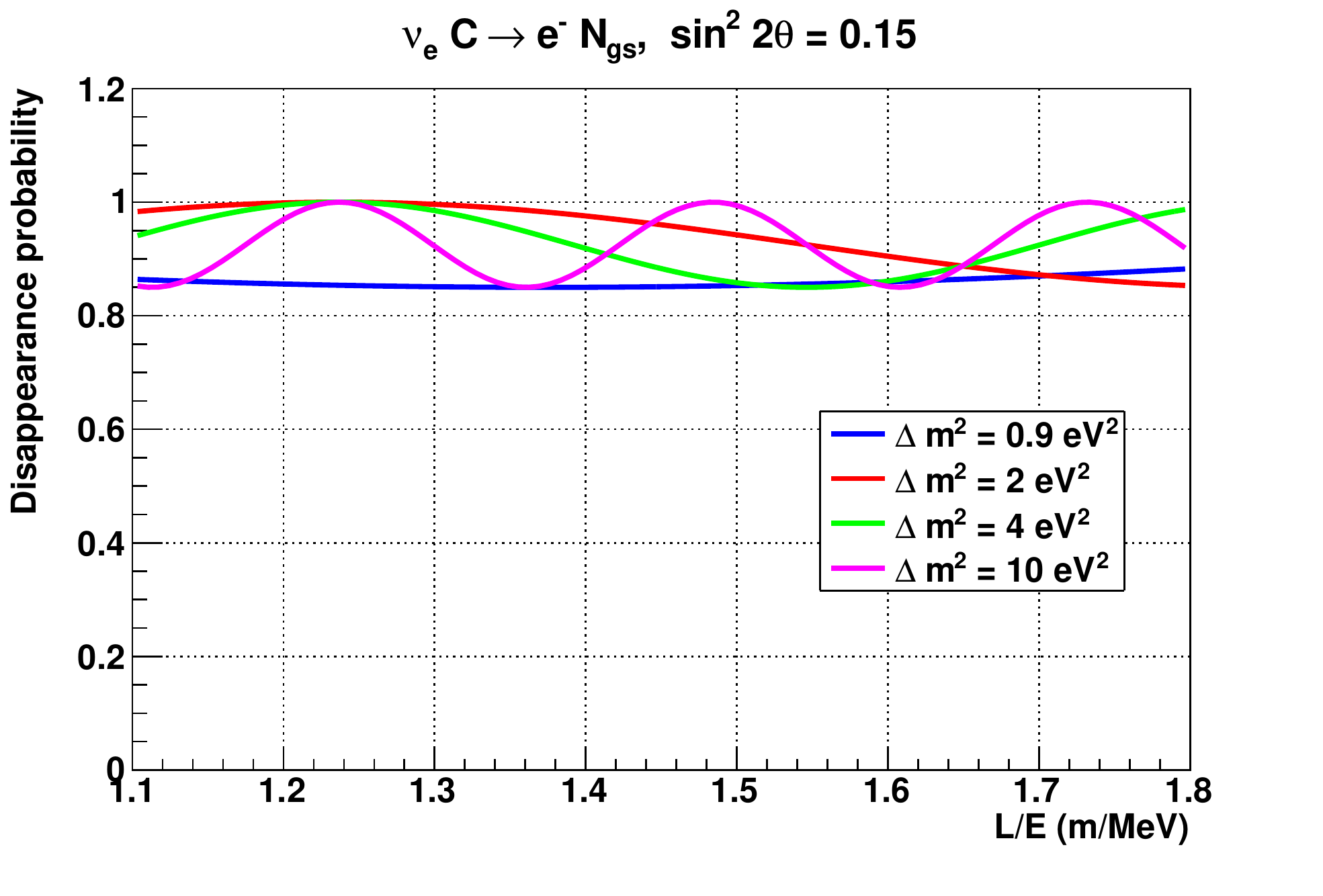}
\vspace{1mm}
\caption{
Probability of $\nu_e$ appearance oscillations as a function for $L/E$, for a range relevant for a 10-m-long OscSNS detector volume and for values of
$\Delta m^2$ around 1 eV$^2$.
}
\label{nue_osc}
\end{figure}

\begin{figure}
\vspace{5mm}
\centering
\includegraphics[width=5.5cm,angle=90,clip=true]{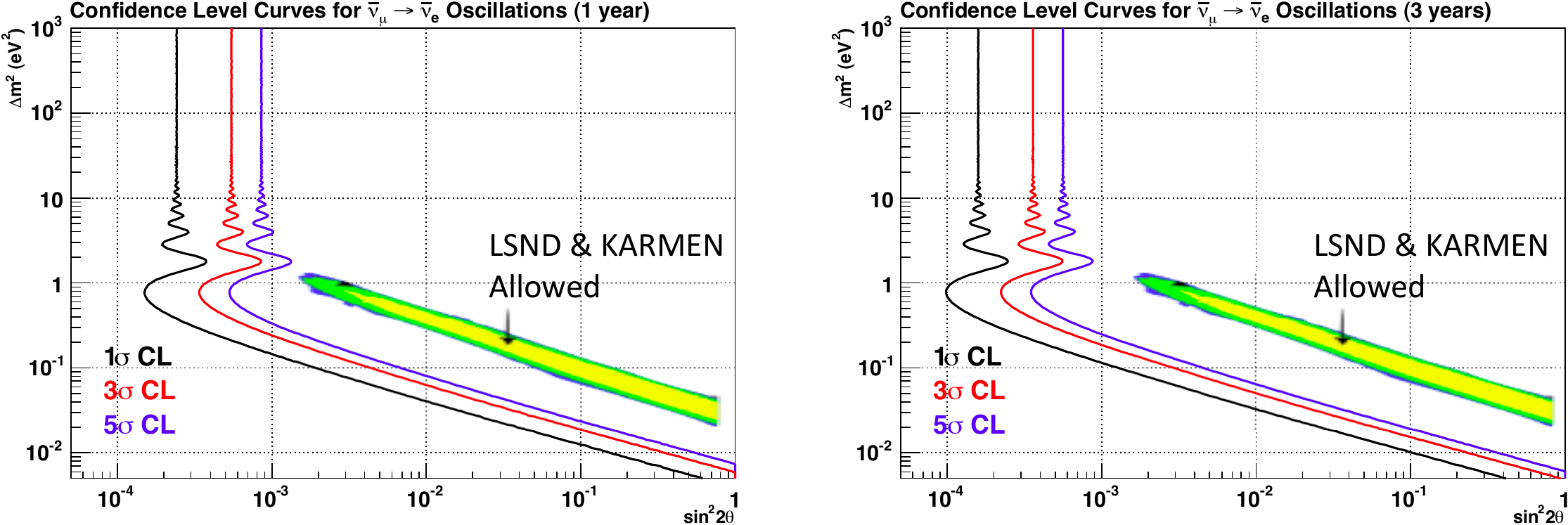}
\vspace{1mm}
\caption{
The OscSNS sensitivity curves for the simulated sensitivity to $\bar \nu_\mu \rightarrow \bar \nu_e$  
oscillations after one (left) and three (right) years of operation. Note that it has more than 5$\sigma$
sensitivity to the LSND result in 1 year.
}
\label{app}
\end{figure}

\begin{figure}
\vspace{5mm}
\centering
\includegraphics[width=5cm,angle=90,clip=true]{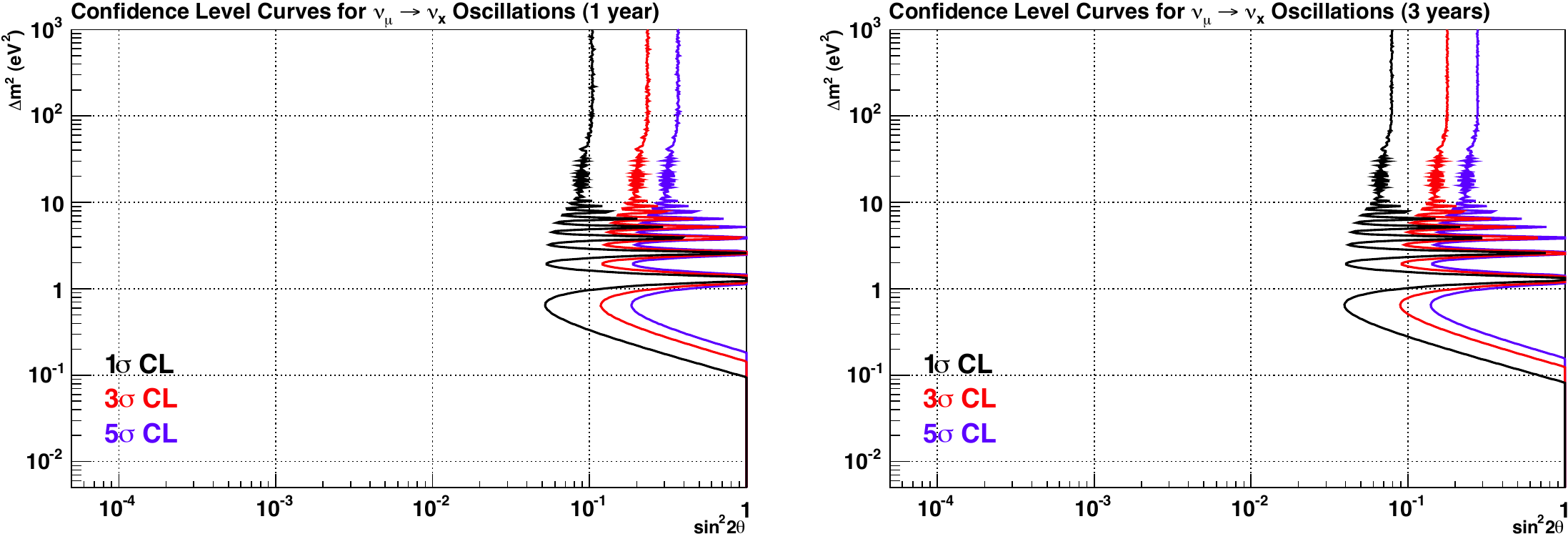}
\vspace{1mm}
\caption{
The OscSNS sensitivity curves for $\nu_\mu$ disappearance for 1 and 3 years, respectively.  
}
\label{disap}
\end{figure}

The SNS neutrino flux is ideal for probing $\bar \nu_\mu \rightarrow \bar \nu_e$ and
$\nu_\mu \rightarrow \nu_e$ appearance,
as well as $\nu_\mu$ disappearance into sterile neutrinos. The appearance searches
both have a two-fold coincidence for the rejection of background. For $\bar \nu_\mu \rightarrow \bar \nu_e$
appearance, the signal is an $e^+$ in coincidence with a 2.2 MeV $\gamma$: $\bar \nu_e p \rightarrow e^+ n$,
followed by $n p \rightarrow D \gamma$. For $\nu_\mu \rightarrow \nu_e$ appearance, the signal is an $e^-$ 
in coincidence with an $e^+$ from the $\beta$ decay of the ground state of $^{12}N$: $\nu_e~^{12}C \rightarrow
e^-~^{12}N_{gs}$, followed by $^{12}N_{gs} \rightarrow ~^{12}C e^+ \nu_e$. The disappearance search will detect 
the prompt 15.11 MeV $\gamma$ from the NC reaction $\nu_\mu C \rightarrow \nu_\mu C^*$(15.11). 
This reaction has been measured by the KARMEN experiment, which has determined a cross section that is 
consistent with theoretical expectations. However, the KARMEN result was measured in a sample of 86 events, 
and carries a 20\% total error. OscSNS will be able to greatly improve upon the statistical and systematic 
uncertainties of this measurement. If OscSNS observes an event rate from this NC reaction that 
is less than expected, or if the event rate displays a sinusoidal dependence with distance, then this will be 
evidence for $\nu_\mu$ oscillations into sterile neutrinos.
Figures~\ref{nuebar_osc}, \ref{numu_osc}, and \ref{nue_osc} show how neutrino oscillations could actually be observed within the
detector itself for $\bar \nu_e$ appearance, $\nu_\mu$ disappearance, and $\nu_e$ disappearance, respectively.
Figures~\ref{app} and \ref{disap} show the $\bar \nu_e$ appearance and $\nu_\mu$ disappearance parameter sensitivities.

In addition to the neutrino oscillation searches, OscSNS will also make precision cross section measurements of 
$\nu_e C \rightarrow e^- N$ scattering and $\nu e^- \rightarrow \nu e^-$ elastic scattering. The former reaction 
has a well-understood cross section and can be used to normalize the total neutrino flux, while the latter 
reaction, involving $\nu_\mu$, $\nu_e$, and $\bar \nu_\mu$, will allow a precision measurement of 
$\sin^2 \theta_W$ at low energy.

\subsection{CC and Inelastic NC Cross Section Measurement Experiments}\label{ccncmeas}

We describe here some possible experimental setups for measuring CC and NC neutrino-nucleus interactions, for which physics motivations are discussed in Section~\ref{ccncphys}.  One of these, NuSNS, has in the past been studied in some detail~\cite{Avignone:2003ep,nusns}; others (argon and lead) are at a more conceptual stage.  We emphasize that these are just examples; other targets and detector configurations could be deployed.

\subsubsection{NuSNS: A Multi-Target Facility}

We describe here a possible shielded neutrino detector enclosure at a closest possible proximity from the SNS target, where the short-pulse neutrino flux will be several times greater than has been achieved at any previous facility. The enclosure can hold two independently operable detectors, designed so that measurements with several different targets can be performed with little modification to the detectors. The anticipated neutrino flux at the SNS facility will allow measurement of the CC neutrino-nucleus cross section for any selected nuclear target to a statistical accuracy of better than 10\%. We anticipate that this can allow double-differential cross section measurements (vs. energy and angle), and that NC measurements may also be possible. 
As described in Section~\ref{ccncphys}, measurements with this level of precision will provide a unique test of fundamental questions in nuclear structure (allowing the resolution of the forbidden component of the strength distribution) and will validate the complex nuclear structure models required to compute these cross sections for nuclei that will not be measured. Armed with measured rates and improved nuclear structure theory, we will be able to improve our understanding of supernovae, important links in our cosmic chain of origins.

Two possible detector options could cover a wide range of nuclear targets that can be studied at the SNS.
 The first is a segmented detector:  this is a flexible universal detector in which neutrino interactions with a variety of targets can be studied. To achieve this, the detector must:
contain ten fiducial tons of target material and fit inside the fixed shielding volume,
minimize the mass of non-target material, have the capability to easily replace targets without rebuilding the sensitive part of the detector,
have relatively good time resolution (few ns) to facilitate operation in the SNS background environment, have particle identification capability and allow three-dimensional track reconstruction for further background discrimination,
have good energy resolution to allow differential cross section measurements, and
be affordable (minimize channel count).

A detector satisfying these requirements is a highly segmented detector with strawtubes separated by thin-walled corrugated sheets of the target material. Signals can be read from both sides of each strawtube’s anode wire in order to provide time information and three-dimensional position information by charge division. A similar concept was used by the 1000-tonne Soudan-II proton decay experiment~\cite{Allison:1996wp}. However, the measurements here require a detector with much finer segmentation and significantly improved time resolution.

The energy of detected particles can be determined by the number of strawtubes hit, which is closely related to the particle’s range. For electrons in the energy range of a few tens of MeV, the energy resolution obtained by a measurement of the track length is comparable to that obtained by energy sampling~\cite{nusns}.

Gas-based detectors have a number of advantages over other detector technologies for this application because they are less expensive than other detectors (\textit{e.g.}, scintillator) and do not require an expensive readout system. In addition, the low detector mass eliminates the necessity to statistically separate interactions in the target from interactions in the detector.

\begin{figure}
\vspace{5mm}
\centering
\includegraphics[width=7.5cm]{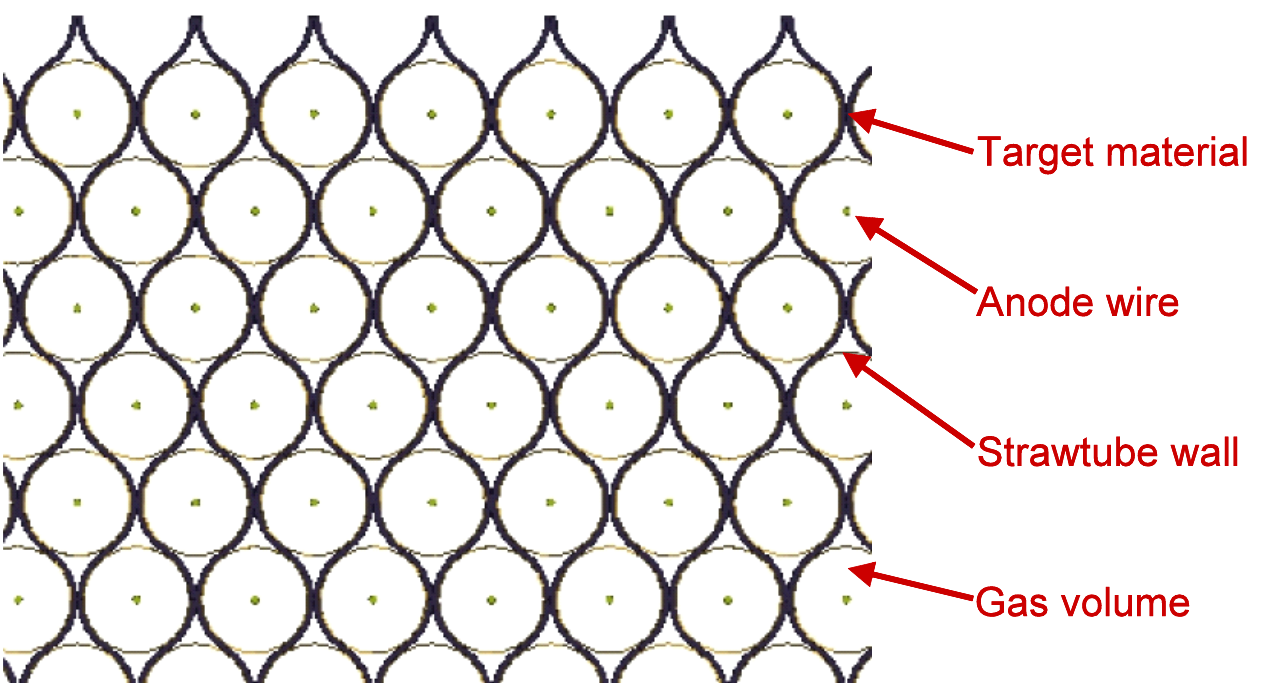}
\vspace{1mm}
\caption{Schematic cross cut view of the segmented detector. }
\label{fig:segmented}
\end{figure}

With such a segmented detector one can study a wide range of nuclei that can exist in metallic form. For example, an initial set of targets could consist of Al, Cu, Fe, Pb.

Another possible detector is a
homogeneous detector,  a flexible universal detector in which neutrino interactions with a variety of liquid targets can be studied. To achieve this, the detector must:
contain at least ten fiducial tonnes of liquid target material and fit inside the fixed shielding volume,
have the capability to easily replace targets without rebuilding the sensitive part of the detector, have relatively good time resolution (few ns) to facilitate operation in the SNS background environment,
have good energy and angular resolution, even with non-scintillating target materials, in order to allow differential cross section measurements,
have sufficient pixelization to allow for electron identification through detection of the Cherenkov ring in the presence of scintillating light, and be affordable.

\begin{figure}
\vspace{5mm}
\centering
\includegraphics[width=7.5cm]{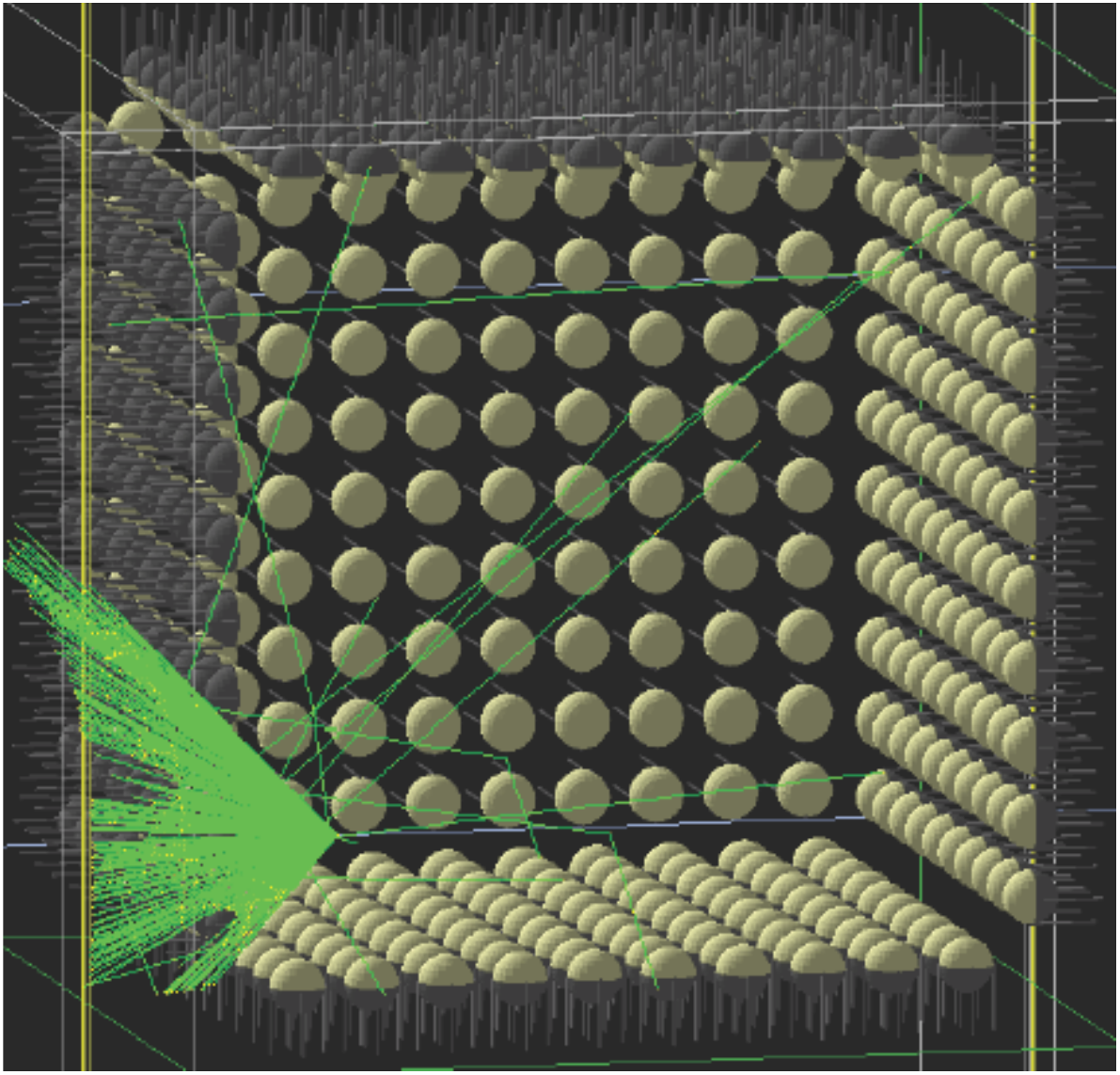}
\vspace{1mm}
\caption{
Schematic view (Geant) of the $\nu$SNS homogeneous detector showing response to a 1.5~MeV electron. The beige surfaces show the active area of the photocathodes. The front wall and its PMTs have been removed for clarity.
}
\label{fig:homogeneous}
\end{figure}

A candidate homogeneous detector consists of a $3.5 \times 3.5 \times 3.5$ m$^3$ steel vessel with 600 8'' photomultiplier tubes (PMTs) mounted on the inner walls, to provide approximately 41\% photocathode coverage. A schematic drawing of the detector is shown in Fig.~\ref{fig:homogeneous}. The actual distribution and orientation of the PMTs can be optimized using Monte Carlo simulations. It is expected that at least the edge and corner PMTs should be angled such that the light collection efficiency is maximized. The 41\% surface coverage allows the detector to have good event reconstruction and particle identification when operating with a variety of fluids as active media (\textit{e.g.} mineral oil, water, heavy water), independent of the amount of scintillator doping (\textit{i.e.} operation as a pure Cherenkov imaging detector).

With such a detector one could study any nuclei that can exist in liquid transparent form. For example, an initial set of targets could be O, C, d, I.

 \subsubsection{Argon Cross Section Measurement Experiments}   
A direct measurement of the $\nu_e(^{40}{\rm Ar},^{40}{\rm K}^*)e^-$ cross section at proton beam dump facilities has been repeatedly advocated~\cite{Raghavan:1986hv,Kolbe:2003ys,SajjadAthar:2004yf}.
Electron neutrino reactions in $^{40}$Ar can be experimentally studied
with  muon decay-at-rest neutrinos, whose
energy spectrum, with maximum energy E$_\nu^{max}$=53 MeV, is similar to the SNS neutrino spectrum.

The absorption reaction signature from  the SNS (as well as from supernova neutrinos) in a LArTPC detector is well defined. It consists of one primary, prompt energetic electron track (with electromagnetic shower activity when above $\sim$30 MeV, the critical energy in LAr) surrounded by a cluster of secondary electron tracks in a volume ($r \sim 50$ cm) around the primary vertex in the MeV range and below from Compton conversion of K$^*$ deexcitation $\gamma$s (in LAr $X_0=14$ cm). 
The energy of the prompt recoil electron yields the neutrino incident energy. The total energy of the secondary photons
(partly prompt and partly delayed) corresponds, when all the Compton electrons are detected, to the K$^*$ level above the ground state of the nuclear transition.

\begin{figure}[ht]
\begin{centering}
\begin{tabular}{c c}
\includegraphics[height=2.in]{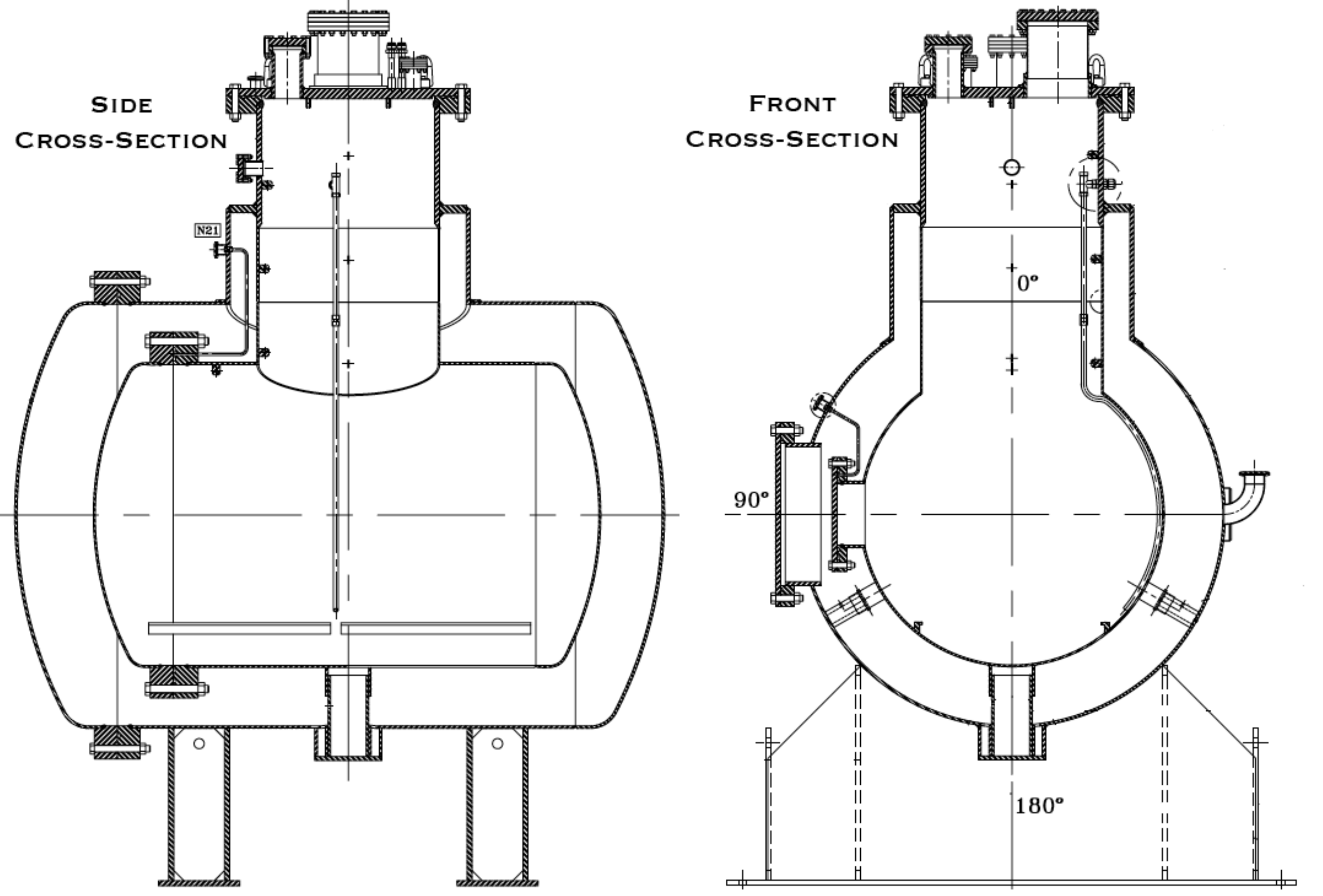} 
&
\includegraphics[height=3.0in]{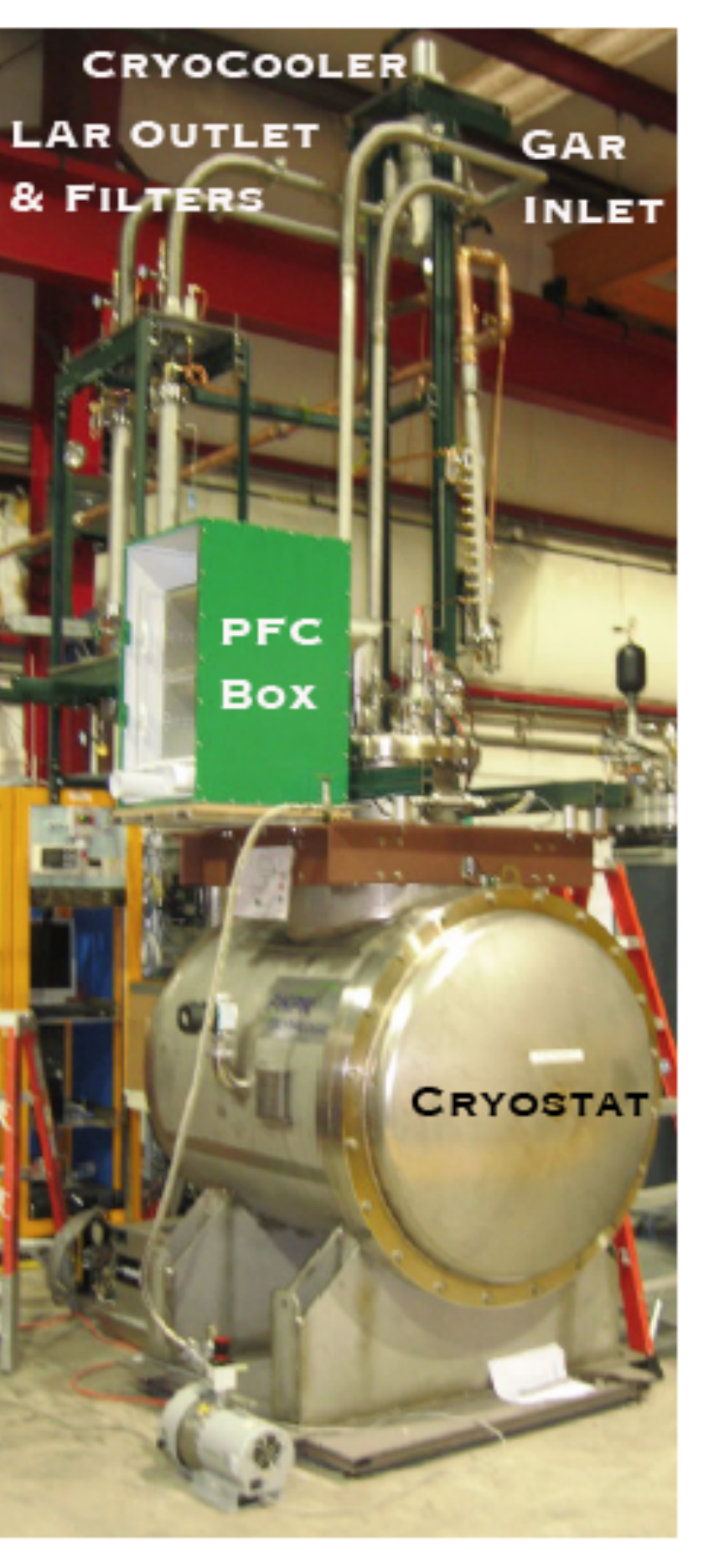} 
\end{tabular}
\caption{Left: side and front cross-sectional views of the {\sf ArgoNeuT} cryostat. The inner and outer vessels, the chimney on the top and the removable end-caps at one side are visible in the drawings. Right: picture taken during assembly with details of the cooling and recirculation system (the 4-pipe pathways of the recirculation circuit and the cold head of the cryocooler located inside a vacuum-insulated containment vessel).}
\label{fig:argoneut-det}
\end{centering}
\end{figure}

The neutrino fluxes  at the SNS
facility \cite{Avignone:2003ep}, with a nominal (time-integrated) $\nu_e$ fluence $F_{\nu e}=6\times 10^{14}~\nu/{\rm cm}^2$ at a distance of 20 m,
allows the collection of a large number of absorption events of the order of $N_{evt}\simeq1800$ per tonne of LAr per yr (assuming the theoretical cross section from \cite{SajjadAthar:2004yf} or \cite{Kolbe:2003ys}).
A $1\times 1\times 2~{\rm m}^3$ ($\sim$3 tonne) LArTPC detector looks thus definitively adequate, considering that for LArTPCs the detection efficiency $\epsilon_{Det}$ can be assumed close to unity and the fiducial volume cut for containment of the final state event topology should be small (${\rm V}_{Fid}\simeq {\rm V}_{Active}$).
The neutron background (faking the $\nu_e$ absorption signature) is small, due to the high rejection capability of the imaging LArTPC technology.  Cosmic ray background and trigger efficiency need to be
understood, however, and carefully evaluated (LArTPC is a slow read-out detector, \textit{e.g.} 600 $\mu$s for 1~m drift).

In principle, a smaller detector, such as the existing {\sf ArgoNeuT} LArTPC detector \cite{Anderson:2012vc}, can also be used. The {\sf ArgoNeuT} detector, shown in Fig.\ref{fig:argoneut-det}, made of a 550~liter cryostat and a TPC of  $40~{\rm height}\times 47~{\rm width}\times 90~{\rm length}~$cm$^{3}$, corresponding to about $\sim$170~liters of LAr active volume, can detect about 350~events per year in ArgoNeuT at the SNS, if a suitable location at 20~m distance from the target is made available. Though the statistics are limited, a first direct measurement of the $\nu_e$ absorption reaction on argon could be possible.\footnote{The {\sf ArgoNeuT} detector is presently committed for an extended run at the Fermilab Test Beam Facility (2012-14). A subsequent transportation and use at the ORNL SNS facility would require an agreement with the {\sf ArgoNeuT} Collaboration and the institutions who own different components of the detector and readout system.}

\begin{figure}[ht]
\begin{center}
\includegraphics[width=14cm]{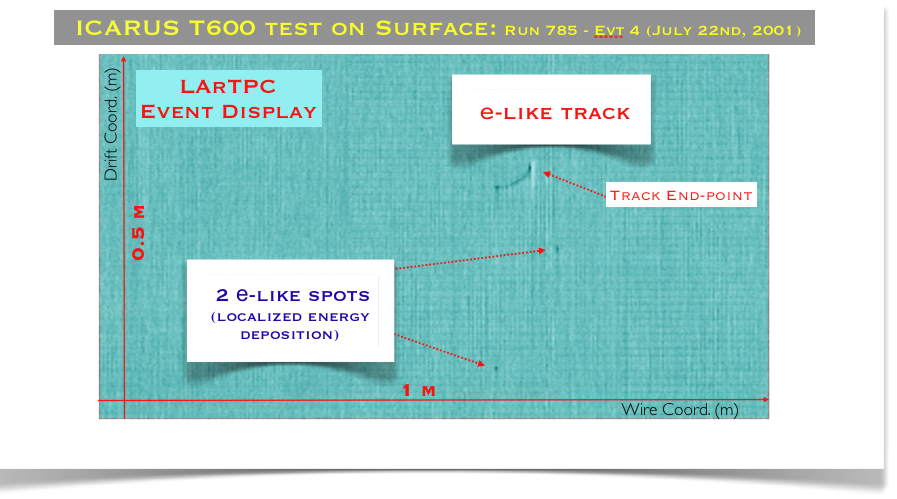}
\caption{
Cosmic-ray-background-induced event from ICARUS T600. An electron-like track of about 11~cm length is visible with a reconstructed energy of about 24 MeV, with two nearby 
 spots of 1 and 1.5~MeV.
}
\label{fig:SN-like-ICA}
\end{center}
\end{figure}
An event detected by the ICARUS T600 detector during the commissioning test run on surface in 2001 has been reported \cite{Cavanna:2001gp}. Its topological signature, shown in Fig.~\ref{fig:SN-like-ICA},  with an electron-like track and a couple of 
localized energy  depositions in the surrounding volume, is similar to the expected signature for a supernova or an SNS neutrino interaction in LAr.
The event was very likely induced by a cosmic ray interaction in the material surrounding the LAr volume. 

With LArTPC detectors, only a modest cut of about 8~MeV threshold in the neutrino energy spectrum is required. This threshold is determined by  
a minimum energy around 5~MeV for the leading electron track detection, in addition to the 
Q-value of the $(^{40}{\rm Ar},^{40}{\rm K}^*)$ nuclear transition (2.3 MeV to the lowest-lying GT level), but an accurate calibration of the detector energy response is necessary.

A LArTPC detector of modest mass, of the order of a tonne (or even less) in a suitable location at the shortest possible distance from the proton target (\textit{e.g.} 20~m), can detect thousand events per year, allowing for a first and indispensable measurement of the so-far-unmeasured neutrino cross section of  the absorption reaction, the relevant  channel for supernova detection with LAr-based detectors.
The experimental result in comparison with the theoretical
results available from various models will be extremely helpful in understanding and constraining the theoretical uncertainties in the neutrino-nuclear cross sections due
to nuclear structure in the supernova energy region. 
The final goal of reducing the cross section uncertainty to the order of 10\% (or less, possibly) will improve sensitivity in the parameter space of current supernova models, for the lucky circumstance of a stellar core-collapse event in our Galaxy detected by a large mass LArTPC detector at Earth.

 \subsubsection{Lead Cross Section Measurement Experiments}  

\begin{figure}[ht]
\begin{center}
\includegraphics[width=9cm]{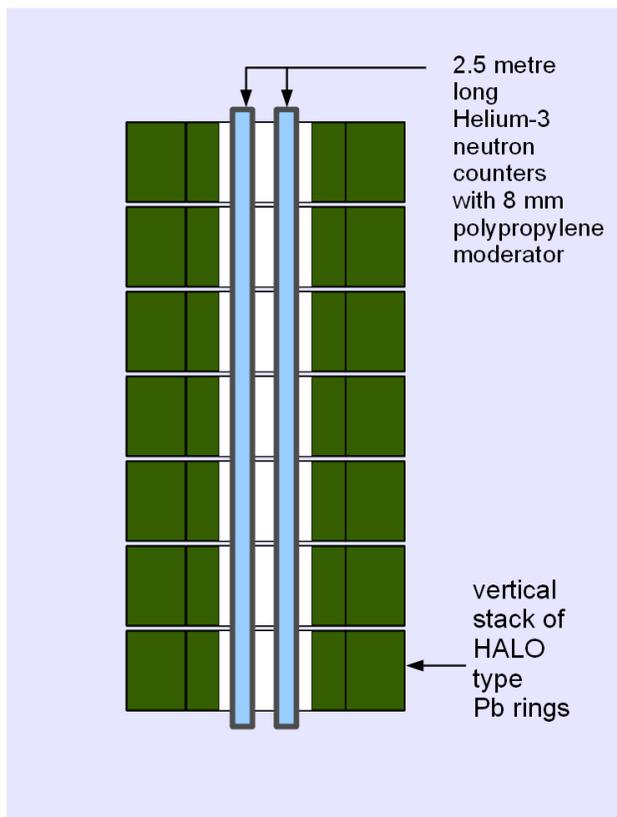}
\caption{
Lead target consisting of lead rings and $^3$He counters.
}
\label{fig:pbtarget}
\end{center}
\end{figure}

Experiments to measure cross sections of neutrinos on lead could consist of lead bricks or plates combined with neutron detectors, with or without detectors (\textit{e.g.} scintillators) to measure $e^-$ from CC $\nu_e$ interactions.  
One possible concept is illustrated in Fig.~\ref{fig:pbtarget}, making use of the type of lead rings and remaining leftover $^{3}$He counters from SNO.  This example would be especially relevant to HALO, but other detector configurations are also possible.  Detector materials such as lead perchlorate~\cite{Elliott:2000su} or lead acetate~\cite{yen} are also conceivable. 


\subsection{Experiments to Measure Coherent Elastic $\nu$A Scattering}\label{cohmeas}

The physics that can be addressed by coherent elastic scattering experiments is described in Section~\ref{coherent_phys}.   One can imagine approximately three experimental phases with different experimental scales that will address different physics:

\begin{itemize}
\item[] Phase 1: a few to few tens of kg of target material (depending on distance to the source) could make the first measuremenet.
\item[] Phase 2: a few tens to hundreds of kg of target material could set significant limits on non-standard neutrino interactions, and could also begin to address sterile neutrino oscillations, depending on configuration.
\item[] Phase 3: a tonne-scale or more experiment could begin to probe neutron distributions.
\end{itemize}

Various technologies are suitable at different scales.   We highlight two possibilities here (a single-phase Ar and/or Ne detector and a LXe TPC), for which specific scenarios have been relatively well fleshed out.  However we emphasize that other technologies could also be suitable.

We note also that the measured neutrino flux will be a valuable input to the proposed OscSNS experiment~\cite{oscsns} (see Section~\ref{oscsns}) as well.

\subsubsection{Experiments for Coherent Scattering on Argon or Neon}

One possiblity is a low-energy-threshold single-phase liquid argon detector to observe the coherent elastic $\nu$A scattering at the SNS~\cite{Scholberg:2005qs}. The proposed noble liquid neutrino detector is conceptually similar to dark matter detectors using liquid argon. This kind of detector will utilize pulse-shape discrimination of scintillation light between nuclear recoil and electron recoil interactions (and ionization yield) in the liquid argon to identify coherent elastic $\nu$A interactions from  background events. The majority of electromagnetic and neutron backgrounds can be rejected using the standard active and passive shielding methods together with self-shielding fiducialization. 

\begin{figure}[!ht]
  \centering
    \includegraphics[height=2.0in]{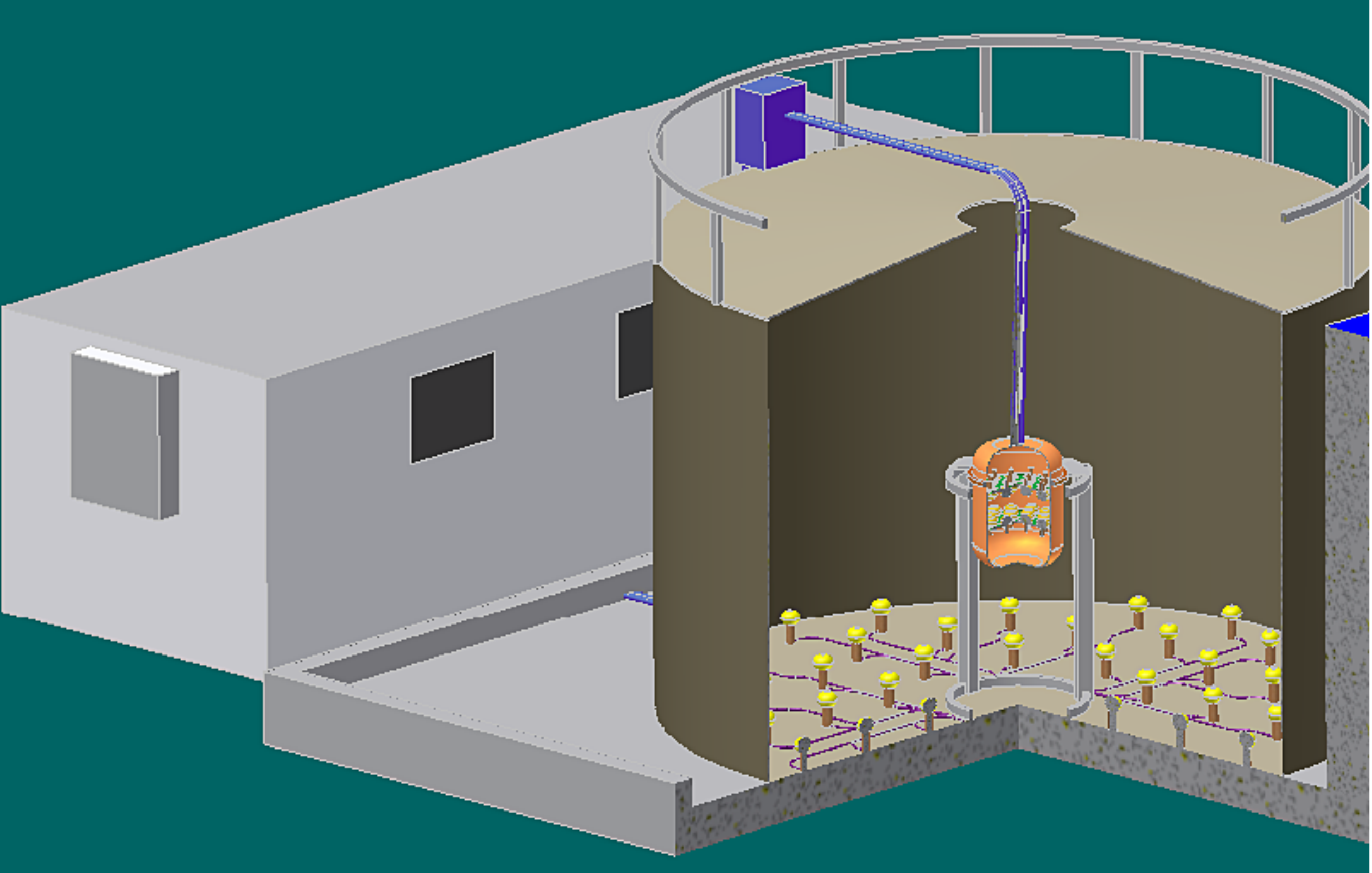}
  \caption{CLEAR experiment concept.  The cryogenic inner
    detector enclosed in a vacuum vessel will be positioned inside a tank
    of water, which provides neutron shielding and an active muon veto
    by detection of Cherenkov radiation with an array of PMTs. Image credit: J. Fowler.
   }\label{fig:overview}
\end{figure}

A specific detector proposed to accomplish these goals was designed,
called CLEAR (Coherent Low
Energy A (Nuclear) Recoils~\cite{Scholberg:2009ha}).  This concept employed a single-phase design to allow
interchangeable noble liquid target materials.  Multiple targets are desirable to test
for physics beyond the SM. 
This design comprises an inner noble-liquid detector
placed inside a water tank. The water tank instrumented with
PMTs acts as a cosmic ray veto. An overview diagram
of the experiment is shown in Fig.~\ref{fig:overview}.

\begin{figure}[!ht]

  \centering
    \includegraphics[height=2.0in]{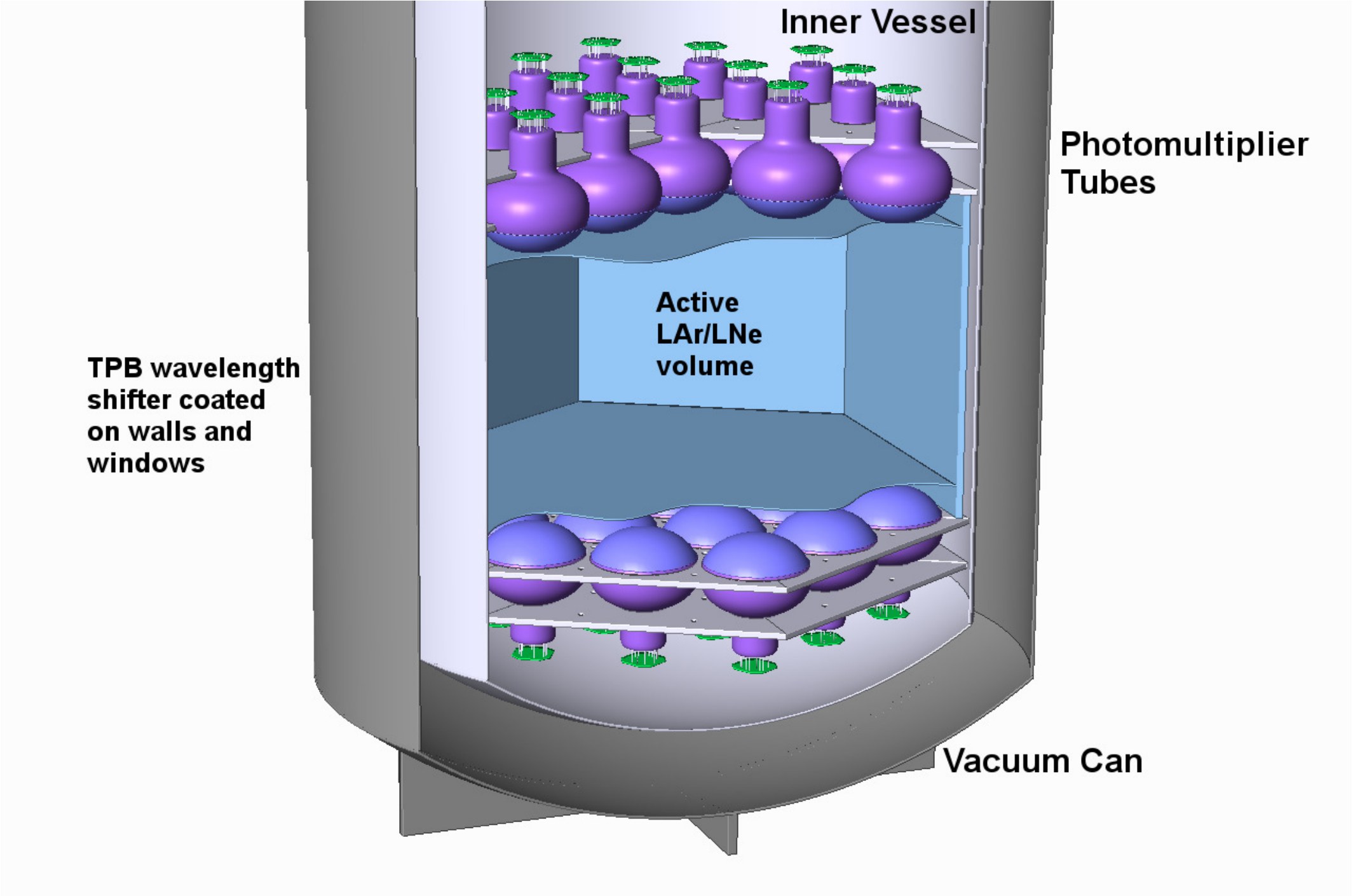}
  \caption{The inner detector, containing an active target of LAr or 
   LNe viewed by photomultipliers, as described in the text.  Image credit: J. Nikkel.
   }\label{fig:inner_det} 

\end{figure}

The CLEAR design uses liquid argon (LAr) and/or liquid neon (LNe) as the
detector materials.  LAr and
LNe are bright scintillators, comparable in light yield to NaI but
with a faster response. Several properties of LAr and LNe make this overall
approach attractive.  First,
LAr and LNe scintillate strongly in the vacuum ultraviolet and
are transparent to their own scintillation light, allowing for event
detection with a low energy threshold. LAr and LNe are dense
enough (1.4 and 1.2 g $\rm cm^{-3}$, respectively) to allow
significant target mass in a modest detector volume.  Pulse
shape discrimination (PSD) 
to select nuclear recoils is possible because both LAr and LNe have
two distinct mechanisms for the emission of scintillation light. 
These two scintillation channels, resulting from singlet molecule
decay and triplet molecule decay, have very different fluorescence
lifetimes and are populated differently for electron recoils than for
nuclear recoils.  This allows nuclear recoils and electron recoils to
be distinguished on an event-by-event basis.  This approach to
electron recoil discrimination has been proposed
for liquid neon~\cite{McKinsey:2004rk}, and 
for liquid argon~\cite{Boulay:2006}.  Demonstrations of 
discrimination in the energy window of interest have been
accomplished in the MicroCLEAN~\cite{Nikkel:2008}, DEAP-I~\cite{Boulay:2009zw}, 
and WARP~\cite{Brunetti:2005} experiments.
The ability to exchange LAr with LNe, with different sensitivities to
coherent neutrino scattering and fast neutrons, would allow both event
populations to be distinguished and characterized.  
Finally, argon and neon are relatively inexpensive detector materials.  

The CLEAR design, summarized from reference~\cite{Scholberg:2009ha}, is cylindrical LAr/LNe scintillation
detector, with an active LAr (LNe) mass of 456 (391)~kg.  
The active volume is about 60~cm in diameter, and 44~cm tall.
A schematic of the active detector is shown in
Fig.~\ref{fig:inner_det}.  The central active mass is viewed by
38 Hamamatsu R5912-02MOD PMTs divided into two
arrays, one on the top of the active volume
facing down, and the second array on the bottom 
facing up.  All PMTs are completely immersed in the
cryogenic liquid.  A cylinder of polytetrafluoroethylene (PTFE) will define the outer radius of the
active volume. The bottom and top of the active volume is defined by 
two fused silica or acrylic plates.  
Ionizing radiation
events in the liquid cryogen will cause
scintillation in the vacuum ultraviolet (80~nm in LNe
or 125~nm in LAr), which is too short to pass through 
the PMT glass.  
The inner surface of the PTFE walls and end plates must be
be coated with a thin film 
of tetraphenyl butadiene (TPB) wavelength shifter.
The ultraviolet scintillation light is
absorbed by the wavelength shifter and re-emitted at
a wavelength of 440~nm. The photon-to-photon conversion efficiency is about 100\% for LAr scintillation
and about 130\% for LNe scintillation~\cite{McKinsey:1997}. The wavelength-shifted light is then detected
by the PMTs.   Work with the MicroCLEAN detector has verified that the chosen PMT model can be used immersed in
LAr or LNe.

In this design, the detector is contained in a stainless steel vacuum cryostat. 
A pulse-tube refrigerator, mounted near the water tank, provides
cooling power to maintain the active fluid at the desired temperature
value.  The noble gas is continuously circulated, boiled,
purified and re-liquefied during operation to maintain a sufficiently
large light yield and triplet molecule lifetime.  Molecular impurities
that affect light collection are removed using gas-phase
recirculation through a commercial heated getter.  
This is 
the same approach used in the XENON~\cite{Angle:2008we} and LUX~\cite{LUX} experiments.

The tank used can be a standard agricultural water tank. 
Instrumented with 
PMTs, the tank can also
serve as a cosmic ray muon veto.  
Geant4~\cite{Agostinelli:2002hh,Allison:2006ve} simulations for CLEAR showed that
excellent cosmic veto efficiency is obtained with at least 20~PMTs,
and a configuration in which
all PMTs are placed on the bottom of the tank is near-optimal.

	Beam coincident neutrons are the most worrisome, as they cannot be measured off-beam-pulse.  In order to properly understand the neutron fluxes at the potential experimental sites, an extensive simulation of neutron flux at 20~m away from the target was carried out~\cite{Scholberg:2009ha}. The expected neutron flux is less than 0.1 neutrons/cm$^2$/s at 1~MW of beam operation.  However extensive measurements will be needed to validate the simulations.   For an initial R\&D effort, we propose to carry out background estimations of the beam induced neutrons (1) using a kg-scale commercial liquid-scintillator for an initial test which will provide operational experience of running a detector at the SNS site, and (2) 10-kg size of prototype liquid argon detector to measure neutron fluxes in the energy range of the region of interest.  

\begin{figure}[!htbp]

\centering
\includegraphics[height=2.6in]{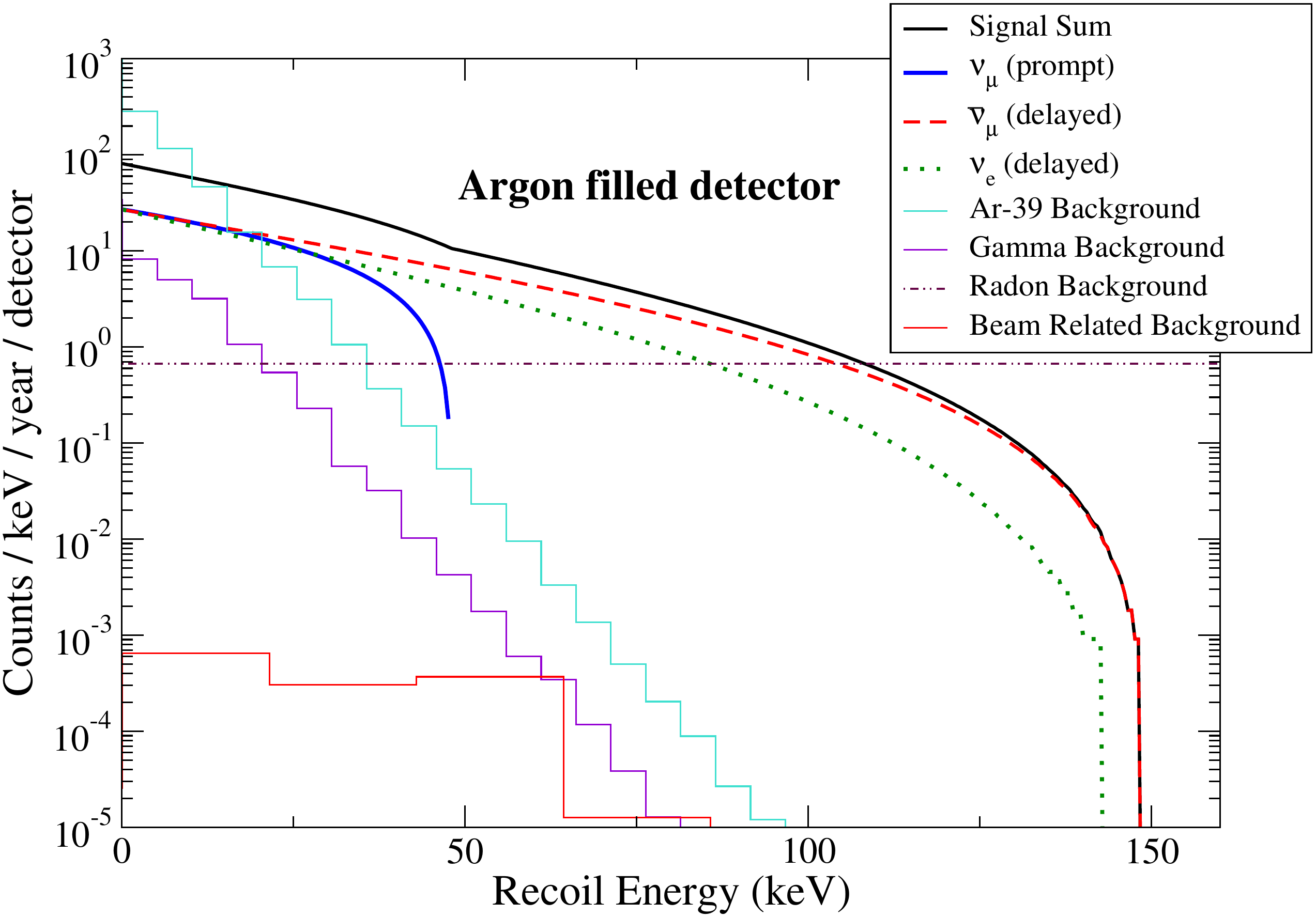}
\caption{Number of events in both neutrino signals along with beam-
and detector-related backgrounds, for a LAr-filled detector (456~kg).}
\label{fig:Ar_sig_bg}
\end{figure}

\begin{figure}[!htbp]
\centering
\includegraphics[height=2.6in]{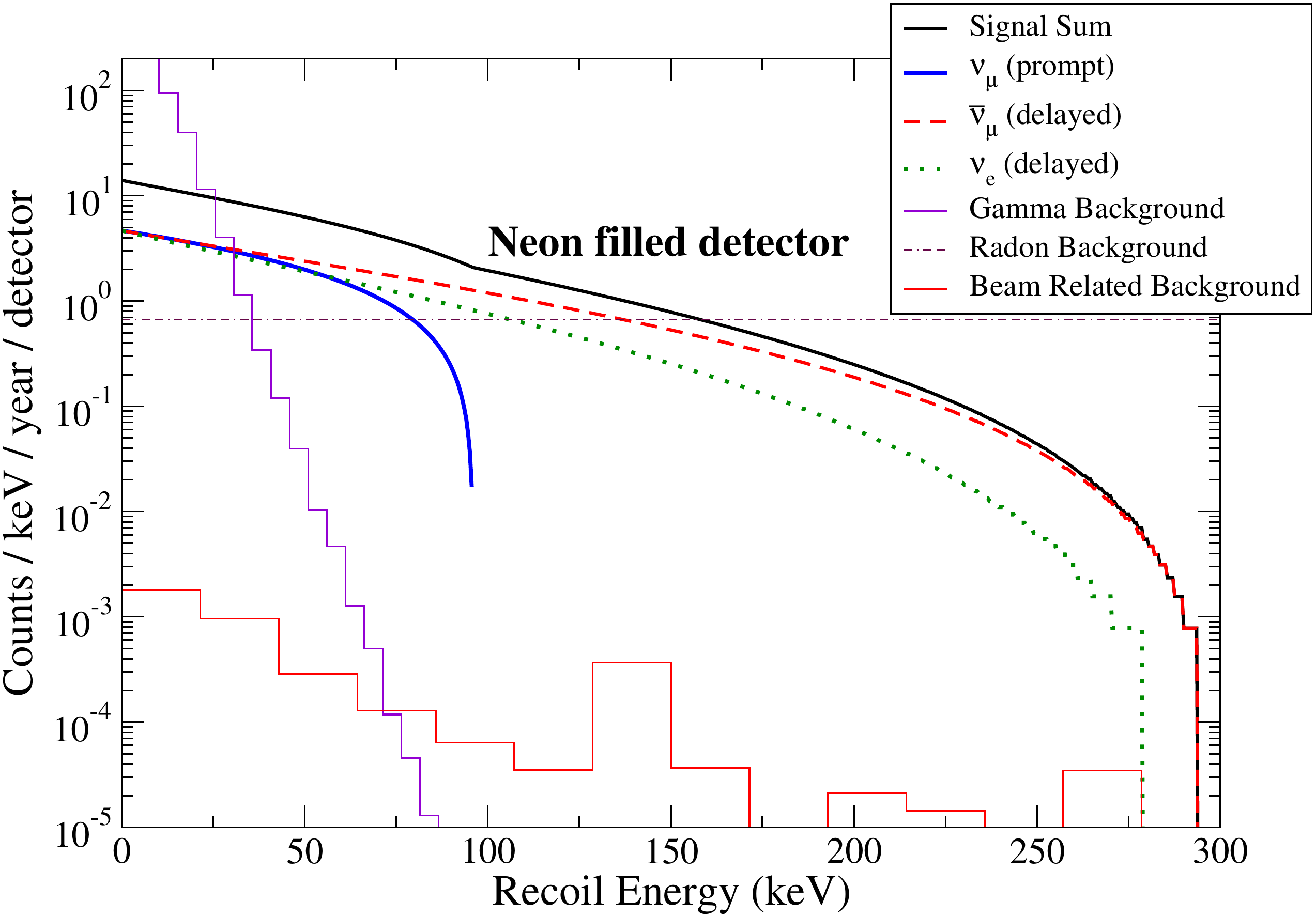}
\caption{Same as Fig.~\ref{fig:Ar_sig_bg} for a LNe-filled detector (391 kg).}
\label{fig:Ne_sig_bg}
\end{figure}

Assuming
the SNS is
running at its full 1.4 MW power, a live running time of
2.4$\times10^7$ s/yr for each of LAr and LNe,
a nuclear recoil energy window between
20-120~keV (30-160~keV), and a 456 (391) kg LAr (LNe) target, we will
have about 890 (340) signal events from the muon decay flux, and about
210 (110) signal events from the prompt $\nu_\mu$ flux.  Backgrounds for
the $\nu$ signal detection come from neutrons (cosmic and SNS-related) and
misidentified $\gamma$s: these are detailed in reference~\cite{Scholberg:2009ha} and are summarized in Figs.~\ref{fig:Ar_sig_bg} and~\ref{fig:Ne_sig_bg}.

\vspace{0.1in}

\vspace{0.1in}

\subsubsection{Experiments for Coherent Scattering on Xenon}

Since 1992, the employment of a liquid xenon detector has been considered for probing anti-neutrino/electron scattering cross sections for deposited energies below 100 keV, where the cross section for elastic scattering of a neutrino with a magnetic moment of 10$^{-11}\mu_B$ may exceed the electroweak cross section, due to additional magnetic dipole-dipole scattering~\cite{Cline:1991xi}. Such an experiment could be performed with a moderately-sized ($\sim$1 tonne mass)  LXe emission detector~\cite{BaldoCeolin:1992yj} and with an artificial neutrino source  could achieve a sensitivity of $< 3\times10^{-12}\mu_B$. 
The emission method of particle detection invented 40~years ago at the Moscow Engineering Physics Institute (MEPhI) Department of Experimental Nuclear Physics~\cite{Dolgoshein:1970} allows an arrangement of a ``wall-less'' detector sensitive to single ionization electrons~\cite{Bolozdynya:1995}. 

\begin{figure}
\vspace{5mm}
\centering
\includegraphics[width=7.5cm]{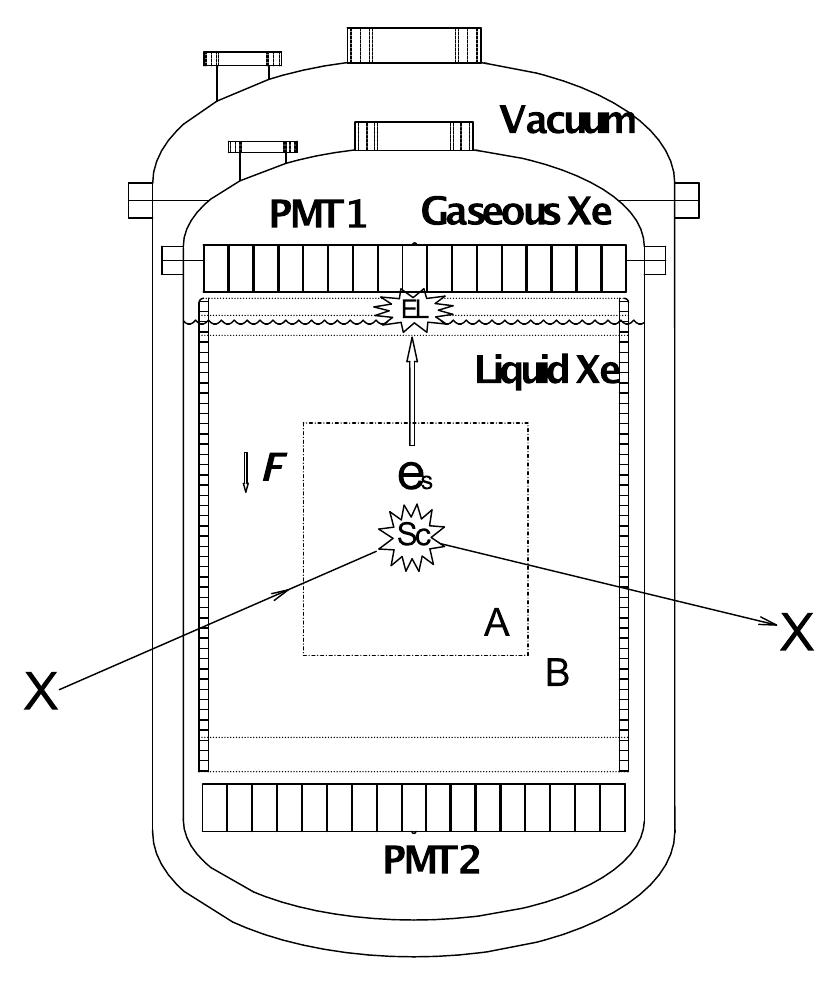}
\vspace{1mm}
\caption{
Principle of operation for a “wall-less” liquid xenon emission detector detecting hypothetical weakly-interacting particle X. Sc:  scintillation flash generated at the point of primary interaction between X and Xe atoms; EL: electroluminescence flash of gaseous Xe excited by electrons extracted from liquid Xe by electric field F and drifting through the gas at high electric fields ($>$1 kV/cm/bar); PMT1 and PMT2: arrays of photodetectors detecting Sc and EL signals; A: the fiducial volume where events considered to be useful occur; B: the shielding layer of LXe. The active volume of the detector is surrounded with highly reflective cylindrical PTFE reflector embodied with drift electrode structure providing a uniform field F. The detector is enclosed in a vacuum cryostat made of low-background pure titanium.
}
\label{fig:xenon1}
\end{figure}

\begin{figure}
\vspace{5mm}
\centering
\includegraphics[width=9.5cm]{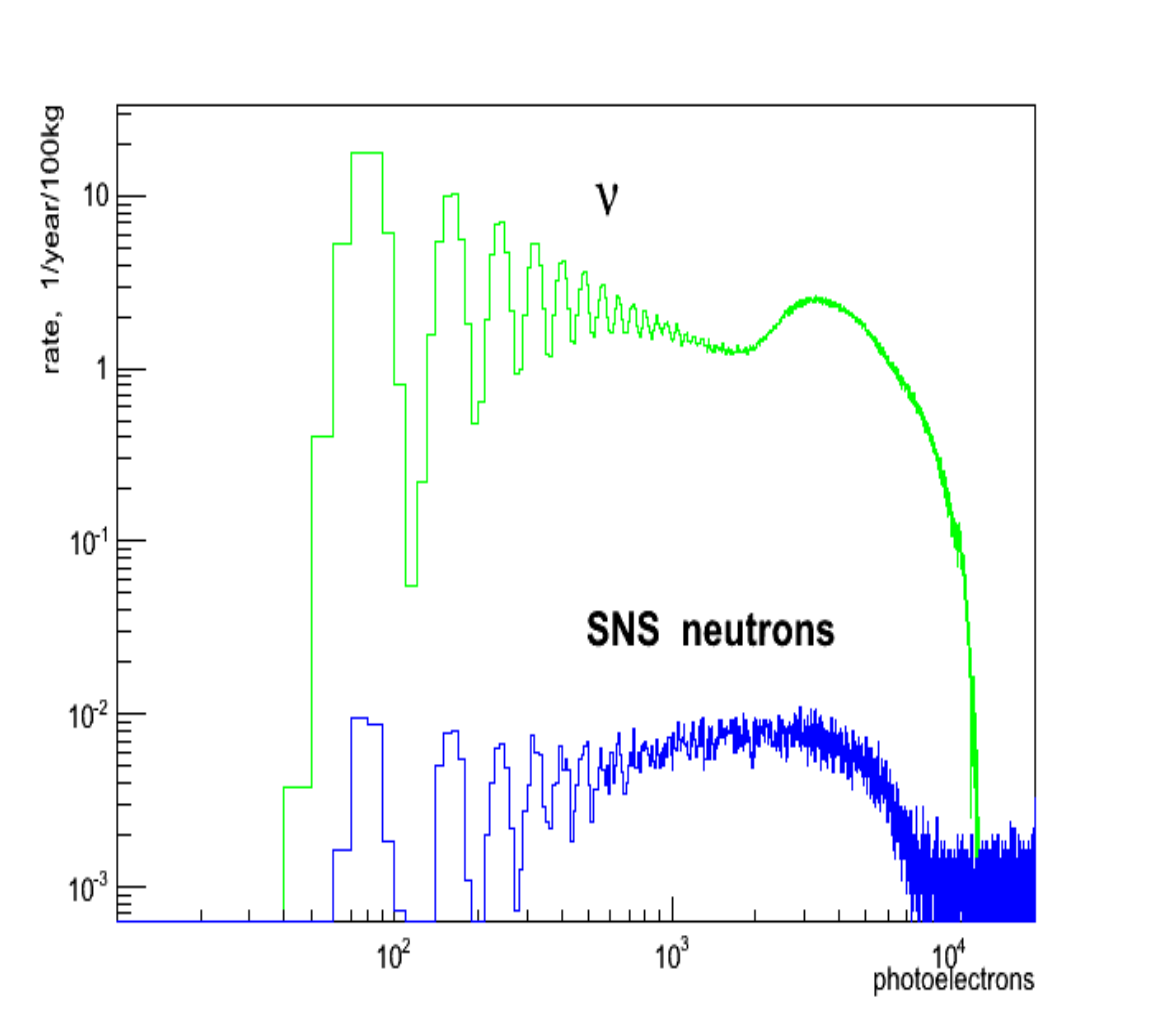}
\vspace{1mm}
\caption{
Signal and background for the RED100 LXe detector at the SNS.
}
\label{fig:xenon2}
\end{figure}

The wall-less detector works as follows (Fig.~\ref{fig:xenon1}): 
first, radiation interacts with the condensed target medium, exciting and ionizing atoms; this process generates a prompt signal that manifests itself in the form of scintillation in noble liquids and solids, phonons in crystals, and rotons in superfluid helium. This signal may serve as a trigger.
Next, in response to the applied external electric field, ionization electrons drift to the surface of the condensed medium and then escape into the rarefied gas or vacuum region (or superconductive collector, for cryogenic crystal targets) and generate a second, amplified signal. Different processes can be used for signal amplification: electroluminescence of the gas phase, electron avalanche multiplication in a low-density gas, acceleration of electrons in vacuum, the Trofimov--Neganov--Luke effect in cryogenic crystals, breaking of Cooper pairs and generation of a pulse of quasi-particles in superconductors, \textit{etc.}~\cite{Bolozdynya:2010zz}. An array of sensors is used to measure the two-dimensional distribution of the secondary particles and to determine the coordinates of the original event on the plane of the sensor array. Since the second signal is delayed with respect to the first one, the third coordinate of the original interaction is also uniquely determined. 
Next, from the three-dimensional position reconstruction, a fiducial volume can be defined (A, Fig.~\ref{fig:xenon1}). Events originating in the vicinity of the detector walls can be eliminated as being potentially associated with radioactive background radiated from the surrounding materials. By making the detector sufficiently large and choosing a target medium with a high stopping power for nuclear radiation, the fiducial volume is effectively shielded by the outer detector medium layer (B, Fig.~\ref{fig:xenon2}). Layer B can be used as active shielding to reject events in fiducial volume A correlated with detection interactions in layer B.  This configuration allows rejection of events associated with multiple scattering background particles.  
Next, analysis of the redistribution of energy deposited by detected particles between ionization (EL signal in Fig.~\ref{fig:xenon1}), photon- and phonon excitations (Sc signal in Fig.~\ref{fig:xenon1}) improves the efficiency of background suppression. 
The above described features, along with the availability of super-pure noble gases in large amounts, make condensed noble gases the most attractive media for emission detectors of rare events.  

We note that there are other detector technologies that can be used to construct ``wall-less'' detectors. For example, bulk scintillators viewed by a photo-detector array totally surrounding the ``crystal-ball'' have been considered as ``wall-less'' detectors for such experiments as XMASS~\cite{Liu:2012et} and CLEAN~\cite{McKinsey:2005b}. However, emission detectors based on pure noble gases require fewer readout channels and allow detection in different channels: the first signal is proportional to the excitation of the condensed medium; the second is proportional to the ionization. Since the efficiency of different modes of dissipation of the deposited energy depends on the nature of the interactions, multi-mode readout helps distinguish events of different origin and effectively suppresses the background.

To measure the ionization yield of heavy nuclear recoils, an experiment is underway that will model the detection of xenon nuclear recoils through a study of the elastic scattering of a monochromatic filtered beam of neutrons from the IRT MEPhI 
reactor with a 5-kg LXe emission detector recently used for detection of single electrons~\cite{enpl}.  Quasi-monochromatic neutron beams with a half-width energy 1.5-3 keV in the range from 1.86 to 133.3 keV will be formed by transmitting reactor neutrons through 2-2.5~m thick materials with deep interference minima in the total neutron cross sections to imitate nuclear recoils from coherently-scattered reactor neutrinos.  In this experiment, the method of detection of single-electron events and their separation from the single-electron background will be tested. If successful, the low-energy thresholds of current dark matter experiments can be improved (for comparison, the detection threshold of XENON10 in the most successful run was about 4 keV).

For observation of neutrino coherent scattering at the SNS we consider the emission detector RED100 with a 100-kg LXe working medium. The basic parameters for construction of this detector are the following:
a low-background titanium cryostat, a readout system based on two arrays of low-background PMTs (\textit{e.g.} Hamamatsu R11410) located in the gas phase and in the liquid below the grid cathode, and a PTFE light-collection system embodied with a  drift electrode system as shown in Fig.~\ref{fig:xenon1},
a cryogenic system based on thermo-siphon technology similar to that used in the LUX detector, and
a location in a borehole 10~m underground at 40~m distance from the target.
For these conditions, the expected count rates for neutrino signals and major background associated with scattered neutrons are shown in Fig.~\ref{fig:xenon2}.  The total event rate is 1470 events/year.

\subsection{Experiments for Hidden Sector Measurements}\label{hidden_experiments}

\begin{figure}
\centering
\includegraphics[width = .8\linewidth]{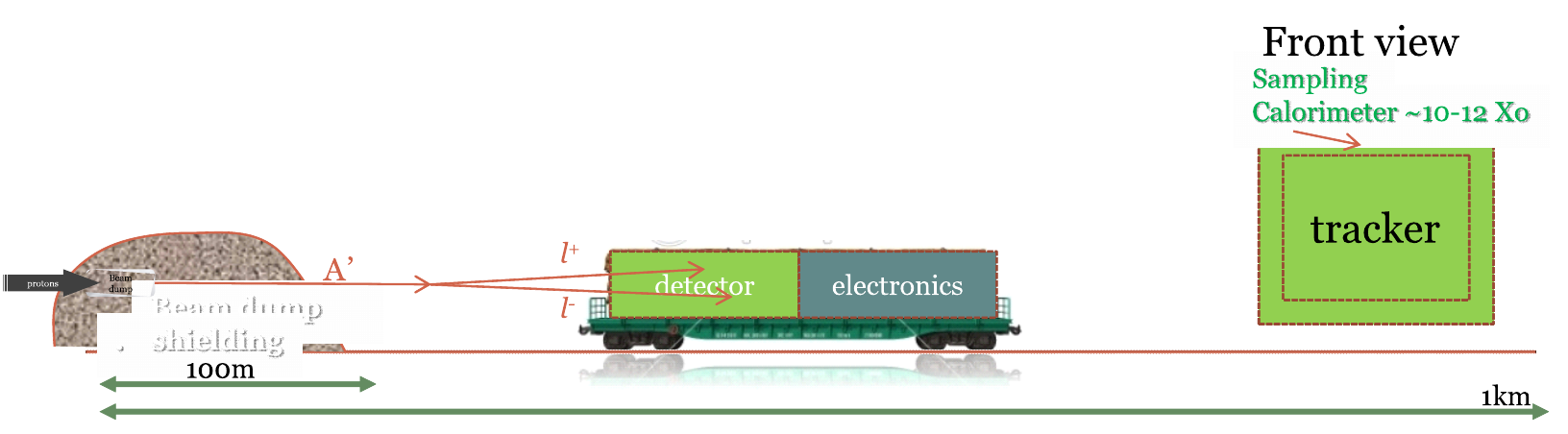}
\caption{ Conceptual design of a movable detector  behind a beam-dump, but in the beam direction.  Included is a cross-sectional view of the detector concept with a tracking system in the center and a sampling calorimeter of 10 to 12 radiation lengths surrounding it.} 
\label{fig:hidden2}
\end{figure}

We describe in this section a concept for an experiment to address the physics described in Section~\ref{hidden_physics}.  An SNS-based detector at any distance in the practical range will cover the intermediate lifetime range of HS particles. A movable detector would further improve sensitivity for each mass by scanning more lifetime ranges instead of a single value.  A combination of a tracker with a surrounding calorimeter has good vertex and mass reconstruction capabilities (see Fig.~\ref{fig:hidden2}) that allow the scan of decay vertices close to the interaction point as well as within the detector itself, extending the sensitivity of the system.  Such a configuration also enables reconstruction of slower/heavier particles, further increasing the scanned phase space. The signatures of these particle decays in the detector can produce an excess of NC scattering events as well as di-particle signals with a vertex in the line of the incoming initial proton of the beam-line. The pairs can be either electron-positron, muon, or pion pairs depending on the mass of the paraphoton generated at the proton target/beam-dump material interaction. The analysis of these signatures is model independent. The mass of the paraphoton can be reconstructed from the invariant mass of the pair. Independently, the measured signal excess at a given reconstructed paraphoton mass can provide its coupling strength to ordinary particles; this gives a measure of the particle lifetime.

The relatively small size of the proposed detector  of Fig.~\ref{fig:hidden2}  implies that it can be magnetized by a large constant magnetic field, or preferably  a magnetic field pulsed at the accelerator frequency,  producing higher-than-few-Tesla fields within the detector's fiducial volume. Funding (material choices) and technology limits define how closely we can reach the B-L conditions existing in the much higher density and temperature of the Sun's core, in an attempt to force these particles to interact rather than simply waiting for them to decay within the detector. 
 A reasonable first step forward is an accurate Monte Carlo calculation that  will incorporate the required branching, production cross sections, kinematics, and decay-length distributions of the mediating particles and include a detailed detector acceptance representation. The calculation will then determine optimum distances from the target/beam-dump for a specialized near-detector type, taking into account intermediate energies and the tight angles at which the possible hidden sector particles may be emitted.  Moreover, various models from the literature must be reviewed in order to determine reasonable coupling strengths of these particles to ordinary matter within the target medium, as opposed  to the Sun medium, and to evaluate the use of high magnetic fields to increase the interaction probability of paraphotons in the detector fiducial volume. 

\section{Conclusion}

We have outlined in this document a number of physics motivations for taking advantage of the extremely high-quality stopped-pion neutrino source available at the Spallation Neutron Source available at Oak Ridge National Laboratory, as well as some specific experimental configurations that could address the physics.

\section*{Acknowledgements}

We thank the hosts of the workshop,  the
Fundamental Symmetries Consortium in coordination with the ORNL Physics Division.  We also thank all of the workshop participants.

\bibliographystyle{vitae}
\bibliography{refs,apj_journals,add_journals,supernova,rspn_process}

\end{document}